\documentclass[Journal]{IEEEtran}
\IEEEoverridecommandlockouts
\usepackage{graphicx, tikz, cite}
\usepackage{verbatim}
\usepackage{amsmath}
\usepackage{makeidx}
\usepackage{amsthm}
\usepackage{amssymb}
\usepackage{epstopdf}
\usepackage{multirow}
\usepackage{array}
\usepackage{makecell}
\usepackage{color}
\usepackage{hyperref}
\usepackage{graphicx, psfrag, minibox, mathrsfs, bigints, stfloats, color, upgreek, gensymb, textcomp, euscript, calligra, xcolor, pict2e}
\date{}
\usepackage{rotating}
\usepackage{diagbox}
\usepackage{slashbox}
\usepackage{tikz}
\usetikzlibrary{shapes,arrows}
\usetikzlibrary{automata,chains}
\usepackage{textcomp}
\usepackage{algorithmicx}
\usepackage{algpseudocode}
\usepackage[ruled]{algorithm2e}
\usepackage{setspace}
\usepackage{float}
\usepackage[font=scriptsize]{caption}
\usepackage{subcaption}
\usepackage{enumerate}
\usepackage{eso-pic}
\definecolor{red}{rgb}{0.8,0.1,0.1}
\definecolor{blue}{rgb}{0,0,1}
\definecolor{green}{rgb}{.1,.6,.3}
\newcommand{\red}{\color{red}}
\newcommand{\blue}{\color{blue}}
\newcommand{\green}{\color{green}}
\makeatletter
\newcommand{\leqnomode}{\tagsleft@true}
\newcommand{\reqnomode}{\tagsleft@false}
\makeatother
\begin{document}
\title{On Distributed Detection in EH-WSNs With  Finite-State Markov Channel  and Limited Feedback}
\author{\IEEEauthorblockN{Ghazaleh Ardeshiri, 
		Azadeh Vosoughi~\IEEEmembership{Senior Member,~IEEE}}
	\IEEEauthorblockA{University of Central Florida\\  Email:gh.ardeshiri@knights.ucf.edu,  azadeh@ucf.edu} }
\maketitle

\begin{abstract}

We consider a network,  tasked with solving binary distributed detection, consisting of $N$  sensors, a fusion center (FC), and a feedback channel from the FC to sensors. Each sensor is capable of harvesting energy and is equipped with a finite size battery  to store randomly arrived energy. Sensors process their observations and transmit their symbols to the FC over orthogonal fading channels. The FC fuses the received symbols and makes a global binary decision. We aim at developing adaptive channel-dependent transmit power control policies such that $J$-divergence based detection metric is maximized  at the FC, subject to total transmit power constraint. Modeling the quantized fading channel, the energy arrival, and the battery dynamics as time-homogeneous finite-state Markov chains,
and the network lifetime as a geometric random variable, we formulate our power control optimization problem as a discounted infinite-horizon constrained Markov decision process (MDP) problem, where sensors’ transmit powers are functions of the battery states, quantized channel gains, and the arrived energies. We utilize stochastic dynamic programming and Lagrangian approach to find the optimal and sub-optimal power control policies. We demonstrate that our sub-optimal policy provides  a close-to-optimal performance with a reduced computational complexity and without imposing signaling overhead on sensors. 
\end{abstract}
%%%%%%%%%%%%%%%%%%%%%%%%%%%%%%%%%%%%%%%%
%\begin{comment}
%\end{comment}
%%%%%%%%%%%%%%%%%%%%%%%%%%%%%%%%%%%%%%%%%%
%%%%%%%%%%%%%%%%%%%%%%%%%%%%%%%%%%%%%%%%%%%%%%%%
\section{Introduction}
Event detection for smarter cities, healthcare systems, farming, and  greenhouse environmental monitoring systems is one of the vital tasks in wireless sensor networks (WSNs) \cite{Ciuonzo2021,ahmadi} and the Internet of things (IoT)-based WSNs \cite{kuodesign,aldecision}. 
The classical studies of binary distributed detection in 
\cite{varshney1996distributed,tsitsiklis1988,viswanathan1997} cannot be directly applied to these networks, since the classical works are based on the assumption that the rate-constrained communication channels between distributed sensors and fusion center (FC) are {\it error-free}.
This has motivated researchers to study {\it channel-dependent} local decision rules for sensors and decision fusion rules for the FC, such that the effect of wireless communication channels in WSNs is integrated into the system designs \cite{Varshney2,niu2006fusion,zahra,ahmadi1}. 
%
%Z. Hajibabaei, A. Vosoughi and N. Mastronardea, “Optimal power allocation for M-ary distributed detection in the presence of channel uncertainty,” Elsevier Signal Processing, vol. 169, April 2020.
%
%H. Ahmadi and A. Vosoughi, “Impact of wireless channel uncertainty upon distributed detection systems,” IEEE Transactions on Wireless Communications, vol. 12, no. 6, pp. 2566-2577, June 2013.
%
Even for channel-dependent distributed detection system designs, providing a guaranteed detection performance by a conventional WSN, in which sensors are powered by conventional non-rechargeable batteries and become inactive when the stored energy in their batteries is exhausted, is impossible.
Although adaptive signal transmission strategies, including optimal channel-dependent power control \cite{vin,Goodman,ahmadi2018power} 
%
% H. Ahmadi, N. Maleki and A. Vosoughi, “On power allocation for distributed detection with correlated observations and linear fusion,” IEEE Transactions on Vehicular Technology, vol. 67, no. 9, pp. 8396-8410, September 2018.
%
and censoring \cite{maleki1,cheng2010}, can enhance the energy efficiency and increase the lifetime of a conventional WSN, they cannot change the fact that  the network lifetime  is inherently limited.
This limited lifetime  disrupts the event detection task
%due to discontinuities in sensor-FC communication, 
and drastically degrades the detection performance. 
%{\blue Motivated by this challenge,
%researchers have developed novel adaptive signal transmission
%strategies that improve the energy efficiency and hence extend
%the lifetimes of these networks. }

%Thus, energy harvesting (EH) of such tiny IoT devices will provide solutions to the energy problem and all its inherent limits %\cite{khelifieh}.

%from abstract: Energy harvesting is a promising paradigm in the Internet of things (IoT)-based wireless sensor networks (WSNs) for emerging applications such as 
%
%Although the amalgamation of energy harvesting in WSNs (EH-WSNs) fostered many new opportunities, they still face challenging requirements to achieve high sustainability.
%
%The Internet of Things (IoT) envisages billions of tiny
%devices with sensing, computation, and communication
%capabilities to be used in everyday life and currently represents a game-changing technology for the wireless communications and sensing sector \cite{kuodesign,aldecision}. 
%
%However, the growing popularity of these IoT devices is hampered by their limited lifetime suffering from their restricted energy sources. Batteries are considered to be their primary power source.
%Nevertheless, changing these energy-limited batteries might be difficult especially in distant and inaccessible areas, due to issues related to environmental hazards and maintenance complexity. Thus, energy harvesting (EH) of such tiny IoT devices will provide solutions to the energy problem and all its inherent limits \cite{khelifieh}.

Energy harvesting (EH) from the environment is a promising solution
to address the energy constraint problem in conventional WSNs,
and to render these networks to self sustainable networks
with perpetual lifetimes. The new class of EH-powered WSNs, where nodes have EH capability and are equipped with rechargeable batteries, will be also important for the development of IoT-based WSNs. 
In EH-powered WSNs power/energy management is necessary, in order to balance the rates of energy harvesting and energy consumption for transmission.
If the energy
management policy is overly aggressive, sensors may stop
functioning, due to energy outage. 
On the other hand, if
the policy is overly conservative, sensors may fail to utilize
the excess energy, due to energy overflow, leading into a
performance degradation. 
EH has been also considered in the contexts of  data communications \cite{wang,advances}, cognitive radio systems \cite{yazdani2020}, 
distributed estimation \cite{nourian}, and distributed detection  \cite{GengJun,Gupta,Tarighati, Duan,Ardeshiri,Ard2,Ard3}.
The body of research on EH-enabled communication systems can be grouped into two main categories, depending %zenaidi2016,
on the adopted energy arrival model \cite{advances,Ozel}: 
%
%{\red I removed reference \cite{zenaidi2016} from here} %
in deterministic models the transmitter has full (causal and non-causal) knowledge of energy arrival instants and amounts at the beginning of transmission, in {\it stochastic models}, suitable for modeling ambient RF and renewable energy sources that are intrinsically time-variant and sporadic, the transmitter only has causal knowledge of energy arrival.  In addition, wireless communication channels change randomly in time due to fading. 
These together prompt the need for developing new  power control/energy scheduling strategies for an EH-enabled transmitter that can best exploit and adapt to the random energy arrivals and time-varying fading channels. 
%
%{\blue 
%In practice, the energy arrival of ambient energy sources, including ambient RF signal sources, is intrinsically time-variant and often sporadic. 
%This natural factor degrades the performance of the battery-free EH-enabled communication systems in which a “harvest-then transmit” strategy is adopted, i.e., users can only transmit
%when the energy harvested in one time slot is sufficient
%for data transmission \cite{yazdani2020}. }
%
%
%%%%%%%%%%%%%%%%%%%%%%%%%%%%%%%%%%%%%%%%%%%%%
%
%
%There are two approaches regarding energy management policy designs, depending on whether the knowledge of channel state information (CSI) and energy state information (ESI) is available non-causally or causally at the beginning of transmission.

Designing power control/energy scheduling schemes corresponding to random energy arrivals and time-varying fading channels can be viewed as a {\it multistage stochastic optimization problem}, where the goal is to find a sequence of decisions a decision maker has to make, such that a specific metric over a horizon spanning several
time slots is optimized (e.g., optimizing transmission completion time, data throughput, outage probability, or symbol error rate in a point-to-point EH-powered wireless communication system \cite{advances}). A common approach to solve this sequential decision making problem is to adopt the mathematical framework of Markov decision processes (MDP). 
%
%With the stochastic energy harvesting models
%to acquire the statistical knowledge, stochastic optimization techniques, 
%e.g., Markov decision processes (MDP), are appealing
%solutions to maximize the long-term utilities of relevant optimization problems \cite{advances,Ozel}.
%
%Several metrics (objectives) have been considered in the literature for designing point-to-point EH-powered wireless communication systems, including transmission completion time, data throughput, outage probability, etc \cite{advances}. 
%
%Authors view the energy management policy design as a sequential decision making problem, 
%and hence, they adopt the MDP framework to solve the problem. 
%The goal is typically optimizing a specific metric over a horizon spanning several time slots. 
%The obtained solutions (using dynamic programming) are dependent across time slots, and also depend on the initial state of the battery.
%
The main ingredients of the MDP are states,
%$s \in \mathcal{S}$, 
actions,
%$a \in \mathcal{A}$,$R_a(s)$ $\Pr(s'|s,a)$.
rewards,  and state transition probabilities. 
The state could be a composite states of fading channel, energy arrival,  and battery condition.
The action is  the transmit power level or the amount of energy to be consumed, and 
%The affordable action at the states is limited to the corresponding battery condition. 
 the reward is a function of the states and the actions.
%, which
%could be data throughput, outage probability, symbol error rate (SER), etc., 
The state transition probability describes the transition probability from the current state to the next state with respect to each action. The goal is to find the optimal policy, which specifies the optimal action in the state and maximizes the long-term expected discount infinite-horizon reward  starting from the initial state  \cite{advances}.
In the context of distributed detection in WSNs, there are only few
studies that consider EH-powered sensors \cite{GengJun,Gupta,Tarighati, Duan,Ardeshiri,Ard2,Ard3}. In the following we provide a concise review of these works, highlight how our present work fills the knowledge gap in the literature, and how it is different from our previous works in \cite{Ardeshiri, Ard2,Ard3}.

\subsection{Related Works and Knowledge Gap}
%
%
%{\red the list of references for EH in DD here should match the one in the bold list of previous page, in the blod list it says [13]-[20].}
%
%{\blue It is worth pointing out that, while  the studies in \cite{vin,Goodman} on optimal {\blue channel-dependent transmit}  power control strategies  for distributed detection problem can be applied to  conventional WSNs, 
%they cannot be applied to EH-enabled WSNs. These works have not considered the new challenges related to energy management imposed by the random nature of the energy arrival and  harvesting}
%
Considering an EH-powered node, that is deployed to monitor the change in its environment, the authors in \cite{GengJun} formulated a quickest change detection problem, where the goal is to detect the time at which the underlying distribution of sensor observation changes. 
Considering an EH-WSN and choosing deflection coefficient as the detection performance metric, the authors in \cite{Duan} formulated an adaptive transmit power control strategy based on PHY-MAC cross-layer design. 
%With the competition probability of each sensor, the number of observed samples, the average power of transmitted signals and the average power of detection as decision variables, 
%the optimization of the transmission control scheme was formulated as the maximization of the deflection coefficient that correlates detection and communication in \cite{Duan}.
%
Considering an EH-WSN and choosing error probability as the detection performance metric, the authors in \cite{Gupta} proposed ordered transmission schemes, that can lead to a smaller average number of transmitting sensors, without comprising the detection performance. 
Modeling the randomly arriving energy units during a
time slot as a Bernoulli process, the battery state as a $K$-state Markov chain, and choosing Bhattacharya distance as the detection performance metric, the authors in \cite{Tarighati} have investigated the optimal local decision thresholds at the sensors, such that the detection performance is optimized.
We note the system model in \cite{Duan} lacks a battery to store the harvested energy. Further, the adopted energy arrival model in \cite{Duan} is deterministic. 
On the other hand, \cite{GengJun,Gupta} assumed sensor-FC channels are error-free and \cite{Tarighati} considered a binary asymmetric channel model for sensor-FC links. The high level communication channel model, combined with a simple stochastic model for random energy arrival is limiting.
Specifically, it does not allow one to study channel-dependent
transmit power control strategies. Such a study requires a more
realistic communication channel model and stochastic energy arrival
model that match the energy needed for a channel-dependent
transmission. This is the knowledge gap that we address in this work. 
%

%{\red [16] and [17] are identical papers.}
%
%
%{\blue An adaptive transmission control strategy based on cross-layer design for EH-WSNs was proposed \cite{Duan2022}. Authors studied the problem of optimal average transmitted power and optimal average detection power such that the deflection coefficient of detection and communication is maximized.}
%
%{\red cross layer of which layers? how is it different from yours? what do mean by "deflection coefficient of detection and communication"? I don't understand your text above}
%
To highlight how our present work is different from our previous works in \cite{Ardeshiri, Ard2,Ard3}, we briefly summarize them in the following.
Modeling the random energy arrival as a Bernoulli process, the dynamics of the battery as a finite-state Markov chain, and considering fading channel model, in \cite{Ardeshiri} we adopted channel-inversion transmit power control policy, where allocated power is inversely proportional to fading channel state information (CSI) in full precision, and we found the optimal  decision thresholds at sensors such that Kullback-Leibler (KL) distance detection metric at the FC is maximized. 
%
%assuming each sensor knows its fading channel state information (CSI) in full precision and adapts its transmit power according to the channel-inversion power control policy in wireless communications (i.e., allocated power is inversely proportional to CSI). We found the optimal  decision thresholds at sensors such that Kullback-Leibler (KL) distance detection metric at the FC is maximized. 
%
Different from \cite{Ardeshiri}, in \cite{Ard2} we modeled the random energy arrival as an exponential process and assumed that each sensor only knows its quantized CSI and adapts its transmit power according to its battery state and its quantized CSI, such that $J$-divergence based detection metric at the FC is maximized. 
%
%, we modeled random energy arrival as an exponential process. We assumed each sensor only knows its quantized CSI and adapts its transmit power according to its battery state and its quantized CSI, such that $J$-divergence based detection metric at the FC is maximized. 
%
Modeling the random energy arrival as a Poisson process in \cite{Ard3}, 
we proposed a novel transmit power control strategy that is parameterized in terms of the channel gain quantization thresholds and the scale factors corresponding to the quantization intervals, and found the jointly optimal quantization thresholds and the scale factors such that $J$-divergence based detection metric at the FC is maximized. 
%
%subject to the average transmit power per sensor constrain. 
%We showed that the proposed hybrid search methods in \cite{Ard3} have the lowest computational complexity and provide a close-to-optimal performance. 

Our present work is different from our prior works in \cite{Ardeshiri, Ard2,Ard3} in several aspects. The transmit power control strategies in these works are intrinsically different from our present work, since in \cite{Ardeshiri, Ard2,Ard3}  we have assumed that the battery operates at the steady-state and the energy arrival and channel models are independent and identically distributed (i.i.d) across transmission blocks.  Consequently, the  power optimization problem in \cite{Ardeshiri, Ard2,Ard3} became a deterministic optimization problem, in terms of the optimization variables,  and  the obtained solutions are different. In this work, the battery is not at the steady-state. Also, { both the channel and the energy arrival are modeled as homogeneous finite-state Markov chains (FSMCs)}. Therefore,  the power control optimization problem at hand becomes a  multistage stochastic optimization
problem, and can be solved via the MDP framework. To the best of our knowledge, this is the first  work that develops MDP-based channel-dependent power control policy for distributed detection in EH-WSNs. The MDP framework has been utilized before in \cite{banerjee2021bayesian,premkumar2008optimal} to address a quickest change detection problem. 
%

%=======================
%\\{\blue \cite{Ard4} Asilomar paper}
%-------------------------------------
\subsection{Our Contribution}
%
%
%MDP-based power allocation for DD? {\blue Ghazaleh: I couldn't find any paper for MDP DD. I only found MDP for Change Detection. 1. A Bayesian Theory of Change Detection in
%Statistically Periodic Random Processes, 2. Optimal Sleep–Wake Scheduling for Quickest Intrusion Detection using Sensor Networks, 3. Optimal Detection Task Allocation: A Reinforcement Learning Approach (last one is not change detection!). Do you think we need to mention these papers in the introduction?} {\red yes, when we say we do MDP, we mention that nobody has done MDP for DD and they did for change detection only} {\blue \cite{banerjee2021bayesian,premkumar2008optimal} these are the papers 1,2,3}
%
%{\red what is being optimized in \cite{huang2017optimal}} {\blue It seems that this paper is about Detection task allocation in military. they consider a detection system. In the detection system, there are a set of detection
%equipments and the protected buildings, such as hospital, power plant, etc. Each detection equipment has a detection range based on its property. the objective function is the weighted sum
%of how many detection equipments are
%assigned to detect the same target in average at time t 
%
%I don't think this paper is relevant to detection that we are familiar with!!!!! }

Given our adopted WSN model (see Fig. \ref{Fig-1-network}), we aim at developing  an adaptive channel-dependent transmit power control policy for sensors such that a detection performance metric is optimized. 
We choose the $J$-divergence between the distributions of the detection statistics at the FC under two hypotheses, as the detection performance metric. Our choice is motivated by the fact that $J$-divergence is a widely adopted metric for designing distributed detection systems \cite{vin,Guo,Goodman,Ard3}. We note that $J$-divergence and $P_e$ are related through $P_e>\Pi_0 \Pi_1 e^{-J/2}$, where $\Pi_0, \Pi_1$ are the a-priori probabilities of the null and the alternative hypotheses, respectively \cite{vin,Guo,Goodman}. 
Hence,  maximizing the $J$-divergence is equivalent to minimizing
the lower bound on $P_e$.  
%
%Furthermore, when sensors' observations conditioned on the true hypothesis are statistically independent, the total $J$-divergence of a parallel network at the FC is the summation of the individual $J$-divergence corresponding to individual sensors. Therefore, maximizing  the total $J$-divergence at the FC is equivalent to maximizing the individual $J$-divergence.
%
Modeling the quantized fading channel, the energy arrival, and 
the dynamics of the battery as homogeneous FSMCs, and the network lifetime as a
random variable with geometric distribution, we formulate $J$-divergence-optimal transmit power control problem, subject to total transmit power constraint, 
as a {\it discounted infinite-horizon constrained MDP} optimization problem,
where the control actions (i.e., transmit powers) are functions of the battery state, quantized CSI, and the arrived energy. 
We obtain the optimal and sub-optimal policies
and propose  two algorithms based on value iterations
in the MDP.
%
%
%We model channel state as time-homogeneous finite-state Markov chain (FSMC) and harvested energy as a set of independent stationary first-order homogeneous Markov process. 
%
%
%Our proposed power control policy is parameterized in terms of the each sensor battery state, channel state and harvested energy state. We seek the optimal actions such that the $J$-divergence at the FC is maximized,  subject to total transmit power constraint in each time slot. 
%
Our main contributions can be summarized as follow:

$\bullet$  
Given our adopted system model, we develop the optimal power control policy, using dynamic programming and utilizing the Lagrangian approach to transform the constrained MDP  problem into an equivalent unconstrained MDP  problem.
For the optimal policy, the local action (i.e., a sensor's transmit power) depends on the network state (i.e., all sensors' battery states, quantized CSIs, and the arrived energies), and the computational complexity of the algorithm grows exponentially in number of sensors $N$. Implementing this solution requires each sensor to report its battery state and arrived energy to the FC, which imposes a significant signaling overhead to the sensors. 

%propose the centralized optimal algorithm and by using Lagrangian multiplier, transform the constrained MDP optimization problem into an equivalent unconstrained MDP optimization problem. Then solving the optimization problem by the value iteration in MDP.
   
$\bullet$ To eliminate this overhead, we develop a sub-optimal power control policy, using a uniform Lagrangian multiplier to  transform the constrained MDP problem into $N$ unconstrained MDP  problems. For the sub-optimal policy, the local action  depends on only the local state (i.e., a sensor's battery state, quantized CSI, and the arrived energy), and the computational complexity of the algorithm grows linearly in $N$.

%Unlike the conventional centralized solutions to the MDP which have very high computation complexity and unacceptable communication overheads,

$\bullet$  We numerically study the performance of our proposed  algorithms and showed that the sub-optimal policy has a close-to-optimal performance. 
    
$\bullet$ We study how our system setup and proposed solutions can be extended to the scenario where sensors are randomly deployed in the field.

 %--------------------------------------  
 \subsection{Paper Organization}
The paper organization follows: Section \ref{sym} describes our system and observation models, derives a closed-form expression for the total $J$-divergence and
introduces our constrained optimization problem.  Sections \ref{section3} describes the optimal and the sub-optimal policies. Section\ref{random_dep} discusses how our setup can be extended to the scenario where sensors are randomly deployed in the field (i.e., sensors' locations are unknown a-priori).
Section \ref{simulation} illustrates our numerical results. Section \ref{conclu} concludes our work.
 %and discusses our future direction. 
%%%%%%%%%%%%%%%%%%%%%%%%%%%%%%%%%%%%%%%%%%%%%%%%%%%%%%%%%%%%%%%%%%%%%%%%%%%%%%%%%%%%
\section{System Model}\label{sym}
%------------------------------------
%-------------------------------------------------------------
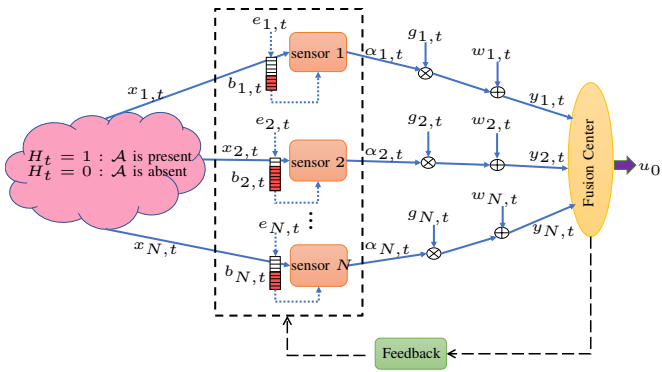
\begin{figure}[!t]
  \centering 
  \hspace{-24mm}
  \vspace{-15mm}
  \scalebox{1.15}{
    \begin{picture}(180,180)
      %\small
      \footnotesize
     
      {    \tiny
        \put(15,50){\includegraphics[width=75mm]{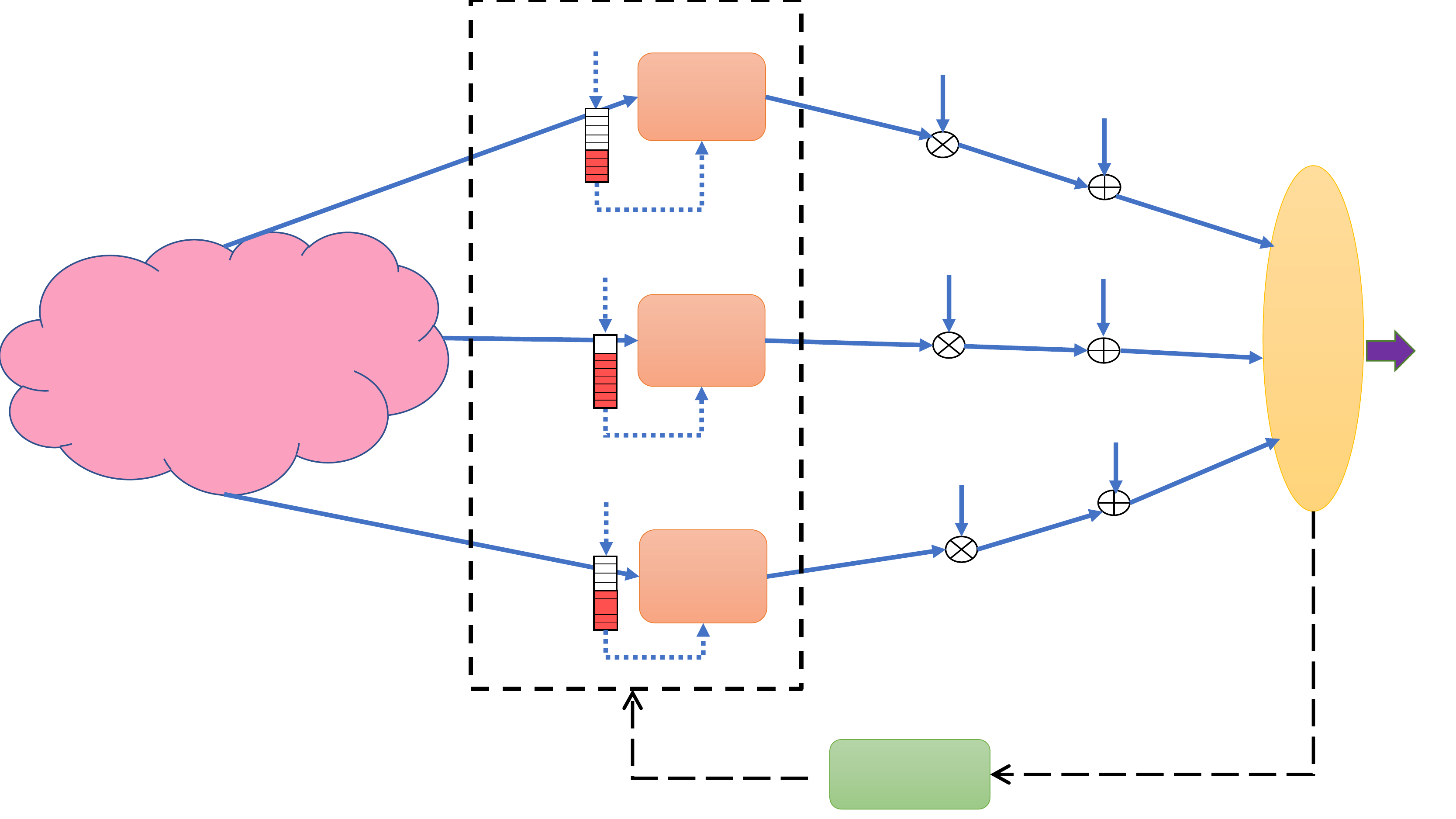}}
        \put(22,120){$H_t=1:  {\cal A}$ is present} 
        \put(22,115){$H_t=0:  {\cal A}$ is absent} 
        \put(205,105){\rotatebox{90} {Fusion Center}} 
        \put(139,55){Feedback}
        \put(223,117){$u_0$}
        
        \put(109,154){sensor $1$}    
        \put(55,142){$x_{1,t}$}
        \put(96,165){$e_{1,t}$}
        \put(88,144){$b_{1,t}$}
        \put(133,155){$\alpha_{1,t}$}
        \put(147,161){$g_{1,t}$}
        \put(168,155){$w_{1,t}$}
        \put(187,140){$y_{1,t}$}
        \put(115,97){\rotatebox{90}{\small ...}}
        
        \put(109,118){sensor $2$}    
        \put(86,123){$x_{2,t}$}
        \put(97,132){$e_{2,t}$}
        \put(89,113){$b_{2,t}$}
        \put(133,122){$\alpha_{2,t}$}
        \put(147,133){$g_{2,t}$}
        \put(168,132){$w_{2,t}$}
        \put(187,121){$y_{2,t}$}
        
        \put(109,84){sensor $N$}    
        \put(57,91){$x_{N,t}$}
        \put(98,98){$e_{N,t}$}
        \put(87,82){$b_{N,t}$}
        \put(133,91){$\alpha_{N,t}$}
        \put(147,101.5){$g_{N,t}$}
        \put(168,107.5){$w_{N,t}$}
        \put(188,97){$y_{N,t}$}
        
      }
    \end{picture}
  }
\caption{Our system model and the schematic of battery state in time slot $t$.} 
\label{Fig-1-network}  
\end{figure}
%-------------------------------------------------------------
%%%%%%%%%%%%%%%%%%%%%%%%%%%%%%%%%%%%%%%%%%%%%%%%%%%%%
\subsection{Observation Model at Sensors}

\begin{figure}[!t]
\centering
\scalebox{0.7}{
\begin{tikzpicture}
\small
\draw [thick, fill=pink!100] (-2,0) rectangle (0.8,0.8);
\node at (-0.5,0.4) {Slot $1$};
\draw [thick, fill=pink!100] (0.8,0) rectangle (3.6,0.8);
\node at (2.2,0.4) {Slot $2$};

\large
\node at (4.2,0.45) {$\ldots\ldots$};

\small
\draw [thick, fill=pink!100] (4.8,0) rectangle (7.6,0.8);
\node at (6.2,0.4) {Slot $t$};

\large
\node at (8.3,0.45) {$\ldots\ldots$};
\draw [-] (-2,0) -- (-2,-0.25);
\draw [-] (0.8,0) -- (0.8,-0.25);
\draw [-] (3.6,0) -- (3.6,-0.25);
\draw [-] (4.8,0) -- (4.8,-0.25);
\draw [-] (7.6,0) -- (7.6,-0.25);
\draw [<->] (-2,-0.15) -- (0.8,-0.15);
\draw [<->] (0.8,-0.15) -- (3.6,-0.15);
\draw [<->] (4.8,-0.15) -- (7.6,-0.15);
\draw [->] (-2,1.3) -- (-2,0.8);
\draw [->] (0.8,1.3) -- (0.8,0.8);
\draw [->] (3.6,1.3) -- (3.6,0.8);
\draw [->] (4.8,1.3) -- (4.8,0.8);
\node at (-0.5,-0.4) {$T_s$};
\node at (2.2,-0.4) {$T_s$};
\node at (6.2,-0.4) {$T_s$};
\node at (-2,1.6) {{$s_{n,1}$}};
\node at (0.8,1.6) { {$s_{n,2}$}};
\node at (3.4,1.6) {{$s_{n,3}$}};
\node at (5.3,1.6) {{$s_{n,t}$}};
\end{tikzpicture}
}
\caption{Our adopted time frame structure for harvesting and transmission.} % title of the Figure
\label{Frame}
\vspace{-4mm}
\end{figure}
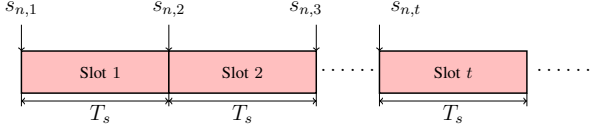

We consider a WSN tasked with
solving a binary distributed detection problem (see Fig. \ref{Fig-1-network}). To describe our signal processing blocks at sensors and the FC as well as energy harvesting model, we divide time horizon into slots of equal length $T_s$. Each time slot is indexed by an integer $t$ for  $t=1,2,...,T$ (see Fig. \ref{Frame}). 
We model the underlying binary hypothesis $H_t$ in time slot $t$ as a binary random variable $H_t \in \{0,1\}$ with a-priori probabilities $\zeta_0=\Pr(H_t=0)$ and $\zeta_1=\Pr(H_t=1)=1-\zeta_0$. We assume that the hypothesis $H_t$ varies over time slots in an independent and identically distributed (i.i.d.) manner. Let $x_{n,t}$ denote the local observation at sensor $n$ in time slot $t$. We assume that  sensors' observations given each hypothesis with conditional distribution $f(x_{n,t}|H_t=h_t)$ for $h_t \in \{0,1\}$ are independent across sensors. This model is relevant for WSNs that are tasked with detection of a known signal in uncorrelated Gaussian noises with the following signal model
%------------------------------------
\begin{align}\label{xk}
&H_{t}=1:~~ x_{n,t} ={\cal A}+v_{n,t},\nonumber\\
&H_{t}=0:~~ x_{n,t} = v_{n,t},~~
\text{for}~n=1,\dots,N,
\end{align}
%------------------------------------
where Gaussian observation noises $v_{n,t} \! \sim \! {\cal N}(0,\sigma_{v_{n}}^2)$ are independent over time slots and across sensors. %Sensor $n$ makes a local binary decision, independent of other sensors, according to a certain local decision rule based on $x_{n,t}$ only.  
Given observation $x_{n,t}$ sensor $n$ forms local log-likelihood ratio (LLR)
%----------------------------------
\begin{equation}\label{lrt_sensor}
 \Gamma_n(x_{n,t})\triangleq\log\left( \frac{f(x_{n,t}|h_{t}=1)}{f(x_{n,t}|h_{t}=0)}\right ),   
\end{equation}
%-------------------------------------
and uses its value to choose its non-negative transmission symbol $\alpha_{n,t}$ to be sent to the FC.
In particular, when LLR is below a given local threshold $\theta_n$, sensor $n$ does not transmit and let $\alpha_{n,t}=0$. When LLR exceeds the given local threshold $\theta_n$, sensor $n$ chooses $\alpha_{n,t}$ according to the available information (will be explained later).
%its battery state, the harvested energy, and the feedback information about its communication channel. 
%
In particular, we have
%------------------------------------------
\begin{align}\label{Pi01}
    &\widehat{\zeta}_{n,0} = \Pr(\alpha_{n,t}\! =\! 0) = \zeta_0(1\!-\!P_{\text{f}_n}) + \zeta_1(1\!-\! P_{\text{d}_n}),\nonumber\\
    &\widehat{\zeta}_{n,1}= \Pr(\alpha_{n,t}\! \neq \! 0) = \zeta_0 P_{\text{f}_n} + \zeta_1 P_{\text{d}_n},
\end{align}
%-----------------------------------
where the probabilities $P_{\text{f}_n}$ and $P_{\text{d}_n}$ can be determined using our signal model in \eqref{xk}
and given the local threshold $\theta_n$
%-----------------------------------
\begin{align}\label{pd_pf1}
 &P_{\text{f}_n} \!= \! \Pr(\alpha_{n,t}\!\neq\!0 |h_t=0)\!=\!Q\Big(\frac{\theta_n+{\mathcal{A}^2}/{2\sigma^2_{v_n}}}{\sqrt{\mathcal{A}^2/{\sigma^2_{v_n}}}}\Big), \nonumber\\
 &P_{\text{d}_n} \!= \!\Pr(\alpha_{n,t}\!\neq \!0 |h_t=1)\!=\!Q\Big(\frac{\theta_n-{\mathcal{A}^2}/{2\sigma^2_{v_n}}}{\sqrt{\mathcal{A}^2/{\sigma^2_{v_n}}}}\Big). 
\end{align}
%---------------------------------------------------
%Suppose $P_{\text{d}_n}$ is required to be fixed at a given value  $P_{\text{d}_n} = \overline{P}_{\text{d}}, \forall n$.  Then the false alarm probability can be written as
Instead of fixing $\theta_n$, one can fix  $P_{\text{d}_n}$ and let $P_{\text{d}_n} = \overline{P}_{\text{d}}, \forall n$. Then the false alarm probability in (\ref{pd_pf1}) can be written as
    $P_{\text{f}_n} = Q\left(Q^{-1}(\overline{P}_{\text{d}})+\sqrt{\mathcal{A}^2/{\sigma^2_{v_n}}}\right)$.
    
Note that sensors are typically deployed in hostile outdoor environments (e.g., for forestry fire and volcano monitoring and detection, and battlefield surveillance) in an unattended and distributed
manner. 
%\cite{akyildiz22002}.  
Therefore,  they are highly susceptible to physical destruction. We include this factor into our modeling by letting $\eta\in[0,1)$ be the probability that  a sensor can survive physical destruction or hardware failure and continue to function in time slot $t$. Defining the network lifetime $T$ as the time until the first sensor fails, we find that
%For simplicity, we assume that $\eta$ is fixed in all allocation intervals \cite{Mao,di2011}. 
 $T$ becomes a geometrically distributed random variable with mean $\mathbb{E}\{T\}\!=\!1/(1-\eta)$ \cite{ Mao}. %{\red can you choose the most relevant one from these three references? we should reduce the number of references}
%---------------------------------------------------
%%%%%%%%%%%%%%%%%%%%%%%%%%%%%%%%%%%%%%%%%%%%
%----------------------------------------

%-----------------------------------
%%%%%%%%%%%%%%%%%%%%%%%%%%%%%%%%%%%%%%%%%%%%%%
\subsection{Battery State, Energy Harvesting and Transmission Models}\label{channel-model}
We assume sensors are equipped with identical batteries of finite size $K$ cells (units), where each cell corresponds to $b_u$ Joules of stored energy. Therefore, each battery is capable of storing at most $K b_u$ Joules of harvested energy. Let $b_{n,t} \in \mathcal{B} \! =\!\{0,1,...,K\}$ denote the energy state of battery of sensor $n$ at the beginning time slot $t$ (also referred to as the battery state). Note that $b_{n,t}\!=\!0$ and $b_{n,t}\!=\!K$ represent energy states of empty battery and full battery, respectively.

Let $e_{n,t}$ be the number of energy units that are harvested and stored at sensor $n$ during time slot $t$, i.e., at the beginning of time slot $t$, sensor $n$
knows the value of $e_{n,t-1}$ but not $e_{n,t}$, and hence
the harvested energy $e_{n,t}$ cannot be used during slot $t$. 
We assume $e_{n,t}$'s are independent across sensors, and
{  model $e_{n,t}$ as a set of independent stationary first-order homogeneous Markov process} with transition probability matrix $\Phi_{\mathcal{E}}$. 
 %{\red Azadeh: what is a "first order" Markov chain?  "stationary Markov chain" and "time-homogeneous Markov chain" synonyms?} {\blue Ghazaleh: A Markov chain with memory (or a Markov chain of order m) the future state depends on the past m states. Time-homogeneous Markov chains are processes where the probability of the transition is independent of t. Stationary Markov chains are processes where $Pr(X_0=x_0, X_1=x_1,...,X_k=x_k)=Pr(X_n=x_0, X_{n+1}=x_1,...,X_{n+k}=x_k)$ for all n and k. Every stationary chain can be proved to be time-homogeneous by Bayes' rule. A necessary and sufficient condition for a time-homogeneous Markov chain to be stationary is that the distribution of $X_0$ is a stationary distribution of the Markov chain.}
%
 For each time slot $t$ we assume that the random variable $e_{n,t}$ takes values from a finite set $\mathcal{E}= \{E_1, E_2,\dots E_M\}$ where $ E_m \! \in \! Z^+$,  $ E_m< E_{m+1}$. %As to the state $e$, the actual value of harvested energy is $E_e$. The transition probability of this independent random variable is denoted as 
Therefore,  matrix $\Phi_{\mathcal{E}}$ is $M \times M$ and its $(i,j)$-th entry is $[\Phi_{\mathcal{E}}]_{i,j}=\Pr (e_{n,t}=E_i | e_{n,t-1}=E_j)$. This modelling for the harvested energy processes is justified by empirical measurements in the case of solar energy \cite{Ku}.
 %Also, the entries of the transition matrix of this Markov process can be written in terms of the transition probability 
%$\Pr (e_{n,t} | e_{n,t-1})$. {\green are the entries of transition matrix different from $\Pr (e_{n,t} | e_{n,t-1})$?} {\green needs a citation}
%which defined according to the corresponding Markov model. {\red Azadeh: please continue like $J_1$, you can define the probability vector equation (7) and the transition matrix and equations (8) and (9).} {\blue Ghazaleh: As we discussed about this comment, eq 7,8,9 from $J_1$ will not use in this paper.}
 %%%%%%%%%%%%%%%%%%%%%%%%%%%%%%%%%%%%%%%%%%%%
 \par Let $g_{n,t}$ indicate the narrow-band (flat) fading channel gain between sensor $n$ and the FC during time slot $t$. 
%We assume block fading model and 
We assume $g_{n,t}$'s are independent across sensors. We consider a coherent FC with the knowledge of all channel gains. The FC quantizes $g_{n,t}$ to $\bar{g}_{n,t}$ and informs sensor $n$ of $\bar{g}_{n,t}$, through a bandwidth-limited feedback channel from the FC to sensor $n$. Suppose the quantizer has $L$ quantization thresholds $\bar{\mathcal{G}} \!=\! \{\mu_{1},\mu_{2},...,\mu_{L}\}$, where $0\! = \mu_{1}\! < \mu_{2}\! < \dots \!< \mu_{L}\! = \infty$, and  $\mathcal{I}_{l}=[\mu_{l}, \mu_{l+1})$ for $l=1,\ldots,L$ denote the corresponding quantization intervals. 
%$L$ intervals obtained from partitioning the positive real line with this quantizer. 
%A bandwidth-limited feedback channel from the FC to sensor $n$ conveys the information about which interval $g_{n,t}$ belongs to.
%
{ Suppose $\bar{g}_{n,t}=Q(g_{n,t})$ indicates the input-output relationship of the quantizer. If $g_{n,t} \! \in \! \mathcal{I}_{l}$ then  $\bar{g}_{n,t}\!=\!\mu_{l}$.
%The channel gain quantization rule at the FC for sensor $n$ follows: if $g_{n,t} \in \mathcal{I}_{n,l}$ then $g_{n,t}$ is quantized to $\mu_{n,l}$.
We define the probability
$\phi_{n,l}\!=\!\Pr(\bar{g}_{n,t} \! =\! \mu_{l})$,} which can be found based on the distribution of fading model in terms of the two quantization thresholds $\mu_{l}$ and $\mu_{l+1}$. 

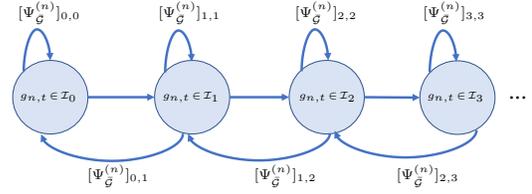
\begin{figure}[!t]
  \centering 
  \hspace{-45mm}
  %\vspace{-10mm}
  \scalebox{.77}{
    \begin{picture}(150,150)
      %\small
      %\footnotesize
      { \scriptsize
        \put(15,20){\includegraphics[width=90mm]{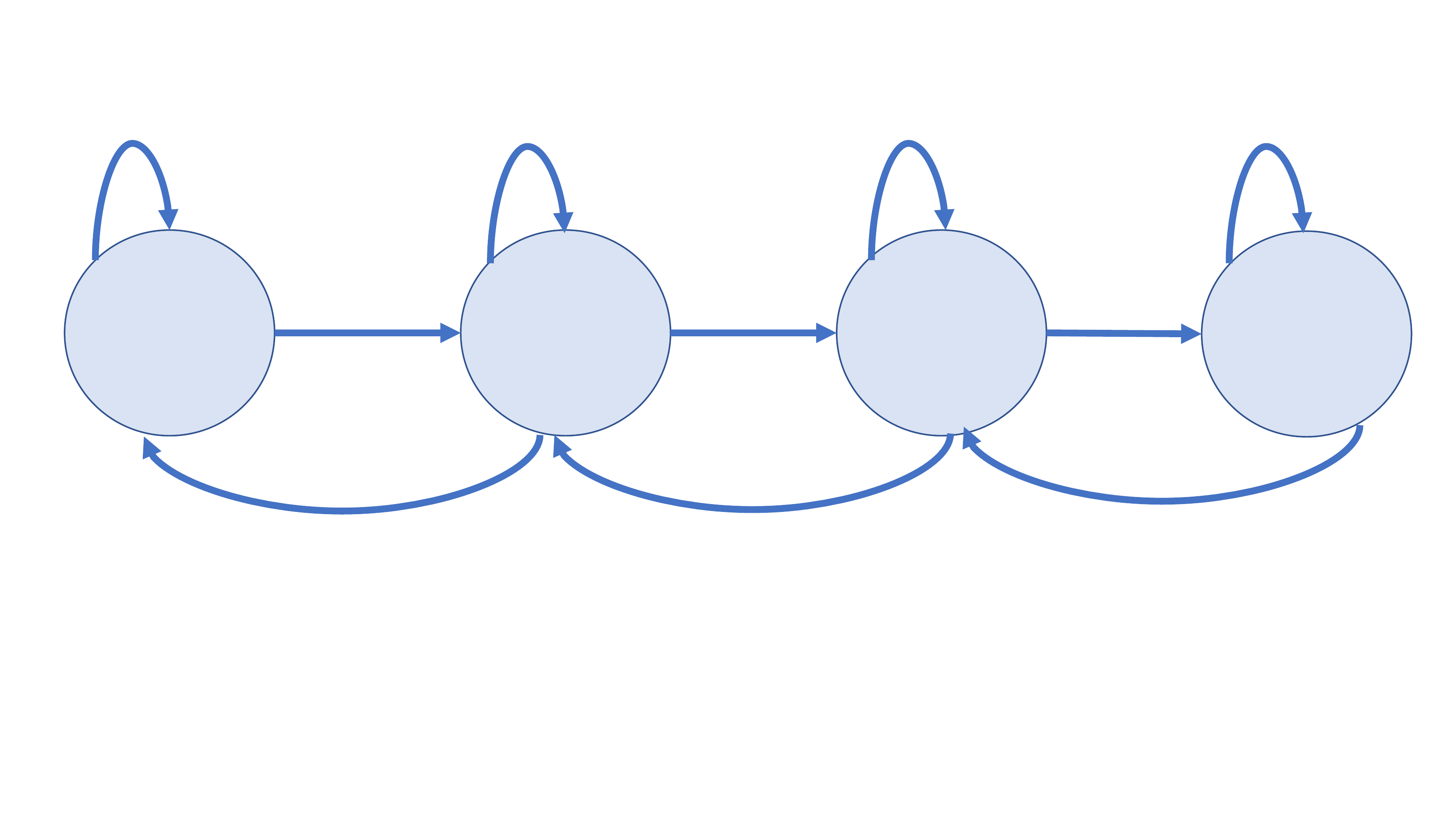}}
        \put(63,65){$[\Psi_{\bar{\mathcal{G}}}^{(n)}]_{0,1}$} 
        \put(145,65){$[\Psi_{\bar{\mathcal{G}}}^{(n)}]_{1,2}$} 
        \put(215,65){$[\Psi_{\bar{\mathcal{G}}}^{(n)}]_{2,3}$} 
        
        \put(30,105){\tiny $g_{n,t}\!\in\!{\cal I}_{0}$} 
        \put(100,105){\tiny $g_{n,t}\!\in\!{\cal I}_{1}$} 
        \put(167,105){\tiny $g_{n,t}\!\in\!{\cal I}_{2}$} 
        \put(230,105){\tiny $g_{n,t}\!\in\!{\cal I}_{3}$}
        \put(270,105){\large ...}
        
        \put(29,145){$[\Psi_{\bar{\mathcal{G}}}^{(n)}]_{0,0}$} 
        \put(98,145){$[\Psi_{\bar{\mathcal{G}}}^{(n)}]_{1,1}$} 
        \put(165,145){$[\Psi_{\bar{\mathcal{G}}}^{(n)}]_{2,2}$} 
        \put(228,145){$[\Psi_{\bar{\mathcal{G}}}^{(n)}]_{3,3}$} 
        
      }
    \end{picture}
  }
\vspace{-15mm}
\caption{Our adopted FSMC model for channel fading process. } 
\label{channel} 
\vspace{-4mm}
\end{figure}

{ We assume $\bar{g}_{n,t}$  is a  homogeneous finite-state Markov chain (FSMC)} \cite{Fan} with an $L \times L$ transition probability matrix $\Psi_{\bar{\mathcal{G}}}^{(n)}$ and its   
 $(k,l)$-th entry is $[\Psi_{\bar{\mathcal{G}}}^{(n)}]_{k,l}=\Pr (\bar{g}_{n,t}=\mu_{k} | \bar{g}_{n,t-1}=\mu_{l})$. %$\bar{g}_{n,t}$ fluctuates due to Doppler \cite{Fan} \cite{Amirnavaei}. 
%The channel state is assumed to remain unchanged in each time slot, and the fluctuation of channel is slow. 
%
Fig. \ref{channel} is the schematic
representation of this $L$-state Markov chain.
Suppose that the channel fluctuation due to Doppler  is slow enough such that the transition in $\bar{g}$ only happens between adjacent channel states \cite{Ku}. For Rayleigh fading model,  $g_{n,t}^2$ is modeled as an exponential random variable with the mean $\mathbb{E}\{g_{n,t}^2\}=\gamma_{g_n}$ and we have
%----------------------------------------
\begin{equation}\label{gaz}
\phi_{n,l}=\Pr(\bar{g}_{n,t}=\mu_{l})= e^{\frac{-\mu_{l}^2}{\gamma_{g_n}}}-e^{\frac{-\mu_{l+1}^2}{\gamma_{g_n}}}.
\end{equation}
Furthermore
%------------------------------------------
\begin{align}\label{trans_gg}
    [\Psi_{\bar{\mathcal{G}}}^{(n)}]_{k,l}=\begin{cases}
    \frac{G(\mu^2_{l+1})}{\phi_{n,l}},&~k = l\!+\!1,~l = 1,...,L\!-\!1\\
    \frac{G(\mu^2_{l})}{\phi_{n,l}},&~k = l-1,~l = 2,...,L\\
    1-\frac{G(\mu^2_{l})+G(\mu^2_{l+1})}{\phi_{n,l}}
    ,&~k = l,~l = 2,...,L\!-\!1\\
     1-\frac{G(\mu^2_{2})}{\phi_{n,1}},&~k = 1,~l=1\\
     1-\frac{G(\mu^2_{L})}{\phi_{n,L}},&~k = L,~l=L \\
     0, &~~ \mbox{O.W.}
    \end{cases}
\end{align}
%---------------------------------------
%---------------------------------------
%-------------------------------------------------------------
where $G(x)=\sqrt{2\pi x/\gamma_{g_n}}f_D T_s\exp{(-x/\gamma_{g_n}})$ is the level crossing rate, and $f_D$ is the maximum Doppler
frequency. %normalized by $1/T_s$ \cite{Shen_Wang}.
 We assume the feedback channel from the FC to sensor $n$ has a delay, i.e., 
%that the FC sends delayed
%channel state information (CSI) of the previous allocation interval back to the sensor node. 
at the beginning of time slot $t$, sensor $n$ only knows $\bar{g}_{n,t-1}$ but not $\bar{g}_{n,t}$.

Let $s_{n,t}$ denote the state of sensor $n$ during time slot $t$. We characterize $s_{n,t}$ %$s_{n,t}=(b_{n,t},\bar{g}_{n,t-1},e_{n,t-1}),$ and $s_{n,t}\in \mathcal{S}$.  The state of sensor $n$ at time slot $t$ 
by a three-tuple $s_{n,t} = (b_{n,t}, \bar{g}_{n,t-1}, e_{n,t-1})$.
%which includes the battery energy state $b$ for the current time slot and channel state $\bar{g}$ and the energy harvesting state $e$ in the previous time slot.  $b,\bar{g},e$ take discrete values and are all bounded.
%
We denote the state space as $\mathcal{S}\!=\!\mathcal{B}\times\bar{\mathcal{G}}\times\mathcal{E}$, where $\mathcal{B}$ is the set of battery energy states, $\bar{\mathcal{G}}$ is the set of communication channel states and $\mathcal{E}$ is the set of energy harvesting states.
Let $\boldsymbol{s}_t\!=\!(s_{1,t},s_{2,t},...,s_{N,t})$ denote the network  state during time slot $t$ and $ \bar{\mathcal{S}}\!=\!\mathcal{S}\times\mathcal{S}\times \dots \mathcal{S}$ denote the network state space, where $\mathcal{S}\!=\!\mathcal{B}\times\bar{\mathcal{G}}\times\mathcal{E}$. 
We refer to $s_{n,t} \! \in \! \mathcal{S}$ and $\boldsymbol{s}_t \in \bar{\mathcal{S}}$ as the {\it local state} and the {\it global (network) state}, respectively. Clearly, $\mathcal{S}, \bar{\mathcal{S}}$ are discrete and finite.
Let the dimensions of $\mathcal{S}, \bar{\mathcal{S}}$ be denoted as  $|\mathcal{S}|, |\bar{\mathcal{S}}|$. We have  $|\bar{\mathcal{S}}|=|\mathcal{S}|^N=((K+1)LM)^N$.

In time slot $t$, if LLR exceeds a given local threshold $\theta_n$, sensor $n$ chooses its non-negative transmission symbol $\alpha_{n,t}$ according to the {\it available information} (either the local state $s_{n,t}$ or the global state $\boldsymbol{s}_t$).
%
%to its battery state $b_{n,t}$,  the harvested energy $e_{n,t-1}$  and the feedback information $\bar{g}_{n,t-1}$. 
%
Note that the amount of energy consumed for transmitting non-negative symbol $\alpha_{n,t}$ cannot be more than the energy stored in the battery, i.e.,   $\alpha_{n,t}$ must satisfy the inequality $\alpha^2_{n,t}T_s/b_u \leq b_{n,t}$. This implies that
$\alpha_{n,t} \in {\cal U}_{n,t}$ where the feasible set 
%
%At time slot $t$ sensor $n$ chooses an action $\alpha_{n,t}$ from the feasible set ${\cal U}_{n}$.  Hence, the feasible set is
%
%we have
%--------------------------------------
%\begin{equation}\label{fess_set}
%    a(s_{n,t}) = \{\alpha_{n,t}|\alpha^2_{n,t}T_s/b_u \leq b_{n,t}\}
%\end{equation}
%--------------------------------------------
${\cal U}_{n,t} \!=\! \big\{0,\sqrt{b_u /T_s},...,\sqrt{b_{n,t}b_u /T_s}\big\}$  is discrete and finite. Let $\boldsymbol{\alpha}_t=\big(\alpha_{1,t},\alpha_{2,t},...,\alpha_{N,t}\big)$ contains transmission symbols by all sensors. We have  $\boldsymbol{\alpha}_t \in \bar{\mathcal{U}}_{t}=\mathcal{U}_{1,t}\times\mathcal{U}_{2,t}\times\dots\mathcal{U}_{N,t}$.
Further, we assume that the nodes in the network must satisfy a total transmit power constraint. Such power constraint can be translated into  $\sum^N_{n=1}\alpha_{n,t}^2\ \leq {\cal P}_{tot}$. 
%which specifies to which interval the channel gain $g_{n,t}$ belongs to. 
%In fact, $\alpha_{n,t}\in\{0,1,2,...,b_{n,t}\}$. 
%-------------------------------------------------------
%
%{\green the rest of this sub-section should be similar to Asilomar, definition of system state should be in this sub-section} 
%Assuming the consumed energy for sensing is negligible, 
%
%-------------------------------------
%------------------------------------------------
%\subsubsection{States}
%
{\it Our goal is to develop (sub-)optimal adaptive power control strategy such that the detection performance at the FC is optimized.} 

In Section \ref{section3}  we formulate the constrained optimization of transmission symbol $\alpha_{n,t}$ as a  discounted infinite-horizon constrained MDP problem. In this problem formulation, 
$\alpha_{n,t}$ is the action taken by sensor $n$, and $\boldsymbol{\alpha}_t$ is the collection of actions taken by all sensors, during time slot $t$.  We refer to $\alpha_{n,t}$ as the {\it local action} and $\boldsymbol{\alpha}_t$ as the {\it  global (network) action}, respectively. 
We use dynamic programming to solve the problem and provide two types of solutions: 
%
%by specifying the system state, the action set, the state transition probability, and the reward function.
%
%
(i) {\it the optimal policy}, in which local action $\alpha_{n,t}$ depends on the global state ${\boldsymbol{s}}_t\!=\!(s_{1,t},s_{2,t},...,s_{N,t})$ where $s_{n,t}\!=\!(b_{n,t}, \bar{g}_{n,t-1}, e_{n,t-1})$, i.e., during time slot $t$ sensor $n$ has access to  the global state ${\boldsymbol{s}}_t$ and determines its action $\alpha_{n,t}$  according to ${\boldsymbol{s}}_t$, and (ii) {\it the sub-optimal policy}, in which  local action $\alpha_{n,t}$ depends on the local state $s_{n,t}$ only, i.e., during time slot $t$ sensor $n$ has access to the local state ${{s}}_{n,t}$ only and determines its action $\alpha_{n,t}$  according to ${{s}}_{n,t}$.
 
 %its battery state, the harvested energy, and the feedback information about its communication channel. 

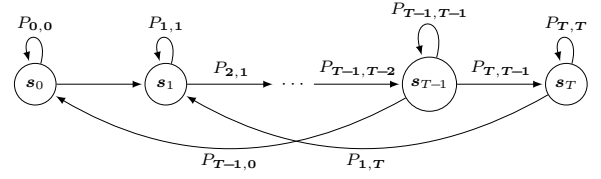
\begin{figure}[!t]
 \centering
 \hspace{-0mm}
\scalebox{0.9}{
  \begin{tikzpicture}[start chain=going left,node distance=1.3cm]   
     \scriptsize
     \node[state, on chain]                 (K) {$\boldsymbol{s}_{T}$};
     \node[state, on chain]                 (K-1) {$\boldsymbol{s}_{T\!-\!1}$};
     \node[on chain]                   (g) {$\cdots$};
    \node[state, on chain]                 (1) {$\boldsymbol{s}_{1}$};
    \node[state, on chain]                 (0) {$\boldsymbol{s}_{0}$};
   \draw[
    >=latex,
        every node/.style={above,midway},% either
     auto=right,                      % or
     loop above/.style={out=75,in=105,loop},
     every loop,
     ]
     (0)   edge[loop above] node {$P_{\boldsymbol{0},\boldsymbol{0}}$}   (0)
      edge             node { } (1)
      (1)   edge[loop above] node {$P_{\boldsymbol{1},\boldsymbol{1}}$}   (1)
      edge             node {$P_{\boldsymbol{2},\boldsymbol{1}}$}   (g)
      (g)  edge[] node {$P_{\boldsymbol{T\!-\!1},\boldsymbol{T\!-\!2}}$}   (K-1)
     (K-1)   edge[loop above] node {$P_{\boldsymbol{T\!-\!1},\boldsymbol{T\!-\!1}}$}   (K-1)
      edge             node {$P_{\boldsymbol{T},\boldsymbol{T\!-\!1}}$}   (K)
      (K-1)   edge[bend left] node[below] {$P_{\boldsymbol{T\!-\!1},\boldsymbol{0}}$}   (0)
      (K)   edge[loop above] node {$P_{\boldsymbol{T},\boldsymbol{T}}$}   (K-1)
      (K)   edge[bend left] node[below] {$P_{\boldsymbol{1},\boldsymbol{T}}$}   (1);
    \end{tikzpicture}
  }
  \caption{ Schematics of Markov chain corresponding to the global state $\boldsymbol{s}_{t}$. In this figure we have $P_{\boldsymbol{t},\boldsymbol{t+1}}=\Pr\big(\boldsymbol{s}_{t+1}|\boldsymbol{s}_{t},\boldsymbol{\alpha}_{t}\big)$.} 
  \label{fig_battery}
\end{figure}

%The action performed by each sensor is $a(s_{n,t})$.
%to consume $\alpha^2_{n,t}T_s/b_u$ units of stored energy for data transmission. That is, sensor $n$ takes an action $a(s_{n,t})=\alpha_{n,t}$ from the feasible set  ${\cal U}_{s_{n,t}} = \big\{0,\sqrt{b_u /T_s},...,\sqrt{b_{n,t}b_u /T_s}\big\}$, which is discrete and finite. 
%The action performed by each sensor is $a(s_{n,t})$.
%=========================================================
%\subsubsection{State Transition}
%
Let the global state transition probability $\Pr\big(\boldsymbol{s}_{t+1} |\boldsymbol{s}_t, \boldsymbol{\alpha}_t\big)$ denote  the probability of entering  network state $\boldsymbol{s}_{t+1}$ if network action $\boldsymbol{\alpha}_t$ is taken at network state $\boldsymbol{s}_{t}$. Define $\boldsymbol{b}_t\!=\!(b_{1,t},b_{2,t},...,b_{N,t})$, $\bar{\boldsymbol{g}}_t\!=\!(\bar{g}_{1,t},\bar{g}_{2,t},...,\bar{g}_{N,t})$, $\boldsymbol{e}_t\!=\!(e_{1,t},e_{2,t},...,e_{N,t})$. We can simplify the global state transition
probability as the product of three conditional probabilities (see Fig. \ref{fig_battery})
%\in \mathcal{U}_{s_{n,t}}$. 
%
%---------------------------------------------------------------------
\begin{align}\label{trans1}
  &\Pr\big(\boldsymbol{s}_{t+1}|\boldsymbol{s}_{t},\boldsymbol{\alpha}_{t}\big) \nonumber\\ 
  =& \Pr\big(\boldsymbol{b}_{t+1},\bar{\boldsymbol{g}}_t,\boldsymbol{e}_t|\boldsymbol{b}_t,\bar{\boldsymbol{g}}_{t-1},\boldsymbol{e}_{t-1},\boldsymbol{\alpha}_{t}\big) \nonumber\\ 
  =& \Pr\big(\boldsymbol{b}_{t+1}|\boldsymbol{b}_{t},{\bar{\boldsymbol{g}}}_t,\boldsymbol{e}_{t},\boldsymbol{\alpha}_{t}\big)\Pr(\boldsymbol{e}_{t}|\boldsymbol{e}_{t-1})\Pr({\bar{\boldsymbol{g}}}_t|{\bar{\boldsymbol{g}}}_{t-1}).
\end{align}
%------------------------------------------------------------------
The second and third conditional probabilities in (\ref{trans1}) can be decomposed across sensors, since  $\bar{g}_{n,t}$'s and  $e_{n,t}$'s are independent across sensors. In other words, we have
\begin{eqnarray}
 \Pr({\boldsymbol{e}}_t|{\boldsymbol{e}}_{t-1})\!\!&\!\!=\!\!&\!\!\prod_{n=1}^N \Pr( e_{n,t}| e_{n,t-1}), \nonumber\\ \Pr({\bar{\boldsymbol{g}}}_t|{\bar{\boldsymbol{g}}}_{t-1})\!\!&\!\!=\!\!&\!\!\prod_{n=1}^N \Pr( \bar{g}_{n,t}|\bar{g}_{n,t-1} )
\end{eqnarray}
in which $\Pr( e_{n,t}| e_{n,t-1}), \Pr( \bar{g}_{n,t}|\bar{g}_{n,t-1}) $ are
the transition probabilities of $e_{n,t}$ and $\bar{g}_{n,t}$ Markov chains,
respectively. To find the first conditional probability in (\ref{trans1}), we need to know the dynamic battery state model. 
The battery state at the beginning of time slot $t+1$ depends on the battery state at the beginning of time slot $t$, the harvested energy during time slot $t$, and the transmission symbol $\alpha_{n,t}$, i.e.,
%-------------------------------------------------------
\begin{equation}\label{b_n,t}
    b_{n,t+1} = \min\big\{[b_{n,t} + e_{n,t}-\alpha^2_{n,t}T_s/b_u]^+,K \big \},
    %\vspace{-1mm}
\end{equation}
%-------------------------------------------------------
where $[x]^+=\max \{0,x\}$. 
%\begin{equation}
%   \alpha^2_{n,t}T_s/b_u \leq b_{n,t}
%\end{equation}
%=========================================================
%\subsubsection{Action Set}
Considering the dynamic battery state model in \eqref{b_n,t} we notice that, conditioned on $e_{n,t}$ and $\alpha_{n,t}$ the value of $b_{n,t+1}$ only depends on the value of $b_{n,t}$ (and not the battery states before time slot $t$). Hence, the process $\boldsymbol{b}_{t}$ can be modeled as a Markov chain and the first conditional probability in (\ref{trans1}) becomes
%----------------------------------------------------------------------
\begin{align}\label{trans2}
   \Pr\big(\boldsymbol{b}_{t+1}|\boldsymbol{b}_{t},{\bar{\boldsymbol{g}}}_t,\boldsymbol{e}_{t},\boldsymbol{\alpha}_{t}\big) 
  = \begin{cases}
  1~~~\text{if (\ref{b_n,t}) is satisfied}~\forall n\\
  0~~~\text{otherwise,}
  \end{cases}
\end{align}
%
% \Pr\big(b_{n,t+1}|b_{n,t},\bar{g}_{n,t},e_{n,t},a(s_{n,t})\big)
%
%-------------------------------------------------------------------------
%as $b_{n,t+1}$ is a deterministic value. 
%Further, $\Pr\big(s_{n,t+1}|s_{n,t},a(s_{n,t})\big)$ for all $s_{n,t}, s_{n,t+1} \! \in \!\mathcal{S}, a(s_{n,t})\in\mathcal{U}_{s_{n,t}}$ are  independent across sensors. 
We define the reward function in Section \ref{opt_problem}.
%=========================================================
%%%%%%%%%%%%%%%%%%%%%%%%%%%%%%%%%%%%%%%%%%%%%%%%%%%%%%%%%%%%%%%%
%-----------------------------------
%-------------------------------------------------------------
%-------------------------------------------------------------
%%%%%%%%%%%%%%%%%%%%%%%%%%%%%%%%%%%%%%%%%%%%%%%%%%%%%%%%%%%%%%%
\subsection{Received Signals at FC and Optimal Bayesian Fusion Rule}
In each time slot, sensors send their data symbols to the FC over orthogonal fading channels.   
 The received signal at the FC from sensor $n$ corresponding to time slot $t$ is
%----------------------------------------------------
\begin{equation}\label{y_n,t}
   y_{n,t} = g_{n,t}\, \alpha_{n,t} + w_{n,t}, ~~~~\text{for}~n=1,\dots,N
\vspace{-1mm}
\end{equation}
%--------------------------------------------------
where $w_{n,t} \sim {\cal N} (0,\sigma_{w_n}^2)$ is the additive Gaussian noise. We assume $w_{n,t}$'s are i.i.d. over time slots and independent across sensors. 
 Let $\boldsymbol{y}_t=[y_{1,t},y_{2,t}, \ldots ,y_{N,t}]$ denote the vector that includes the received signals at the FC from all sensors in time slot $t$. The FC applies the optimal Bayesian fusion rule $\Gamma_0(.)$ to the received vector $\boldsymbol{y}_t$ and obtains a global decision $u_{0,t}=\Gamma_0(\boldsymbol{y}_t)$, where $u_{0,t} \in \{0,1\}$. In particular, we have
%----------------------------
 \begin{equation}\label{u0_lrt}
 u_{0,t}= \Gamma_0(\boldsymbol{y}_t)= \\
 \begin{cases}
 1,~~~~~~\Delta_t > \tau, \\
 0,~~~~~~ \Delta_t < \tau,
 \end{cases}
\end{equation}
 %---------------------------------------
 where the decision threshold is $\tau =\log( \frac{\zeta_0}{\zeta_1})$ and 
 %------------------------------------------
\begin{equation}\label{lrt1}
\Delta_t=\log\left (\frac{f(\boldsymbol{y}_t|h_t=1)}{f(\boldsymbol{y}_t|h_t=0)}\right),%\gtrless\begin{matrix}
%{\scriptstyle {H}_{1}}\cr{\scriptstyle {H}_{0}}
%\end{matrix} \tau
\end{equation}
%----------------------------------------
and $f(\boldsymbol{y}_t|h_t)$ is the conditional probability density function (pdf) of the received vector $\boldsymbol{y}_t$ at the FC. %{\green like Asilomar, define $P_e$ in this sub-section and motivate the use of J-divergence. }
From Bayesian perspective, the natural choice to measure the detection performance corresponding to the global decision $u_{0,t}$ at the FC is the error probability, defined as
%-----------------------------------------
\begin{align}\label{pe1}
 P_e &=\zeta_0 \Pr (u_{0,t}=1|h_t=0)+\zeta_1 \Pr(u_{0,t}=0|h_t=1)\nonumber\\
&=\zeta_0 \Pr(\Delta_t > \tau |  h_t=0) + \zeta_1 \Pr(\Delta_t < \tau |  h_t=1).   
\end{align}
%------------------
However, finding a closed form expression for $P_e$ is mathematically intractable. 
Instead, we choose the total $J$-divergence between the distributions of the detection statistics at the FC under different hypotheses, as our detection performance metric. This choice allows us to provide a tractable analysis. Next, we define the total $J$-divergence and derive a closed-form expression for it, using Gaussian approximation. 
%%%%%%%%%%%%%%%%%%%%%%%%%%%%%%%%%%%%%%%%%%%%%
 \subsection{Total J-Divergence Derivation and Reward Function}\label{opt_problem}
%====================================================
\begin{comment}
{\red
Our goal is to find the optimal $\alpha_{n,t}$ for all sensors such that the total $J$-divergence at the FC is maximized,  subject to total transmit power constraint. We design the optimal transmit power map  with the points $\alpha_{n,t}$ for $n = 1,...,N$  and $s_{n,t} \in {\cal S}$ . Therefore in each time slot, sensor $n$  will assign its transmitted power $\alpha_{n,t}$  to $\alpha_{n,t}$  according to its system
state. The optimal map can be found offline, via solving the constrained optimization problem and the map is shared with all sensors for distributed implementation.
 We formulated the our optimization problem in three different ways and obtain optimal power map from each method. In simulation and result section, we investigate the performance of these power maps. 
\begin{itemize}
    %\item {\blue Remove} First method, which can be consider as upper bound of system, we assume that we have non-casual information of finite time. We obtain optimal transmission powers by maximizing total value of $J$-divergence finite time horizon given future information. In this section, we assume total power constraint in every time-slot.  
    \item Second method, we model the system as infinite horizon constrain Markov decision process (CMPC) and imply dynamic programming.
    \item We will solve infinite horizon CMPC by Q-learning.
\end{itemize}}
\end{comment}
%===========================================

%We start with the definition of $J$-divergence. 
Consider two pdfs of a continuous random variable $x$, denoted as $\eta_1(x)$ and $\eta_2(x)$. By definition \cite{vin}, the $J$-divergence between $\eta_1(x)$ and $\eta_0(x)$, denoted as $J(\eta_1,\eta_0)$, is $J(\eta_1,\eta_0) = D(\eta_1||\eta_0)+D(\eta_0||\eta_1)$,
%-------------------------------------------------------------------
%\begin{equation}\label{first_j}
%    J(\eta_1,\eta_0) = D(\eta_1||\eta_0)+D(\eta_0||\eta_1),
%\end{equation}
%---------------------------------------------------------------------
where $D(\eta_i||\eta_j)$ is the non-symmetric Kullback-Leibler (KL) distance between $\eta_i(x)$ and $\eta_j(x)$. The KL distance $D(\eta_i||\eta_j)$ is defined as
%\vspace{-2mm}
%----------------------------------------------------------------------
\begin{equation}\label{kl}
    D(\eta_i||\eta_j) = \int_{-\infty}^{\infty} \log \left(\frac{\eta_i(x)}{\eta_j(x)}\right)\eta_i(x) dx.
    %\vspace{-1mm}
\end{equation}
%----------------------------------------------------------
%Substituting \eqref{kl} into \eqref{first_j} 
Therefore, we obtain
%--------------------------------------------
\begin{equation}\label{j23}
J(\eta_1,\eta_0)=\int_{-\infty}^{\infty} \left[\eta_1 (x)-\eta_0 (x)\right] \log \left(\frac{\eta_1(x)}{\eta_0(x)}\right) dx. 
\end{equation}
%------------------------------------------
In our problem setup, the two conditional pdfs $f(\boldsymbol{y}_t |h_t\!=\!1)$ and $f(\boldsymbol{y}_t |h_t\!=\!0)$ in (\ref{lrt1}) play the role of $\eta_1(x)$ and $\eta_0(x)$, respectively. Let  $J_{tot,t}$ denote the $J$-divergence between $f(\boldsymbol{y}_t| h_t=1)$ and  $f(\boldsymbol{y}_t|h_t=0)$. The pdf of vector $\boldsymbol{y}_t$ given $h_t$ is
%------------------------------------------------------------------
\begin{align}\label{y/H}
f(\boldsymbol{y}_t|h_{t})=\prod_{n=1}^N f(y_{n,t}|h_t)~~~\text{for}~h_t=0,1.
\end{align}
%--------------------------------------------------------------------
where the equality in \eqref{y/H} holds since the received signals from sensors at the FC, given $h_t$, are conditionally independent.
%, equality ($b$) in \eqref{y/H} is obtained from Bayes' rule, and equality ($c$) in \eqref{y/H} is found noting that $H$, $\alpha_n$, $y_n$ satisfy the Markov property, i.e.,  $H \rightarrow \alpha_n \rightarrow y_n$ \cite{vin}, \cite{zahra} and hence $y_n$ and $H$, given $\alpha_n$, are conditionally independent.
%Using \eqref{kl} we can write  the $J$-divergence
%$J\big(f(\boldsymbol{y}|{\mathcal H}_1),f(\boldsymbol{y} |{\cal H}_0)\big)$ 
%\vspace{-1mm}
%By using \eqref{kl} and given
Let $J_{n,t}$ represent the $J$-divergence between the two conditional pdfs $f({y}_{n,t}|h_t=1)$ and $f({y}_{n,t}|h_t=0)$. We have
%Using \eqref{j23} we can express $J_n$ as
%For sensor $n$ the $J$-divergence between $f(y_{n,t}|h_t=1)$ and $f(y_{n,t}|h_t=0)$ is
%----------------------------------------------
\begin{equation}\label{J-n-t-actual}
\begin{split}
&J_{n,t} =\\
&\!\!\int_{-\infty}^{\infty} \!\!\Big [\!f(y_{n,t}| h_t\! =\! 1)\!-\! f(y_{n,t}| h_t\! =\! 0)\!\Big]\!  \log\!\left(\! {\frac{f(y_{n,t}| h_t\! =\! 1)} {f(y_{n,t}|h_t\! =\! 0)}}\! \right)\!dy_{n,t}.
   \end{split}
\end{equation}
%----------------------------------------------------------------
Based on (\ref{y/H}), the total $J$-divergence, denoted as $J_{tot,t}$, is $J_{tot,t}=\sum_{n=1}^N J_{n,t}$. 
Note that $f(y_{n,t}|h_t=0)\! =\! f(y_{n,t} |\alpha_{n,t} \neq 0) P_{\text{f}_n}\! +\! f(y_{n,t}|\alpha_{n,t}\! = \!0) (1\!-\!P_{\text{f}_n})$ and $f(y_{n,t}|h_t=1)\!=\!f(y_{n,t}|\alpha_{n,t} \neq 0) P_{\text{d}_n}\! +\! f(y_{n,t} |\alpha_{n,t} \!= \!0) (1\!-\!P_{\text{d}_n})$ are Gaussian mixtures and the $J$-divergence between two Gaussian mixture densities does not have a closed-form expression \cite{vin,Ard3}. We approximate $J_{n,t}$ in (\ref{J-n-t-actual}) using the Gaussian densities $f^G(y_{n,t}|h_t) \sim {\cal N}(m_{n,h},\Upsilon_{n,h}^2)$, where $m_{n,h}$ and $\Upsilon_{n,h}^2$ are obtained from matching the first and second order moments of the actual and the approximate distributions. % \cite{vin,Ard3}. 
For our problem setup,  $m_{n,h}$ and $\Upsilon_{n,{h}}^2$ are 
%-------------------------------------------
 \begin{align}\label{G_par}
&m_{n,0}={{g_{n,t}}\alpha_{n,t}}P_{\text{f}_n}\nonumber,~~~ \Upsilon_{n,0}^2\!=\!g_{n,t}^2\alpha_{n,t}^2 P_{\text{f}_n}(1\!-\!P_{\text{f}_n})\!+\!\sigma_{w_n}^2,\\ 
&m_{n,1}\!=\!{{g_{n,t}}\alpha_{n,t}}P_{\text{d}_n},~~~~ \Upsilon_{n,1}^2\!=\!g_{n,t}^2\alpha_{n,t}^2 P_{\text{d}_n}(1\!-\!P_{\text{d}_n})\!+\!\sigma_{w_n}^2.
\end{align}
%------------------------------------------
The $J$-divergence between two Gaussian densities, represented as $J\big(f^G(y_{n,t}|h_t=1),f^G(y_{n,t}|h_t=0)\big)$, in terms of their means and variances is \cite{vin, Ard3}
%----------------------------------------------------------------
\begin{align}\label{g_J}
    J\big(f^G(y_n&|h=1),f^G(y_n| h=0)\big)= \nonumber\\
     &\frac{\Upsilon_{n,1}^2\!+\!( m_{n,1}\!-\! m_{n,0})^2}{\Upsilon_{n,0}^2}
     +\frac{\Upsilon_{n,0}^2\!+\!( m_{n,0}\!-\! m_{n,1})^2}{\Upsilon_{n,1}^2}. 
\end{align}
%----------------------------------------------------------------------
Substituting $m_{n,h}$ and $\Upsilon_{n,h}^2$ of (\ref{G_par}) into \eqref{g_J} we approximate $J_{n,t}$ in (\ref{J-n-t-actual}) as the following\vspace{-2mm}
%-----------------------------------------------------------------
\begin{align}\label{sim_j}
    J_{n,t}(g_{n,t},\alpha_{n,t})=\frac{\sigma_{w_{n}}^2 + A_n g_{n,t}^2\alpha_{n,t}^2}{\sigma_{w_{n}}^2 + B_n g_{n,t}^2\alpha_{n,t}^2}&+\frac{\sigma_{w_{n}}^2  + C_n g_{n,t}^2\alpha_{n,t}^2}{\sigma_{w_{n}}^2  +  D_n g_{n,t}^2\alpha_{n,t}^2},
\end{align}
 %-------------------------------------
 where
%---------------------------------------------------------------------
\begin{align} \label{ABCD}
A_n =&~P_{\text{f}_n}(1\!-\!P_{\text{d}_n}) + P_{\text{d}_n}(P_{\text{d}_n}\!-\!P_{\text{f}_n}),\nonumber \\
 C_n=&~P_{\text{d}_n}(1-P_{\text{f}_n}) - P_{\text{f}_n}(P_{\text{d}_n}-P_{\text{f}_n}),\nonumber \\
 B_n =& ~P_{\text{d}_n}(1-P_{\text{d}_n}), ~~D_n = P_{\text{f}_n}(1-P_{\text{f}_n}).
 \end{align}
%---------------------------------------------------
The notation $J_{n,t}(g_{n,t},\alpha_{n,t})$ in \eqref{sim_j} is to emphasize that $J$-divergence depends on both transmission symbol $\alpha_{n,t}$ and fading channel gain $g_{n,t}$.
%
%We note that $J_{n,t}$ in \eqref{sim_j} depends on $g_{n,t}$ value. 
The dependency on $g_{n,t}$ stems from the fact that the FC has full knowledge of all channel gains $g_{n,t}$'s, and the optimal Bayesian fusion rule utilizes this full information to make the binary decision. 
On the other hand, sensor $n$ only knows $\bar{g}_{n,t-1}\!=\!Q(g_{n,t-1})$.
Hence, $\alpha_{n,t}$ can only depend on $\bar{g}_{n,t-1}$.
To resolve this issue, we take the average of $J_{n,t}$  over $g_{n,t}$, conditioned on $\bar{g}_{n,t-1}$. 
%Given state $s_{n,t}$ and action $\alpha_{n,t}$, 
Let $\mathbb{E}_{g_{n,t}|\bar{g}_{n,t-1}}\{J_{n,t}(g_{n,t},\alpha_{n,t})|\bar{g}_{n,t-1}\}$ denote the average of $J$-divergence over $g_{n,t}$ when action $\alpha_{n,t}$ is taken according to the available information at sensor $n$, conditioned on $\bar{g}_{n,t-1}$.  %{\red Is the wording better now?}
%%%%%%%%%%%%%%%%%%%%%%%%%%%%%%%%%%%%%%%%%%%%%%%%%%%%
%\subsection{Average of $J$-divergence derivation}
%-----------------
%
%$r\big(s_{n,t},\alpha_{n,t}\big)$
Let $r\big(\alpha_{n,t}\big)$ indicate the immediate reward function of sensor $n$ at time slot $t$. We define the immediate reward function of sensor $n$ as the average of $J$-divergence over $g_{n,t}$ when action $\alpha_{n,t} \neq 0$ is taken according to the available information at sensor $n$, conditioned on $\bar{g}_{n,t-1}$, i.e., %{\red inside J below is it g or g-bar? how about the expectation?} {\blue Based on our conversion and eq 21,  we concluded that both should be g}
%
%to be the expected amount of $J$-divergence in time slot $t$ when $\alpha_{n,t}\neq 0$ 
%-----------------------------------------------------------------------------
\begin{equation}\label{reward}
    r(\alpha_{n,t}) = \widehat{\zeta}_{n,1}\mathbb{E}_{{g}_{n,t}|\bar{g}_{n,t-1}}\{J({g}_{n,t},\alpha_{n,t})|\bar{g}_{n,t-1}\}
\end{equation}
%--------------------------------------------
where $\widehat{\zeta}_{n,1}\!=\!\Pr(\alpha_{n,t}\! \neq \! 0)$ is given in (\ref{Pi01}). Note that when action $\alpha_{n,t} \!= \!0$ is taken from (\ref{sim_j}) we find  $J_{n,t}(g_{n,t},0)=2$. By defining the immediate reward as (\ref{reward}) we neglect the constant term $\widehat{\zeta}_{n,0}\mathbb{E}_{{g}_{n,t}|\bar{g}_{n,t-1}}\{J({g}_{n,t},0)|\bar{g}_{n,t-1}\}=2\widehat{\zeta}_{n,0}$ and do not count it toward the immediate reward function. 
%
%Note that term  shows that in immediate reward, we ignore the constant amount when $\alpha_{n,t} = 0$ .
%
To compute the immediate reward function in (\ref{reward}), first we define the following
%-----------------------------------
\begin{eqnarray}\label{J-bar-J-hat}
\hat{J}_{n,t}^{(l)} &\!=\!&\mathbb{E}_{g_{n,t}|\bar{g}_{n,t}=\mu_l}\{ J_{n,t}\big(g_{n,t},\alpha_{n,t}\big) | \bar{g}_{n,t} =\mu_l\}\\
\bar{J}_{n,t}^{(l)} \!\!&\!\!=\!\!&\!\!\mathbb{E}_{g_{n,t}|\bar{g}_{n,t-1}=\mu_l}\{ J_{n,t}\big(g_{n,t},\alpha_{n,t}\big) | \bar{g}_{n,t-1}=\mu_l\} \nonumber
\end{eqnarray}
%-----------------------------------------
% 
Hence, the immediate reward function in (\ref{reward}) can be rewritten in terms of $\bar{J}_{n,t}^{(l)}$ for $l=1,...,L$ as 
%-------------------------------------------
\begin{eqnarray}\label{reward-2}
     && \!\!\!\!r(\alpha_{n,t}) = \widehat{\zeta}_{n,1} \Big(\sum_{l=2}^{L-1} \underbrace{\Pr(\bar{g}_{n,t-1}=\mu_{l})}_{=\phi_{n,l}} \bar{J}_{n,t}^{(l)} \nonumber\\ && \!\!\!\! + \underbrace{\Pr(\bar{g}_{n,t-1}=\mu_{1})}_{=\phi_{n,1}} \bar{J}_{n,t}^{(1)}+\underbrace{\Pr(\bar{g}_{n,t-1}=\mu_{L})}_{=\phi_{n,L}} \bar{J}_{n,t}^{(L)} \Big)
\end{eqnarray}
%--------------------------------------------
To fully characterize the reward function in (\ref{reward-2}) we need to find $\bar{J}_{n,t}^{(l)}$ defined in (\ref{J-bar-J-hat}) for $l=1,...,L$. 
When $\bar{g}_{n,t-1} \!=\! \mu_{l}$ for $l=2,..,L-1$,  from (\ref{trans_gg}) we have
%
%our transition channel state model in Section \ref{channel-model}  we know
%
%$\bar{g}_{n,t} \in \{\mu_{l-1}, \mu_{l}, \mu_{l+1}\}$. 
%When $\bar{g}_{n,t-1} \!=\! \mu_{l}$ for $l=2,..,L-1$ we have
%$\bar{g}_{n,t}=\mu_{n,i-1}$ or $\bar{g}_{n,t}=\mu_{n,i}$, or $\bar{g}_{n,t}=\mu_{n,i+1}$ with the probabilities 
%$\in \{\mu_{n,i-1}, \mu_{n,i}, \mu_{n,i+1}\}$. 
%then we have $\bar{g}_{n,t}$ can be
%------------------------------------
\begin{equation}
\bar{g}_{n,t}=
   \begin{cases}
  \mu_{l-1} ~~~\mbox{with probability}~~ [\Psi_{\bar{\mathcal{G}}}^{(n)}]_{l-1,l}\\
  \mu_{l}~~~\mbox{with probability}~~ [\Psi_{\bar{\mathcal{G}}}^{(n)}]_{l,l}\\
  \mu_{l+1}~~~\mbox{with probability}~~ [\Psi_{\bar{\mathcal{G}}}^{(n)}]_{l+1,l}
  \end{cases}
\end{equation}
%-----------------------------------
 When $\bar{g}_{n,t-1} \!=\! \mu_{1}$, from (\ref{trans_gg}) we have
%------------------------------------
\begin{equation}
\bar{g}_{n,t}=
   \begin{cases}
  \mu_{1}~~~\mbox{with probability}~~ [\Psi_{\bar{\mathcal{G}}}^{(n)}]_{1,1}\\
  \mu_{2}~~~\mbox{with probability}~~ [\Psi_{\bar{\mathcal{G}}}^{(n)}]_{2,1}
  \end{cases}
\end{equation}
%-----------------------------------
and when $\bar{g}_{n,t-1} \!=\! \mu_{L}$, from (\ref{trans_gg}) we have
%------------------------------------
\begin{equation}
\bar{g}_{n,t}=
   \begin{cases}
  \mu_{L-1}~~~\mbox{with probability}~~ [\Psi_{\bar{\mathcal{G}}}^{(n)}]_{L-1,L}\\
  \mu_{L}~~~\mbox{with probability}~~ [\Psi_{\bar{\mathcal{G}}}^{(n)}]_{L,L-1}
  \end{cases}
\end{equation} 
%-----------------------------------
%------------------------------------
Therefore, $\bar{J}_{n,t}^{(l)}$ and $\hat{J}_{n,t}^{(l)}$  in (\ref{J-bar-J-hat}) become related as the following
%-----------------------------------------
\begin{equation}\label{J_bar}
 \resizebox{1.04 \hsize}{!}{$ \displaystyle{
\bar{J}_{n,t}^{(l)}=
\begin{cases}
[\Psi_{\bar{\mathcal{G}}}^{(n)}]_{l-1,l}\hat{J}_{n,t}^{(l-1)}
    +[\Psi_{\bar{\mathcal{G}}}^{(n)}]_{l,l}\hat{J}_{n,t}^{(l)}
     +[\Psi_{\bar{\mathcal{G}}}^{(n)}]_{l+1,l}\hat{J}_{n,t}^{(l+1)}~ l \neq 1,L\\
 [\Psi_{\bar{\mathcal{G}}}^{(n)}]_{1,1}\hat{J}_{n,t}^{(1)}
     +[\Psi_{\bar{\mathcal{G}}}^{(n)}]_{2,1}\hat{J}_{n,t}^{(2)}~~l=1\\
     [\Psi_{\bar{\mathcal{G}}}^{(n)}]_{L-1,L}\hat{J}_{n,t}^{(L-1)}
    +[\Psi_{\bar{\mathcal{G}}}^{(n)}]_{L,L}\hat{J}_{n,t}^{(L)}~~l=L
\end{cases}}$}
\end{equation}
%------------------------------------
%The $\mathbb{E}_{g_{n,t}|\bar{g}_{n,t-1}}\Big\{J_{n,t}\big(g_{n,t},\alpha_{n,t}\big)|\bar{g}_{n,t-1}\Big\}$ can be written as  
%\begin{align}
%\mathbb{E}_{g_{n,t}|\bar{g}_{n,t-1}}\Big\{J_{n,t}\big(g_{n,t},\alpha_{n,t}\big)|\bar{g}_{n,t-1}\Big\}=
%    \sum_{l=1}^L \underbrace{\Pr(\bar{g}_{n,t-1}=\mu_{l})}_{=\phi_{n,l}} \bar{J}_{n,t}^{(l)} 
%\end{align}
%-----------------------------------------
%{\red (22) is actually more complicated, I think you need to separate terms for $l=1$, $l=2,...,L-1$, and $l=L$} 
%
{  Note that $\hat{J}_{n,t}^{(l)}$ in (\ref{J-bar-J-hat}) can be obtained based on the distribution of fading model. For Rayleigh fading model, we have \cite{Ard3}} 
%{\red so you should change (28) and (29), for l=2,...,L-1 it uses (28) with three terms, for l=1 and l=L it uses (28) with two terms}
%
%
%{\green yes, you should write (41) and $\Omega$ function and $\beta$ function from J1 here. }
%
%----------------------------------------------------------
\begin{align}\label{ave_J}
  \hat{J}_{n,t}^{(l)} = \phi_{n,l}\Big[\Omega(\alpha^2_{n,t},\mu_{n,l+1}^2)-\Omega(\alpha^2_{n,t},\mu_{n,l}^2)\Big],
\end{align}
%%---------------
%--------------------------------------------
where the two dimensional function $\Omega(x,y)$ in \eqref{ave_J} is
\begin{align}\label{omega-f}
 &\Omega(x,y)\triangleq
  \nonumber\\&\frac{1}{B_n x}\Big[\sigma^2_{w_n}\beta_1(x,y)-\frac{A_n}{B_n}\sigma^2_{w_n}\beta_1(x,y) - A_n x  e^{(-y\gamma_{g_n} )}\Big]+
  \nonumber\\&
  \frac{1}{D_n x}\Big[\sigma^2_{w_n}\beta_2(x,y) - \frac{C_n}{D_n}\sigma^2_{w_n}\beta_2(x,y) - C_n x e^{(-y\gamma_{g_n} )}\Big],
\end{align}
%--------------------------------------
%--------------------------------------
Also, $A_n,~B_n,~C_n$ and $D_n$ are given in \eqref{ABCD} and the two dimensional functions $\beta_1(x,y)$ and $\beta_2(x,y)$  in \eqref{omega-f} are
%--------------------------------
\begin{align*}
&\beta_1(x,y)\triangleq\gamma_{g_n}\text{exp}\Big(\frac{\sigma^2_{w_n}\gamma_{g_n}}{x B_n} \Big)\text{Ei}\Big(-\gamma_{g_n}y-\frac{\sigma^2_{w_n}\gamma_{g_n}}{x B_n}\Big),\nonumber
\end{align*}
%%%%%%%%%%%%%%%%%%%%%%%%%
\begin{align*}
&\beta_2(x,y)\triangleq\gamma_{g_n}\text{exp}\Big(\frac{\sigma^2_{w_n}\gamma_{g_n}}{x D_n}\Big)\text{Ei}\Big(-\gamma_{g_n}y-\frac{\sigma^2_{w_n}\gamma_{g_n}}{x D_n}\Big). 
\end{align*}
%----------------------------------
%{\green The optimal transmission power corresponding to a range of discretized values of the sensors’ channels, harvested energies and battery levels are stored in a lookup table for use in real-time. These policies are based on the assumption that only causal channel state and energy harvesting information is available at transmitters. We present our optimal power allocation problem as a Constrained Markov Decision Process (CMDP) problem by specifying the decision epochs, state, action set, the state transition and the reward functions. } 
%%%%%%%%%%%%%%%%%%%%%%%%%%%%%%%%%%%%%%%%%%%%%%%%%%%%%%%%%%%%%%%%%%%%%%%%%%%%%5
 %%%%%%%%%%%%%%%%%%%%%%%%%%%%%%%%%%%%%%%%%%%%%%%%%%%%%%%
%
%---------------------------------------------------
%%%%%%%%%%%%%%%%%%%%%%%%%%%%%%%%%%%%%%%%%%%%%%%%%%%
%This completes our characterization  of the reward function. 
In summary, the reward function in \eqref{reward-2} can be written as
%---------------------------------
\begin{equation}\label{reward-3}
    r(\alpha_{n,t}) = \widehat{\zeta}_{n,1} \Big(\sum_{l=2}^{L-1}  \phi_{n,l}  \bar{J}_{n,t}^{(l)}   \!\!\ +\phi_{n,1}  \bar{J}_{n,t}^{(1)}+\phi_{n,L}  \bar{J}_{n,t}^{(L)} \Big)
\end{equation}
%------------------------------
in which $\bar{J}_{n,t}^{(l)} $ is given in \eqref{J_bar}. 
The immediate network reward function  at time slot $t$, denoted as $r(\boldsymbol{\alpha}_t)$, is  the sum of the reward functions of all sensors
%--------------------------------------------------------
\begin{equation}\label{reward22}
r(\boldsymbol{\alpha}_t)=\sum_{n=1}^N r\big(\alpha_{n,t}\big).
\end{equation}
%--------------------------------------------------------
where $r\big(\alpha_{n,t}\big)$ is given in (\ref{reward-3}).

At every time slot $t$, sensor $n$ decides the transmission
symbol $\alpha_{n,t}$ according to the available information (either the local state $s_{n,t}$ or the global state $\boldsymbol{s}_t$)  such that the discounted sum of reward is maximized, 
%
%- I don't see any discount factor in (35)?} {\blue Like Mao's paper, (35) represents the expected total reward between the first allocation interval until the sensor
%stops functioning with policy $\pi$. Then Mao explains based on the geometric distribution of the lifetime T of
%the sensor node with mean $1/(1-\eta)$, (36) is equivalent to the objective function of infinite-horizon MDP with discounted reward given.} 
%
subject to two constraints: (i) the amount of energy consumed for transmission symbol $\alpha_{n,t}$ cannot be more than the energy stored in the battery $b_{n,t}$, i.e., $\alpha_{n,t}^2 T_s/b_u \leq b_{n,t}$, or equivalently, $\alpha_{n,t} \in {\cal U}_{{n,t}},~\forall n,t% \!=\! \big\{0,\sqrt{b_u /T_s},...,\sqrt{b_{n,t}b_u /T_s}\big\}, 
$, (ii)  the nodes in the network must satisfy a total transmit power constraint, i.e., $\sum^N_{n=1}\alpha_{n,t}^2 \ \leq {\cal P}_{tot},~\forall t$. 

%Using the specified system state, the action set, the state transition probability, and the reward function, 

In the next section  
%
%Problem formulation based on these system elements is described in more detail in the next section.}
%
we formulate the constrained optimization of $\alpha_{n,t}$ as a  discounted infinite-horizon constrained MDP problem. We use dynamic programming to solve the problem and   
%
%we formulate our problem of finding the optimal and sub-optimal transmit
%power control strategies, i.e., optimizing $\alpha_{n,t}\!=\!\alpha_{n,t}$, as a discounted infinite-horizon constrained MDP problem 
%
provide two types of solutions: 
%
%by specifying the system state, the action set, the state transition probability, and the reward function.
%
%
(i) the optimal policy, in which the local action $\alpha_{n,t}$ depends on the global state ${\boldsymbol{s}}_t$, and (ii) the sub-optimal policy, in which  the local action $\alpha_{n,t}$ depends on the local state $s_{n,t}$ only.
%
%
%We formulate the constrained optimization of αn,t as a dis-
%counted infinite-horizon constrained Markov decision process
%(MDP) problem by specifying the system state, the action set,
%the state transition probability, and the reward function.
%%%%%%%%%%%%%%%%%%%%%%%%%%%%%%%%%%%%%%%%%%%%%%%%%%%%%%%%
\section{Problem Formulation}\label{section3}
We start our problem formulation by defining the set of feasible policies. 
Let $\delta_t$ 
%{\red $\delta_t$ should be a vector too, } 
denote a general decision rule that describes how a network action  $\boldsymbol{\alpha}_t$ is selected according to the global state $\boldsymbol{s}_t$ in time slot $t$, i.e.,  $\boldsymbol{\alpha}_t \! =\!\delta_t(\boldsymbol{s}_t$), and $\pi \!=\! (\delta_1, \delta_2,...\delta_T)$ be the corresponding policy for $t\!=\!1,...,T$, i.e., $\pi$ is the sequence of decision rules to be employed for time slots $t=1,...,T$ \cite[pp.21]{puterman2014}.  
%In general, a power allocation policy is a sequence of decisions that maps a state to an action at time slot $t$. 
 We say that a 
policy $\pi$ is feasible if it satisfies the two constraints: (i)
$\boldsymbol{\alpha}_t\in\bar{\mathcal{U}}_{t}, ~\forall t$, (ii) $\sum^N_{n=1} \alpha_{n,t}^2 \leq {\cal P}_{tot}, ~\forall t$.
%
%\eqref{fess_set} $\forall n$ and total power constrain, i.e., $\sum^N_{n=1}\alpha_{n,t} \leq \alpha_{tot}$.
Let $\Pi$ be the set of feasible policies $\pi$. Then, for any given global state  $\boldsymbol{s}_1$ 
%where $s_{n,1}=(b_{n,1},g_{n,0},e_{n,0})$ 
at the first time slot $t=1$, the expected network reward between the first time slot until a  sensor
stops functioning with policy $\pi \in \Pi$ is given by
%--------------------------------
%------------------------------------------
\begin{align}\label{valu_iter_time}
   &V_\pi(\boldsymbol{s}_1)= \mathbb{E}\left\{\mathbb{E}_T\bigg\{\sum^T_{t=1} r(\boldsymbol{\alpha}_t)\bigg\}\Big|\boldsymbol{s}_1,\pi\right\}\nonumber\\
    &\text{s.t.}~\boldsymbol{\alpha}_t\in\bar{\mathcal{U}}_{t},~\sum^N_{n=1} \alpha_{n,t}^2 \leq {\cal P}_{tot}, ~\forall t
\end{align}
%-------------------------------------------------
 where the outer expectation $\mathbb{E}\{.\}$ in (\ref{valu_iter_time}) denotes the statistical expectation taken over all
relevant random variables given initial global state $\boldsymbol{s}_1$ and policy $\pi$. The inner expectation $\mathbb{E}_T\{.\}$ in (\ref{valu_iter_time})  denotes the expectation with respect to the random variable $T$.
%which is the lifetime of the sensor node.
Note that with a different initial global state $\boldsymbol{s}_1$ and a different policy $\pi$, a different network action $\boldsymbol{\alpha}_t$ will be selected in time slot $t$, which results in a different state transition probability $ \Pr\big(\boldsymbol{s}_{t+1}|\boldsymbol{s}_t,\boldsymbol{\alpha}_t\big)$ when the outer expectation $\mathbb{E}\{.\}$ in (\ref{valu_iter_time}) is computed. Since $T$ is a geometric random variable 
%Based on the geometric distribution of the lifetime $T$ of the sensor node 
with mean $\mathbb{E}\{T\}\!=\!1/(1-\eta)$, \eqref{valu_iter_time} is equivalent to
the objective function of an infinite-horizon MDP with discounted reward given by \cite[Proposition 5.3.1]{puterman2014}
%
%{\red you wrote T is the sensor lifetime, if we say this, then what if some sensors fail and some continue to function? then (34) becomes much more complicated,  we need to consider a simple network lifetime definition. Network lifetime in the literature has different definitions (for instance, a network is alive until all sensors die). we define network lifetime as the time until the first sensor dies. Then T becomes geometric RV with mean $1/(1-\eta)$ } {\blue ok}
%------------------------------------------
\begin{align}\label{valu_iter}
   &V_\pi(\boldsymbol{s}_1)= \mathbb{E}\Bigg\{\sum^\infty_{t=1}\eta^t r(\boldsymbol{\alpha}_t)|\boldsymbol{s}_1,\pi\Bigg\}
   \nonumber\\
    &\text{s.t.}~\boldsymbol{\alpha}_t\in\bar{\mathcal{U}}_{t},~\sum^N_{n=1} \alpha_{n,t}^2 \leq {\cal P}_{tot}, ~\forall t
\end{align}
%-------------------------------------------------
% {\red in Shaobo Mao's paper in (35) it is $\eta^t$, it is geometric and makes sense to me, it is not  $\eta_t$ } {\blue it was typo. I meant $\eta^t$} 
where $\eta$ in (\ref{valu_iter}) can be interpreted as the discount factor of the model and $ V_\pi(\boldsymbol{s}_1)$ in (\ref{valu_iter}) can be interpreted as the long-term expected network reward starting from an initial global state $ \boldsymbol{s}_1$ and continuing with the policy $\pi$ from then on \cite{puterman2014}. Since the network  will stop functioning at some time in the future, the network reward at time slot $t$ is discounted by factor $\eta^t$. 
The problem in (\ref{valu_iter}) is our discounted infinite-horizon constrained MDP problem. 

%
%in algorithm 1, I define the the initial state as s1 and v(s1)=0
%
%{\red like Lemma 1 in Shaobo Mao's paper you need to argue that (35) is finite}
One can easily show that that the objective function $ V_\pi(\boldsymbol{s}_1)$ in \eqref{valu_iter} 
converges to a finite value \cite[pp. 121]{puterman2014}. The proof follows. First, we note 
\begin{align}\label{proof-obj-finite}
    &\sup_{\boldsymbol{s}_t\in\bar{\mathcal{S}},~\boldsymbol{\alpha}_t\in \bar{\mathcal{U}}_{t}} |r(\boldsymbol{\alpha}_t)| \leq  \sup_{\boldsymbol{s}_t\in\bar{\mathcal{S}},~\boldsymbol{\alpha}_t\in \bar{\mathcal{U}}_{t}}  \sum_{n=1}^N |r(\alpha_{n,t})|  
    %\\
   % &\sum_{n=1}^N |r(s_{n,t},K)|= \sum^{N}_{n=1}\left[ \widehat{\zeta}_{n,1} \Big(   \phi_{n,L}  \bar{J}_{n,t}^{(L)} \Big)\right]<\infty
\end{align}
Next, we examine $r(\alpha_{n,t})$ in (\ref{reward-3}) and we note that $\widehat{\zeta}_{n,1}, \{\phi_{n,l}\}_{l=1}^L$ are probabilities and $\{\bar{J}_{n,t}^{(l)}\}_{l=1}^L$ depend on the two dimensional $\Omega(.,.), \beta_1(.,.), \beta_2(.,.)$ functions, which all take finite values $\forall \boldsymbol{s}_t\in\bar{\mathcal{S}},~\boldsymbol{\alpha}_t\in \bar{\mathcal{U}}_{t}$. Hence the right-hand side of (\ref{proof-obj-finite}) is finite. This completes the proof. 
%{\red
%Ghazaleh: you should update second line of (37) according to the new reward function that I have derived in (32), this reward has multiple terms (in the original version you wrote  the reward function had on term only).}
%{\blue The problem is that, we did not define $\bar{J}_{n,t}^{(l)}$ as a function of $a(s_{n,t})$ directly. Not sure how to show replacing $a(s_{n,t})$ with $K$ in the last part of (37).}

%{\red Also, in (36) you let the action in $J_{n,t}$ to be $K$, so what is the maximization over $\boldsymbol{\alpha}_t$  outside the sum? }
%
% {\blue The main goal of the MDP is to find a decision policy that specifies the optimal action in the state $s$ and maximizes the objective function.} 
 %
{  Due to Markovian property
of MDP problems, it suffices to consider only {\it Markovian policies}.  Hence, our aim is finding an optimal Markovian  policy $\pi \in \Pi$  that maximizes $ V_\pi(\boldsymbol{s}_1)$ in  \eqref{valu_iter}. That is, given the initial global state $\boldsymbol{s}_1$, our goal is to obtain the optimal expected total discounted
network reward $V^*(\boldsymbol{s}_1)$ and the optimal Markovian policy $\pi^* \in \Pi$ defined as follows:

%----------------------------------------------------------
\begin{align}\label{cmdp1}
    &V^*(\boldsymbol{s}_1) = \max_{\pi\in \Pi} ~V_\pi(\boldsymbol{s}_1), ~~\pi^* = \arg\max_{\pi\in \Pi} V_\pi (\boldsymbol{s}_1)  \nonumber \\
    &\text{s.t.}~\boldsymbol{\alpha}_t\in\bar{\mathcal{U}}_{t},~\sum^N_{n=1} \alpha_{n,t}^2 \leq {\cal P}_{tot}, ~\forall t  
\end{align}
%--------------------------------------------------------------
%and
%-------------------------------------
%\begin{align}
%    &\pi^* = \arg\max_{\pi\in \Pi} V(\boldsymbol{s}_1) 
%    \nonumber \\
%    &\text{s.t.}~\boldsymbol{\alpha}_t\in\mathcal{U}_{\boldsymbol{s}_t},~\sum^N_{n=1} a(s_{n,t}) \leq \alpha_{tot}, ~\forall t  
%\end{align}
%-----------------------------------------------
%[ from Mao's paper] A policy $\pi\!=\! (\delta_1, \delta_2,...\delta_T)$ is said to be stationary deterministic if $\delta_t$ is deterministic Markovian  \cite[pp. 21]{puterman2014} and $\delta_t=\delta$ for all time slots such that $\pi = (\delta, \delta,\dots, \delta)$. 
%
A Markovian policy $\pi\!=\! (\delta_1, \delta_2,...\delta_T)$ is said to be stationary deterministic if $\delta_t=\delta$ for all time slots such that $\pi = (\delta, \delta,\dots, \delta)$ and 
$\delta$ is deterministic  \cite[pp. 21]{puterman2014}. %
%
%For discrete-time MDPs with finite numbers of states and actions and bounded rewards, a stationary optimal policy exists.
%
The existence of a stationary deterministic  optimal policy is guaranteed when the network state space $\bar{\mathcal{S}}$ is  discrete and finite  \cite{puterman2014}.
%
%For an infinite-horizon MDP, the only
%case of interest is when an optimal stationary deterministic policy exists. 
%
Thus, {\it our objective is to find an optimal stationary deterministic policy} $\pi \in \Pi$ that maximizes $ V_\pi(\boldsymbol{s}_1)$ in  \eqref{valu_iter}.
%, which maximizes the expected total
%discounted network reward in \eqref{valu_iter}. 
}

\subsection{{ Finding the Optimal} Policy}\label{subsection-A}
%{\blue maybe it's better to move this subtitle before eq 40, where we introduce Lagrangian.}
To maximize $ V_\pi(\boldsymbol{s}_1)$  in (\ref{valu_iter}),
we first utilize the Lagrangian approach \cite{Moghadari,Fu} to transform the constrained MDP optimization problem into an equivalent
unconstrained MDP optimization problem. 
For each global state $\boldsymbol{s}_{t}$ we introduce a Lagrangian multiplier $\lambda_{\boldsymbol{s}_{t}}$ associated with the constraint $\left(\sum^N_{n=1} \alpha_{n,t}^2 - {\cal P}_{tot}\right)$.
%at each state $\boldsymbol{s}_{t}$.
{ We define the Lagrangian value function $\mathcal{L}(\boldsymbol{s}_t,\lambda_{\boldsymbol{s}_{t}})$ using the dynamic programming \cite{Fu}
%
%{\red is it necessary to cite [39][40][41][42]? cannot we use [38]?} {\blue I removed 39, 40 and used 38 instead, but, we need 41 and 42 (new 39 and 40)}
%
%{\red the text and Lagrangian approach below can be found in which paper? it is different form Mao's paper}
%
\begin{align} \label{new_lag}
     \mathcal{L}(\boldsymbol{s}_t, \lambda_{\boldsymbol{s}_t})=\max_{\pi \in \Pi} \Big\{ \underbrace{L(\boldsymbol{s}_t,\boldsymbol{\alpha}_t,\lambda_{\boldsymbol{s}_t})}_{=\mbox{~term 1}}
   +\nonumber \\
   \underbrace{\eta
    \sum_{\boldsymbol{s}_{t+1}}\Pr\big(\boldsymbol{s}_{t+1}|\boldsymbol{s}_t,\boldsymbol{\alpha}_t\big)\mathcal{L}(\boldsymbol{s}_{t+1}, \lambda_{\boldsymbol{s}_{t+1}})}_{=\mbox{~term 2}}
    \Big\},
\end{align}
%--------------------------------------------
where $L(\boldsymbol{s}_t,\boldsymbol{\alpha}_t,\lambda_{\boldsymbol{s}_t})$  is defined as}
%
%
%{ \green After applying Lagrangian, this problem is formulated as an MDP based stochastic control problem and the optimal power allocation policy, that obtained offline by the use of dynamic programming techniques.} 
%
%
%
\begin{equation}\label{La}
L(\boldsymbol{s}_t,\boldsymbol{\alpha}_t,\lambda_{\boldsymbol{s}_t})= r(\boldsymbol{\alpha}_t) -\lambda_{\boldsymbol{s}_t}\left(\sum^N_{n=1} \alpha_{n,t}^2 - {\cal P}_{tot}\right).
\end{equation}
In fact, $L(\boldsymbol{s}_t,\boldsymbol{\alpha}_t,\lambda_{\boldsymbol{s}_t})$ in (\ref{La}) can be interpreted as the modified network reward function at time slot $t$, where the cost of violating the constraint is subtracted from the immediate reward $r(\boldsymbol{\alpha}_t)$ earned in time slot $t$.
On the other hand, term 2 in $\mathcal{L}(\boldsymbol{s}_t,\lambda_{\boldsymbol{s}_{t}})$  is the expected total discounted future
network reward if network action $\boldsymbol{\alpha}_t$ is chosen. Since $\boldsymbol{\alpha}_t$ can be zero or non-zero, term 2 can be expanded as \eqref{cons_long_term}. Note that $\boldsymbol{\alpha}_t=0$ only if $\alpha_{n,t}\! = \! 0, \forall n$, i.e., when  LLR is below the local threshold $\theta_n, \forall n$. Using (\ref{Pi01}) we find 
$\Pr(\boldsymbol{\alpha}_{t} = 0)\!=\! \prod_{n=1}^N \Pr(\alpha_{n,t}\! = \! 0)\!=\! \prod_{n=1}^N \widehat{\zeta}_{n,0}$ and $\Pr(\boldsymbol{\alpha}_{t} \neq  0)\!=\! 1- \Pr(\boldsymbol{\alpha}_{t} = 0)$. 
%
%where $\widehat{\zeta}_{n,0} = \Pr(\alpha_{n,t}\! = \! 0)$ and $\widehat{\zeta}_{n,1}= \Pr(\alpha_{n,t}\! \neq \! 0)$ are defined in (\ref{Pi01}).
%
%{\red \eqref{cons_long_term} describes the trade-off between the current reward and the expected future reward - I don't understand this sentence, is it important to say?}.}

%We observe that $\mathcal{L}(\boldsymbol{s}_t)$ in (\ref{cons_long_term}) consists of two parts: ``term 1'' is the immediate network reward achieved in time slot $t$, 

With fixed $\lambda_{\boldsymbol{s}_{t}}$,
the constrained MDP problem in (\ref{valu_iter}) can be viewed as a non-constrained %\label{cons_long_term}
MDP problem in (\ref{cons_long_term}) with the modified network reward function $L(\boldsymbol{s}_t,\boldsymbol{\alpha}_t,\lambda_{\boldsymbol{s}_t})$ at time slot $t$ given in (\ref{La}). 
%-------------------------------------------
\begin{figure*}
    \begin{eqnarray}\label{cons_long_term}
       \resizebox{1 \hsize}{!}{$ \displaystyle{ \mathcal{L}(\boldsymbol{s}_t,\lambda_{\boldsymbol{s}_{t}})=\max_{\pi \in \Pi} \Bigg\{\!\! \underbrace{L\big(\boldsymbol{s}_t,\boldsymbol{\alpha}_t,\lambda_{\boldsymbol{s}_t}\big)}_{\mbox{=~term 1}}\!+\! \underbrace{\eta\Big(\!\!\Pr(\boldsymbol{\alpha}_{t}\! = \! 0)
    \!\sum_{\boldsymbol{s}_{t+1}} \Pr\big(\boldsymbol{s}_{t+1}|\boldsymbol{s}_t,0\big)\mathcal{L}(\boldsymbol{s}_{t+1},\lambda_{\boldsymbol{s}_{t+1}})\!+\!
 \Pr(\boldsymbol{\alpha}_{t}\! \neq \! 0)\!\!
    \sum_{\boldsymbol{s}_{t+1}}\Pr\big(\boldsymbol{s}_{t+1}|\boldsymbol{s}_t,\boldsymbol{\alpha}_t\big)\mathcal{L}(\boldsymbol{s}_{t+1},\lambda_{\boldsymbol{s}_{t+1}})\Big)}_{=~\mbox{term 2}}\!\!\Bigg\}.}$}
    %&&{\text{s.t.}}~\boldsymbol{\alpha}_t\in\mathcal{U}_{\boldsymbol{s}_t},~\sum^N_{n=1} a(s_{n,t}) \leq \alpha_{tot}, \forall t 
    \end{eqnarray}
     \hrulefill
\end{figure*}
Let  $U (\lambda_{\boldsymbol{s}_t}) $ denote the Lagrangian dual function, where
\begin{equation}\label{U-original}
    U (\lambda_{\boldsymbol{s}_t}) = \max_{\pi}\mathcal{L}(\boldsymbol{s}_t,\lambda_{\boldsymbol{s}_{t}}) 
\end{equation}
Then the Lagrangian dual problem can be written as
\begin{equation}\label{dual1}
    {\min_{\lambda_{\boldsymbol{s}_t}\geq 0} U (\lambda_{\boldsymbol{s}_t})}
\end{equation}
%--------------------------------------------------------
The resulting dual solution has zero duality gap compared to the primary problem in (\ref{new_lag}) \cite[pp.2]{Fu}.
To solve the dual problem in \eqref{dual1}, 
%we first assign
%an arbitrary initial value to $\lambda_{\boldsymbol{s}_t}$, and then 
%
we iteratively solve the following two sub-problems until a pre-specified convergence criterion is reached. The outer minimization sub-problem (the outer loop in Algorithm \ref{Alg1} with iteration index $i$) updates $\lambda_{\boldsymbol{s}_{t}}^i$. The inner maximization sub-problem (the inner loop in Algorithm \ref{Alg1} with iteration index $j$) finds the optimal $\pi_i$ given $\lambda_{\boldsymbol{s}_{t}}^i$. The pseudo code of the algorithm is given in Algorithm \ref{Alg1}. 
\begin{enumerate}
    \item \textbf{the inner maximization sub-problem:} %{\red With fixed $\lambda_{\boldsymbol{s}_t}$, the inner maximization problem can be viewed as a non-constrained MDP problem with instant reward $L(\boldsymbol{s}_t,\lambda_{\boldsymbol{s}_t})$ from \eqref{La} - this is repetition}. The optimality equation of the maximization problem is given as follows
    %
    %{\red (44) is repetition of (41) and should be removed. }
    %--------------------------------------
    \begin{comment}
    \begin{align}\label{valu_itr_lag}
         &\mathcal{L}(\boldsymbol{s}_t)=\max_{\pi} \Big\{L(\boldsymbol{s}_t,\lambda_{\boldsymbol{s}_t})+\nonumber\\
   &\eta\Big(\widehat{\zeta}_{n,0}
    \sum_{\boldsymbol{s}_{t+1}}\prod_{n=1}^N\Pr(s_{n,t+1}|s_{n,t},0)\mathcal{L}(\boldsymbol{s}_{t+1})+
    \nonumber\\
    &\widehat{\zeta}_{n,1}
    \sum_{\boldsymbol{s}_{t+1}}\prod_{n=1}^N\Pr\big(s_{n,t+1}|s_{n,t},\alpha_{n,t}\big)\mathcal{L}(\boldsymbol{s}_{t+1})\Big)\Big\},
    \end{align}
    \end{comment}
    %-------------------------------------------------
    Given $\lambda_{\boldsymbol{s}_t}^i$ we adopt the value iteration algorithm \cite{bertsekas2012} to find the optimal policy $\pi_i$. The convergence criterion is $|\mathcal{L}^j(\boldsymbol{s}_t,\lambda^i_{\boldsymbol{s}_t})-\mathcal{L}^{j-1}(\boldsymbol{s}_t,\lambda^i_{\boldsymbol{s}_t})|<\epsilon_1(1-\eta)/2\eta$, for a given $\epsilon_1$, where  $\mathcal{L}^j(\boldsymbol{s}_t,\lambda^i_{\boldsymbol{s}_t})$ indicates the long-term expected reward in the $j$-th iteration from \eqref{cons_long_term}.
    \item \textbf{the outer minimization sub-problem:} The outer minimization over the Lagrangian multiplier $\lambda_{\boldsymbol{s}_t}$ is a linear programming problem.
    %: $\min  U (\lambda_{\boldsymbol{s}_t})$, which is the dual problem of the \eqref{cons_long_term}. %{\red original mode selection problem - what does this mean?}. 
    We use the sub-gradient method to update $\lambda_{\boldsymbol{s}_t}^i$ as the following.  
    %The updating process of the multiplier is given as follows
    %---------------------------------------------
    \begin{equation}\label{update_lam}
    \lambda_{\boldsymbol{s}_t}^{i+1}=\Bigg[\lambda_{\boldsymbol{s}_t}^{i}+\beta^i\left(\sum_{n=1}^N \alpha_{n,t}^2-{\cal P}_{tot}\right)  \Bigg]^+,
\end{equation}
    %--------------------------------------------
    where 
    %$i$ is the iteration number and 
    $\beta$ is a positive scalar step size satisfying the conditions $\sum _{i=1}^{\infty}\beta^i=\infty$ and $\sum _{i=1}^{\infty}(\beta^i)^2<\infty$.
    %
    %By the sub-gradient algorithm, the multiplier  will converge to the optimal value. 
    %
    The update rule is such that if $\sum_{n=1}^N \alpha_{n,t}^2$ is larger (smaller) than ${\cal P}_{tot}$ then $\lambda_{\boldsymbol{s}_t}$ should increase (decrease).
    Unless the convergence criterion 
 $\frac{|\lambda^{i+1}_{\boldsymbol{s}_t}-\lambda^{i}_{\boldsymbol{s}_t}|}{\lambda^{i}_{\boldsymbol{s}_t}}<\epsilon_2$ is met, for a given $\epsilon_2$, we increase $i$ and solve the inner maximization sub-problem again. 
    
   % { We repeat these two steps for $\forall \boldsymbol{s}\in \bar{\mathcal{S}}$.}
\end{enumerate}
Note that the above sub-gradient method is guaranteed to converge to the optimum $\lambda_{\boldsymbol{s}_t}$,  as long as $\beta$ satisfies the conditions stated above, due to the convexity of the dual problem \eqref{dual1} over $\lambda_{\boldsymbol{s}_t}$.

\begin{algorithm}[!t]
 \caption{optimal power control algorithm}
 \label{Alg1}
 1: Specify $\epsilon_1>0$, $\epsilon_2>0$ set $\mathcal{L}(\boldsymbol{s}_{1},\lambda_{\boldsymbol{s}_1})=0 ,~ \boldsymbol{s}_t\in \bar{\mathcal{S}}$ set $i = 1$  \;
 2: \For{\textnormal{fixed} $\lambda^i_{\boldsymbol{s}_t}$}
{ 
 3: Set $j = 1$, \For{\textnormal{each} $\boldsymbol{s}_t \in \bar{\mathcal{S}}$} 
 {
 \For{\textnormal{each} $\boldsymbol{\alpha}_t \in \mathcal{U}_{\boldsymbol{s}_t}$}{ calculate  
%--------------------------------------  
  \begin{align*}
     &\mathcal{F}(\boldsymbol{s}_t,\boldsymbol{\alpha}_t,\lambda^i_{\boldsymbol{s}_{t}})= L(\boldsymbol{s}_t,\boldsymbol{\alpha}_t,\lambda^i_{\boldsymbol{s}_t})+\nonumber\\
   &\eta \sum_{\boldsymbol{s}_{t+1}}\Pr\big(\boldsymbol{s}_{t+1}|\boldsymbol{s}_t,\boldsymbol{\alpha}_t\big) \mathcal{L}^{j-1}(\boldsymbol{s}_{t+1},\lambda^i_{\boldsymbol{s}_{t+1}})
 \end{align*}
 }
  calculate $\mathcal{L}^j(\boldsymbol{s}_t,\lambda^i_{\boldsymbol{s}_{t}}) = \underset{\pi_i \in \Pi}{\max}\{\mathcal{F}(\boldsymbol{s}_t,\boldsymbol{\alpha}_t,\lambda^i_{\boldsymbol{s}_{t}})\}$\;
 }
 4: If $\underset{\boldsymbol{s}_t \in \bar{\cal S}}{\max} |\mathcal{L}^j(\boldsymbol{s}_t,\lambda^i_{\boldsymbol{s}_{t}})-\mathcal{L}^{j-1}(\boldsymbol{s}_t,\lambda^i_{\boldsymbol{s}_{t}})|<\epsilon_1(1-\eta)/2\eta$,\\ go to Step 5. Otherwise, increase $j$ and go back to Step 3.\\
 5: %For each $\boldsymbol{s}_t \in {\cal S}$, 
 We obtain the optimal policy \\
 \begin{align*}
    \pi_i^* =\arg\max_{\pi_i \in \Pi}\Big\{\mathcal{L}^j(\boldsymbol{s}_t,\lambda^i_{\boldsymbol{s}_{t}})\Big\}.
 \end{align*}
 }
 6: Update $\lambda^i_{\boldsymbol{s}_t}$ using the rule in \eqref{update_lam} and $\pi_i^*$\;
 7: If $\frac{|\lambda^{i+1}_{\boldsymbol{s}_t}-\lambda^{i}_{\boldsymbol{s}_t}|}{\lambda^{i}_{\boldsymbol{s}_t}}<\epsilon_2$. then $\pi^*=\pi_i^*$. Otherwise, increase $i$ and go back to Step 2.
 \end{algorithm}
%%%%%%%%%%%%%%%%%%%%%%%%%%%%%%%%%%%%%%%%%%%%%%%%%%%%%%%
%

\textbf{Remark on the computational complexity of Algorithm \ref{Alg1}:} We switch between solving two sub-problems until the convergence criterion for updating the Lagrangian multiplier is met. Given $\lambda_{\boldsymbol{s}_t}$ we solve the inner maximization sub-problem, i.e., 
we solve \eqref{cons_long_term}  for each $\boldsymbol{s}_t \in \bar{\mathcal{S}}$ (refer to Step 3 of Algorithm \ref{Alg1}), where  $|\bar{\mathcal{S}}|=|\mathcal{S}|^N$.
Our numerical results show that 
%, for the fixed $\lambda_{\boldsymbol{s}_t}$, 
the computational complexity of calculating $\pi^*_i$ (Step 5 of Algorithm \ref{Alg1}) is $\mathcal{O}( |\mathcal{S}|^N K^{1.2 N})$. On the other hand, the complexity order of the gradient-descent algorithm to find the local minimum of
function $U (\lambda_{\boldsymbol{s}_t})$  and
converge to an $\epsilon_2$-accurate solution is $\bar{\epsilon}=1/\epsilon_2$ \cite[p.~232]{luenberger}.
%{\blue I double checked it and fixed it.}.
%
%For step 6 of Algorithm \ref{Alg1}, for each $\lambda_{\boldsymbol{s}_t}$, complexity order for convergence to an $\epsilon$-accurate solution is $\bar{\epsilon}$, where $\bar{\epsilon}=\log(1/\epsilon_2)$ \cite[p.~217]{luenberger}. 
%Thus, for each $\boldsymbol{s}_t$, the computational complexity of step 1 and step 2 is $\mathcal{O}(\bar{\epsilon}{ (K^{1.2})^N})$. 
Hence, the overall the computational complexity of finding the optimal solution using Algorithm \ref{Alg1} is $\mathcal{O}(\bar{\epsilon}|\mathcal{S}|^N K^{1.2 N})$. Note that the complexity order scales {\it exponentially} in $N$. 

\textbf{Remark on implementing the optimal policy}: The optimal policy, a.k.a. centralized solution in the dynamic control literature, requires the knowledge of the global state $\boldsymbol{s}_t$ to determine the network action $\boldsymbol{\alpha}_t=\delta(\boldsymbol{s}_t)$.
This implies that sensor $n$ in the network cannot find its local action $\alpha_{n,t}$ at time slot $t$, unless it knows the global state $\boldsymbol{s}_t$.  Hence, implementing this solution requires all $N$ sensors to report their local states $s_{n,t},\forall n$ to the FC. The FC concatenates all the received local states and forms the global state $\boldsymbol{s}_t$.  Then based on $\boldsymbol{s}_t$, the FC determines and broadcasts the network action $\boldsymbol{\alpha}_t$. This process, however, consumes significant signaling overhead. 

To reduce the signaling overhead, we consider 
finding a sub-optimal policy, a.k.a. decentralized solution in the literature, where sensor $n$ in the network finds its local action $\alpha_{n,t}$ at time slot $t$, only based on its own local state $s_{n,t}$.  
\subsection{Finding the Sub-Optimal Policy}

Let $\delta'$ 
%{\red $\delta_t$ should be a vector too, } 
denote a deterministic decision rule that describes how a local action  ${a}_{n,t}$ is selected according to the local state  ${s}_{n,t}$ in time slot $t$, i.e.,  ${a}_{n,t}\! =\!\delta'({s}_{n,t}$), and $\omega \!=\! (\delta', \delta',...,\delta')$ is the corresponding stationary deterministic policy for $t\!=\!1,...,T$ \cite[pp.21]{puterman2014}.  
%In general, a power allocation policy is a sequence of decisions that maps a state to an action at time slot $t$. 
 We say that a 
policy $\omega$ is feasible if it satisfies the two constraints: (i)
$\alpha_{n,t} \in {\mathcal{U}}_{n,t}, ~\forall t, n$, (ii) $\sum^N_{n=1} \alpha_{n,t}^2 \leq {\cal P}_{tot}, ~\forall t$.
Let $\Omega$ be the set of feasible policies $\omega$.
%
%\eqref{fess_set} $\forall n$ and total power constrain, i.e., $\sum^N_{n=1}\alpha_{n,t} \leq \alpha_{tot}$.
%
%
%
%For an infinite-horizon MDP, the only
%case of interest is when an optimal stationary deterministic policy exists. 
%
Our objective is to find a stationary deterministic policy $\omega \in \Omega$ that maximizes $ V_\omega(\boldsymbol{s}_1)$ in \eqref{valu_iter2}. We refer to this solution as the sub-optimal policy.  
%
%\begin{align}\label{valu_iter2}
%   &V_\omega(\boldsymbol{s}_1)= \mathbb{E}\Bigg\{\sum^\infty_{t=1}\eta^t r(\boldsymbol{\alpha}_t)|\boldsymbol{s}_1,\pi\Bigg\}
%   \nonumber\\
%    &\text{s.t.}~\alpha_{n,t} \in {\mathcal{U}}_{n,t}, ~\forall t, n,~\sum^N_{n=1} \alpha_{n,t}^2 \leq {\cal P}_{tot}, ~\forall t
%\end{align}
%

\begin{align}\label{valu_iter2}
   &V_\omega(\boldsymbol{s}_1)= \mathbb{E}\Bigg\{\sum^\infty_{t=1}\eta^t \sum_{n=1}^N r\big(\alpha_{n,t}\big)|\boldsymbol{s}_1,\pi\Bigg\}
   \nonumber\\
    &\text{s.t.}~\alpha_{n,t} \in {\mathcal{U}}_{n,t}, ~\forall t, n,~\sum^N_{n=1} \alpha_{n,t}^2 \leq {\cal P}_{tot}, ~\forall t
\end{align}
We note that maximizing $ V_\omega(\boldsymbol{s}_1)$ in  \eqref{valu_iter2} with respect to $\omega$ is significantly simpler than maximizing $ V_\pi(\boldsymbol{s}_1)$ in  \eqref{valu_iter} with respect to $\pi$, i.e., finding the sub-optimal policy is much easier than finding the optimal policy. This is due to the fact that, when  the local action  ${a}_{n,t}$ is selected according to the local state  ${s}_{n,t}$ only in time slot $t$, the global state transition probability $\Pr\big(\boldsymbol{s}_{t+1} |\boldsymbol{s}_t, ,\boldsymbol{\alpha}_t\big)$ in (\ref{trans1}) can be completely decomposed across sensors. 
In other words, we have
%
%Since state vector $\boldsymbol{s}_{t+1}$ only depends on state vector $\boldsymbol{s}_{t}$ and action vector $\boldsymbol{\alpha}_t$, vector $\boldsymbol{s}_t$ is a Markov process  with the following state transition probability
%---------------------------------------------------------
\begin{align*}
    & \Pr\big(\boldsymbol{s}_{t+1}|\boldsymbol{s}_t,\boldsymbol{\alpha}_t\big)=\prod_{n=1}^N\Pr\big(s_{n,t+1}|s_{n,t},\alpha_{n,t}\big)~~~\\
    &\!\!=\!\!\prod_{n=1}^N\!\! \Pr\!\big({b}_{n,t+1}|{b}_{n,t},{\bar{{g}}}_{n,t},{e}_{n,t},{\alpha}_{n,t}\big)\!\Pr({e}_{n,t}|{e}_{n,t-1})\!\Pr({\bar{{g}}}_{n,t}|{\bar{{g}}}_{n,t-1}).
\end{align*}
%---------------------------------------------------------
where
\begin{align*}
   \Pr\big({b}_{n,t+1}|{b}_{n,t},{\bar{{g}}}_{n,t},{e}_{n,t},{\alpha}_{n,t}\big)
  = \begin{cases}
  1~~~\text{if (\ref{b_n,t}) is satisfied}\\
  0~~~\text{otherwise,}
  \end{cases}
\end{align*}
The decomposition of the global state transition probability into the product of the local state transition probabilities directly impacts how the outer expectation $\mathbb{E}\{.\}$ in (\ref{valu_iter2}) is computed and allows the objective function in  (\ref{valu_iter2}) to be decoupled across sensors. 
If there were no total transmit power constraint in (\ref{valu_iter2}), the MDP problem in (\ref{valu_iter2}) would have become completely decoupled across sensors.
%, and each sensor could have solved its local MDP problem. 
The challenge imposed by the total transmit power constraint can be addressed via adopting a uniform Lagrangian multiplier. Similar to 
%
%the sensor can solve their local problems in a fully decentralized manner. However, due to the total transmission power constraint, sensors’ decision making is still couple.
%
%To address the challenge of centralized control and huge complexity of MDP problems, we introduce a distributive and low-complexity solution. We impose a uniform  resource price as in
%
\cite{Fu} we let a uniform Lagrangian multiplier $\lambda_{\boldsymbol{s}_{t}}\!=\!\lambda, \forall \boldsymbol{s}_t$ be associated with the constraint $\left(\sum^N_{n=1} \alpha_{n,t}^2 - {\cal P}_{tot}\right)$. This uniform Lagrangian multiplier allows us to decouple the MDP problem in (\ref{valu_iter2}) across sensors and reduces solving (\ref{valu_iter2})  into solving $N$ smaller MDP problems. While the computational complexity of finding the optimal policy scales exponentially in $N$, we will show that the computational complexity of finding the sub-optimal policy scales {\it linearly} in $N$.
%
%
%
\begin{comment}
Like previous section, we introduce a Lagrange multiplier $\lambda_{\boldsymbol{s}_t}$ associated with the total power constraint of network state as $\boldsymbol{s}_t$. By applying Lagrangian relaxation, we have $\lambda_{\boldsymbol{s}_t}=\lambda$, $\forall \boldsymbol{s}_t$. 
\end{comment}
%
%

We define the Lagrangian value function $\mathcal{X}(s_{n,t},\lambda)$ using the dynamic programming  
\begin{comment}
%------------------------------------------------------------
\begin{align}
    \mathcal{X}_{\pi}(\boldsymbol{s}_1) = \mathbb{E}\Bigg[\sum^\infty_{t=1}\eta^t\sum_{n=1}^N X(s_{n,t},\lambda)|\boldsymbol{s}_1,{\blue{\omega}}\Bigg] 
\end{align}
%---------------------------------------------------------
\end{comment}
%-------------------------------------------------------
\begin{align} \label{new_lag2}
     \mathcal{X}(s_{n,t},\lambda)=\max_{\omega \in \Omega} \Big\{ { \underbrace{X(s_{n,t},\alpha_{n,t},\lambda)}_{=~\mbox{term 1}}}
   +\nonumber\\{\underbrace{\eta
    \sum_{s_{n,t+1}}\Pr\big(s_{n,t+1}|s_{n,t},\alpha_{n,t}\big)\mathcal{X}(s_{n,t+1},\lambda)}_{=~\mbox{term 2}}}
    \Big\},
\end{align}
%--------------------------------------------
where the modified reward function $X(s_{n,t},\alpha_{n,t},\lambda)$  is defined as
%-----------------------------------------------
\begin{equation}\label{La_dec}
X(s_{n,t},\alpha_{n,t},\lambda)= r\big(\alpha_{n,t}\big)\!-\!\lambda\left(\alpha_{n,t}^2\!-\!\frac{{\cal P}_{tot}}{N}\right)
\end{equation}
%---------------------------------------------------------
%{\blue Note that since we apply the relaxation, we use notation $\omega$ instead of $\pi$ in order to make distinguish.}
%===============================================
\begin{comment}
Solving the dual problem of \eqref{cons_long_term} with ${\lambda}$, gives as,
%------------------------------------------
\begin{equation}\label{dual2}
    \underset{{\lambda}\geq 0}{\min U ({\lambda})}
\end{equation}
%---------------------------------------
where $U ({\lambda})$ is the Lagrange dual function and define as 
%---------------------------------------
\begin{equation}
    U ({\lambda}) = \max_{{\blue{\omega}}}\mathcal{X}_{{\blue{\omega}}}(\boldsymbol{s}_1) 
\end{equation}
%-----------------------------------------
{\blue We define $\omega=\{\omega_n\}_{n=1}^N$}.
\end{comment}
%-======================================================
%
With fixed $\lambda$,
the constrained MDP problem in (\ref{valu_iter2}) can be viewed { as $N$ non-constrained %\label{cons_long_term}
MDP problems in (\ref{cons_long_term2})} with the modified network reward function $X(s_{n,t},\alpha_{n,t},\lambda)$ at time slot $t$ given in (\ref{La_dec}). 
%
%-------------------------------------------
\begin{figure*}
    \begin{eqnarray}\label{cons_long_term2}
    \resizebox{1 \hsize}{!}{$ \displaystyle{
    \mathcal{X}(s_{n,t},\lambda)=\max_{\omega \in \Omega} \Big\{ {\underbrace{X(s_{n,t},\alpha_{n,t},\lambda)}_{=~\mbox{term 1}}}
   +\underbrace{{\eta
    \big(\Pr(\alpha_{n,t}=0)\!\sum_{s_{n,t+1}}\Pr\big(s_{n,t+1}|s_{n,t},0\big)\mathcal{X}(s_{n,t+1},\lambda)+ \Pr(\alpha_{n,t} \neq 0)\sum_{s_{n,t+1}}\Pr\big(s_{n,t+1}|s_{n,t},\alpha_{n,t}\big)\mathcal{X}(s_{n,t+1},\lambda)\big)}}_{=~\mbox{term 2}}
    \Big\},}$}
     \end{eqnarray}
     \hrulefill
\end{figure*}
Let  $\widehat{U}(\lambda) $ denote the Lagrangian dual function, where
\begin{equation}
    \widehat{U} (\lambda) = \max_{\omega \in \Omega}\mathcal{X}(s_{n,t},\lambda) 
\end{equation}
Then the Lagrangian dual problem can be written as
%-------------------------------------------
\begin{equation}\label{dual2}
    {\min_{\lambda\geq 0} \widehat{U} (\lambda)}
\end{equation}
To solve the dual problem in \eqref{dual2}, 
we iteratively solve the following two sub-problems until a pre-specified convergence criterion is reached. The outer minimization sub-problem 
%(the outer loop in Algorithm \ref{Alg2} with iteration index $i$) 
%
updates $\lambda^i$. The inner maximization sub-problem 
%(the inner loop in Algorithm \ref{Alg2} with iteration index $j$) 
finds the optimal $\pi_i$ given $\lambda^i$. The pseudo code of the algorithm is given in Algorithm \ref{Alg2}. 
%
%Like in previous section, we can solve \eqref{dual2} problem by solving the following two sub-problems:
\begin{enumerate}
    \item \textbf{the inner maximization problem:}
\begin{comment}

With given $\lambda^i$, %the inner maximization problem can be viewed as a non-constrained MDP problem with instant reward $L(s_{n,t},\lambda)$ from \eqref{La_dec}. 
    the optimality equation of the maximization problem for each sensor is given as follows
    %--------------------------------------
    \begin{align}\label{valu_itr_lag_dec}
         &\mathcal{X}(s_{n,t},\lambda) =\max_{\pi} \Big\{X(s_{n,t},\alpha_{n,t},\lambda)+\nonumber\\
   &\eta\Big(\widehat{\zeta}_{n,0}
    \sum_{{s}_{t+1}}\Pr(s_{n,t+1}|s_{n,t},0)\mathcal{X}(s_{n,t+1},\lambda) +
    \nonumber\\
    &\widehat{\zeta}_{n,1}
    \sum_{{s}_{n,t+1}}\Pr\big(s_{n,t+1}|s_{n,t},\alpha_{n,t}\big)\mathcal{X}(s_{n,t+1},\lambda) \Big)\Big\},
    \end{align}
    %-------------------------------------------------
    %where $V(s_{n,t})$ is the value function. 
    The value iteration algorithm \cite{bertsekas2012} can be adopted to find the optimal policy with fixed multiplier until convergence, i.e.,  $|\mathcal{X}^j(s_{n,t},\lambda^i) -\mathcal{X}^{j-1}(s_{n,t},\lambda^i) |<\epsilon_1(1-\eta)/2\eta$. This step can be done either by sensors individually or by FC. 
    \end{comment}
    %
     Given $\lambda^i$ we adopt the value iteration algorithm \cite{bertsekas2012} to find the optimal policy $\pi_i$. The convergence criterion is $|\mathcal{X}^j(s_{n,t},\lambda)-\mathcal{X}^{j-1}(s_{n,t},\lambda)|<\epsilon_1(1-\eta)/2\eta$, for a given $\epsilon_1$, where  $\mathcal{X}^j(s_{n,t},\lambda)$ indicates the long-term expected reward in the $j$-th iteration from \eqref{cons_long_term2}.
    \item \textbf{the outer minimization problem:}
    %The outer minimization over the multiplier $\lambda$ is a linear programming problem: $\min_{{\lambda}}  U ({\lambda})$, which is the dual problem of the original mode selection problem. We use the sub-gradient method to update the Lagrangian multiplier.
    %The updating process of the multiplier is given as follows
    The outer minimization over the Lagrangian multiplier $\lambda$ is a linear programming problem. 
    %-------------------------------
    %
    We use the sub-gradient method to update $\lambda^i$ as the following 
    %---------------------------------------------
    \begin{equation}\label{update_lam2}
    \lambda^{i+1}=\Bigg[\lambda^{i}+\beta^i\left(\sum_{n=1}^N \alpha^2_{n,t}-\mathcal{P}_{tot}\right)  \Bigg]^+,
   \end{equation}
    %--------------------------------------------
    where  $\beta$ is a positive scalar step size satisfying the conditions $\sum _{i=1}^{\infty}\beta^i=\infty$ and $\sum _{i=1}^{\infty}(\beta^i)^2<\infty$.  The update rule is such that if $\sum_{n=1}^N \alpha_{n,t}^2$ is larger (smaller) than ${\cal P}_{tot}$ then $\lambda$ should increase (decrease).
    Unless the convergence criterion 
 $\frac{|\lambda^{i+1}-\lambda^{i}|}{\lambda^{i}}<\epsilon_2$ is met, for a given $\epsilon_2$, we increase $i$ and solve the inner maximization sub-problem again. 
    
   % { We repeat these two steps for $\forall \boldsymbol{s}\in \bar{\mathcal{S}}$.}

\end{enumerate}

Note that the above sub-gradient method is guaranteed to converge to the optimum $\lambda$,  as long as $\beta$ satisfies the conditions stated above.

\textbf{Remark on the computational complexity of Algorithm \ref{Alg2}:} We switch between solving two sub-problems until the convergence criterion for updating the Lagrangian multiplier is met. Given $\lambda$ we solve the inner maximization sub-problem, i.e., 
we solve \eqref{cons_long_term2}  for each $s_{n,t} \in {\mathcal{S}}$ (refer to Step 3 of Algorithm \ref{Alg2}), where the dimension of ${\mathcal{S}}$,  denoted as  $|{\mathcal{S}}|$.
Our numerical results show that 
the computational complexity of calculating $\omega^*_i$ (Step 5 of Algorithm \ref{Alg2}) is $\mathcal{O}( |\mathcal{S}| K^{1.5})$. On the other hand, the complexity order of the gradient-descent algorithm to find the local minimum of
function $\widehat{U} (\lambda)$  and
converge to an $\epsilon_2$-accurate solution is $\bar{\epsilon}=1/\epsilon_2$ \cite[p.~232]{luenberger}.
Hence, the overall the computational complexity of finding the optimal solution using Algorithm \ref{Alg2} is $\mathcal{O}(N|\mathcal{S}|\bar{\epsilon}K^{{ 1.5}})$. Note that the complexity order scales {\it linearly} in $N$. 

\textbf{Remark on implementing the sub-optimal policy}: The sub-optimal policy, a.k.a. decentralized solution in the dynamic control literature, requires the knowledge of the local state ${s}_{n,t}$ only to determine the local action ${\alpha}_{n,t}=\delta'({s}_{n,t}), \forall n$.
This implies that sensor $n$ in the network can find its local action $a_{n,t}$ at time slot $t$ with the knowledge of its local state ${s}_{n,t}$. Hence, implementing this solution, different from the optimal solution, does not require sensors to report to the FC and does not impose signaling overhead to the sensors.
\begin{algorithm}[!t]
 \caption{{sub-optimal} power control algorithm}
 \label{Alg2}
1: Specify $\epsilon_1>0$, $\epsilon_2>0$ set $\mathcal{X}(s_{n,0},\lambda)=0 ,~ s_{n,t}\in \mathcal{S}$ set $i = 1$  \;
 2: \For{\textnormal{fixed} $\lambda^i$}
 { 
 3: Set $j = 1$, \For{\textnormal{each sensor} } 
 {
 \For{\textnormal{each} $\alpha_{n,t} \in \mathcal{U}_{s_{n,t}}$}{ calculate 
 \begin{align*}
     &\mathcal{F}(s_{n,t},\alpha_{n,t},\lambda^i)= X(s_{n,t},\alpha_{n,t}\lambda)+\\
   &\eta\Big(
    \sum_{{s}_{t+1}}\Pr(s_{n,t+1}|s_{n,t},\alpha_{n,t})\mathcal{X}^{j-1}({s}_{n,t+1},\lambda^i)\Big)
 \end{align*}
  }
  calculate 
  %-------------------------------------------
  \begin{align*}
  \mathcal{X}^{j}({s}_{n,t},\lambda^i) = \underset{\omega_i \in \Omega }{\max}\{\mathcal{F}(s_{n,t},\alpha_{n,t},\lambda^i)\}
  \end{align*}
  %-------------------------------------
  }
 4: If 
 \begin{align*}
 \underset{s_{n,t} \in {\cal S}}{\max} |\mathcal{X}^{j}({s}_{n,t},\lambda^i)-\mathcal{X}^{j-1}({s}_{n,t},\lambda^i)|<\epsilon_1(1-\eta)/2\eta
  \end{align*}
 go to step 5. Otherwise, increase $j$ and go back to Step 3.\\
 5: We obtain policy 
 \begin{align*}
    \omega_i^* =\arg\max_{\omega_i \in \Omega}\Big\{\mathcal{X}^{j}({s}_{n,t})\Big\}.
 \end{align*}
 }
 6: Update $\lambda^i$ by using \eqref{update_lam2} and $\omega_i^{*}$\;
 7: If $\frac{|\lambda^{i+1}-\lambda^{i}|}{\lambda^{i}}<\epsilon_2$ then $\omega=\omega_i^*$. Otherwise, increase $i$ and go back to Step 2.
 \end{algorithm}

 \section{Effect of random deployment of sensors}\label{random_dep}
%----------------------------------------------------
%---------------------------------------------------------
%The widely adopted signal model in  \eqref{xk} relies on the basic assumption that the distances between the signal source to be detected and the sensors in the field are known \cite{cao,vin,maleki1,maleki2}. 
%
%%%%%%%%%%%%%%%%%%%%%%%%%%%%%%%%%%%%%%
Our signal model in \eqref{xk} is a widely adopted model in the literature of signal (target) detection \cite{vin,maleki1,maleki2}, in which the
signal source is typically modeled as an isotropic radiator 
%(with a general intensity decay model), 
and the emitted power of the signal source at a reference distance $d_0$ is known  \cite{niu}. Suppose $P_0$ is the emitted power of the signal source at  the reference distance $d_0$, and $d_n$ is the Euclidean distance between the source and sensor $n$. For a general intensity decay model, the signal intensity at sensor $n$, denoted as $z_{n,t}$,  is \cite{niu}
%\cite{cao}
%---------------------------------------
\begin{equation}
    z_{n,t} = \frac{P_0}{(d_n/d_0)^\gamma},
\end{equation}
%--------------------------------------------
where $\gamma $ is the path-loss exponent, e.g., for free-space wave propagation $\gamma=2$. With this model, the problem of noisy signal detection  %(when the signal is corrupted by the additive Gaussian noise) 
is equivalent to the following  binary hypothesis testing problem 
%-----------------------------------------
\begin{equation}\label{xs}
H_t={1}: x_{n,t} = z_{n,t}+v_{n,t}, ~~~~~H_t={0}: x_{n,t} = v_{n,t},
\end{equation}
%------------------------------------------
in which $z_{n,t}$ and variance of $v_{n,t}$,  denoted as $\sigma_{v_n}^2$,  are assumed to be known \cite{vin}, \cite{maleki1}, \cite{maleki2}. Note that the binary hypothesis testing problem in \eqref{xs} can be recast as the problem in \eqref{xk}, by scaling the sensor observation $x_{n,t}$ with $(d_n/d_0)^\gamma$. 
%
%%%%%%%%%%%%%%%%%%%%%%%%%
%
This signal model applies to an arbitrary, but fixed (given) deployment of sensors. For  applications where the sensors are deployed randomly in a field, the sensors' locations are not known a prior. 
This implies that  $\cal A$ in \eqref{xk} is unknown, and consequently,  $P_{\text{f}_n}$ in \eqref{pd_pf1} cannot be determined before deployment.
To expand our optimization method beyond fixed deployment of sensors, we assume that sensors are randomly deployed in a circle field, the signal source is located at the center of this field, and it is at least $r_0$ meters away from any sensor within the field. Let $r_{n,t}$ be the distance of sensor $n$ from the center. We assume $r_{n,t}$ is uniformly distributed in the interval $(r_0,r_1)$, i.e.,
%--------------------------------------
\begin{equation}\label{r0}
  f(r_{n,t}) =\begin{cases}
  \frac{1}{r_1-r_0},~~r_0<r_{n,t}\leq r_1,  \\
  0,~~~~~~~~\text{o.w.}
  \end{cases}
\end{equation}
%-----------------------------------------
Suppose the emitted power of the signal source at radius $r_0$ is $P_0$. Then the signal intensity at sensor $n$ is $z_{n,t}=\frac{P_0}{(r_{n,t}/r_0)^2}$. Given the pdf of $r_{n,t}$ in \eqref{r0}, we obtain the pdf of $z_{n,t}$ as follows

%-------------------------
\begin{equation}\label{omg}
    f(z_{n,t}) = 
    \begin{cases}
    \frac{\sqrt{P_0}}{2z_{n,t}\sqrt{z_{n,t}}(r_1-r_0)},~~~\dfrac{P_0}{r_1^2}<z_{n,t}\leq \dfrac{P_0}{r_0^2}, \\
    0,~~~~~~~~~~~~~~~~\text{o.w.}
    \end{cases}
\end{equation}
%--------------------------
Based on the pdf of $z_{n,t}$ in \eqref{omg} we can recompute  $P_{\text{f}_n}$ in \eqref{pd_pf1} 
%
%-------------------------------------
\begin{align}\label{pd_pf2}
 &P_{\text{f}_n}=\int_{\frac{P_0}{r_1^2}}^{\frac{P_0}{r_0^2}} \left[Q\Big(\frac{\theta_n+\frac{z_{n,t}^2}{2\sigma^2_{v_n}}}{\sqrt{z_{n,t}^2/{\sigma^2_{v_n}}}}\Big)\right]f(z_{n,t})d z_{n,t}, \nonumber\\
 &P_{\text{d}_n}=\int_{\frac{P_0}{r_1^2}}^{\frac{P_0}{r_0^2}} \left[Q\Big(\frac{\theta_n-\frac{z_{n,t}^2}{2\sigma^2_{v_n}}}{\sqrt{z_{n,t}^2/{\sigma^2_{v_n}}}}\Big)\right]f(z_{n,t})d z_{n,t},   
\end{align}
%--------------------------------------
\begin{equation}\label{pf-random-d}
P_{\text{f}_n}=\int_{\frac{P_0}{r_1^2}}^{\frac{P_0}{r_0^2}} Q\left(Q^{-1}(\overline{P}_{\text{d}})+\sqrt{z_{n,t}^2/{\sigma^2_{v_n}}}\right)f(z_{n,t})d z_{n,t}.
\end{equation}
With random deployment of sensors, problem (P1) is still valid, with the difference that, for $J_{n,t}$ in \eqref{sim_j},  $P_{\text{f}_n}$ expression should be replaced with the ones in \eqref{pd_pf2}-\eqref{pf-random-d}.
%To investigate the effect of random deployment on $P_e$ of the optimized system, we consider (P1), with the difference that, for $J_n$  in \eqref{sim_j}, $P_{\text{d}_n}$ and $P_{\text{f}_n}$ expressions are replaced with the ones in \eqref{pd_pf2}. 

%%%%%%%%%%%%%%%%%%%%%%%%%%%%%%%%%%%%%%%%%%%%%%%%%%%
%------------------------------------
\begin{figure*}[!t]
 \begin{subfigure}[t]{0.24\textwidth}
 \centering
  \includegraphics[width=47mm]{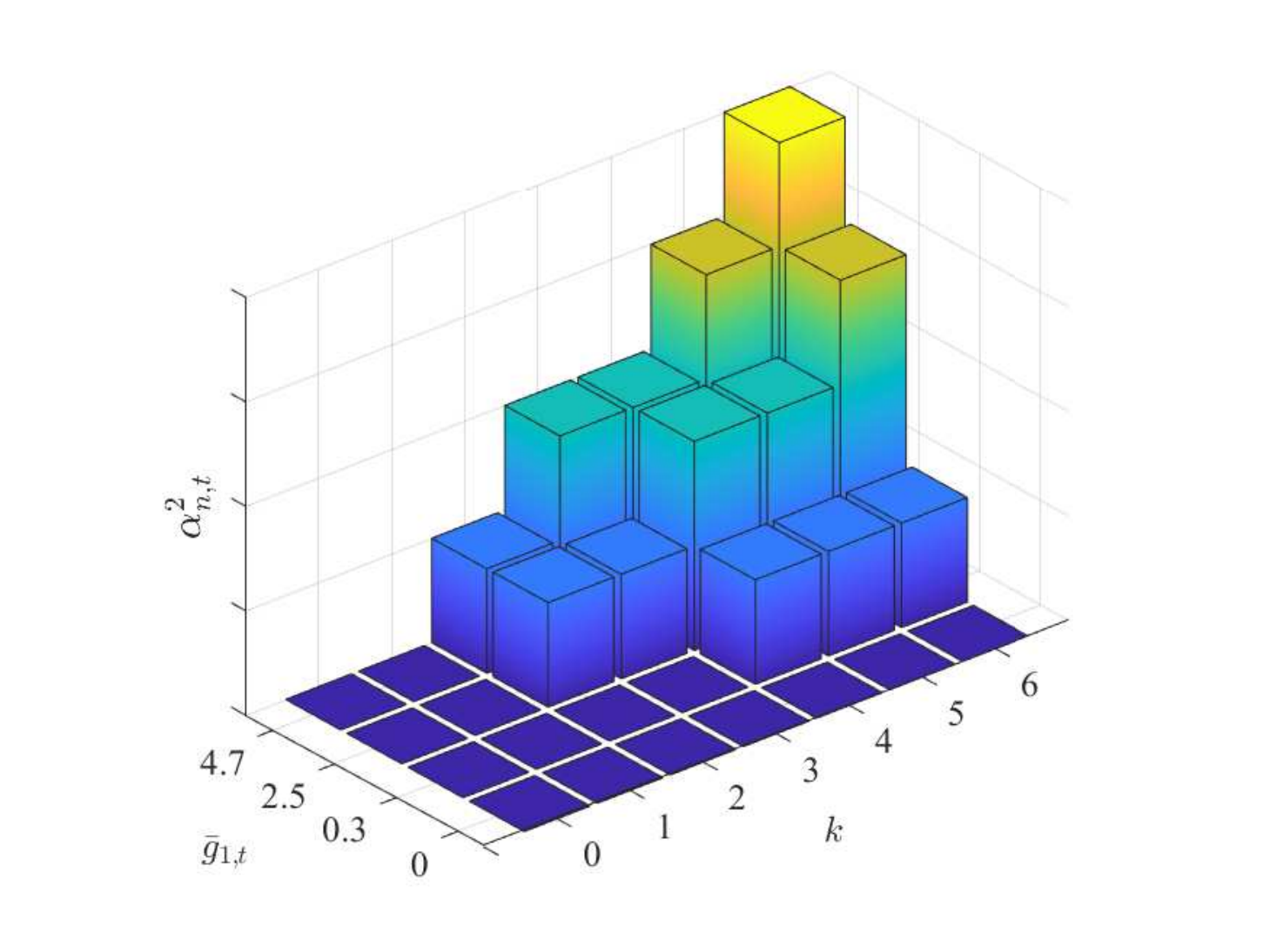}
 \caption{Optimal policy, sensor 1}
\label{f0}
\end{subfigure}
\begin{subfigure}[t]{0.24\textwidth}
  \centering
  \includegraphics[width=47mm]{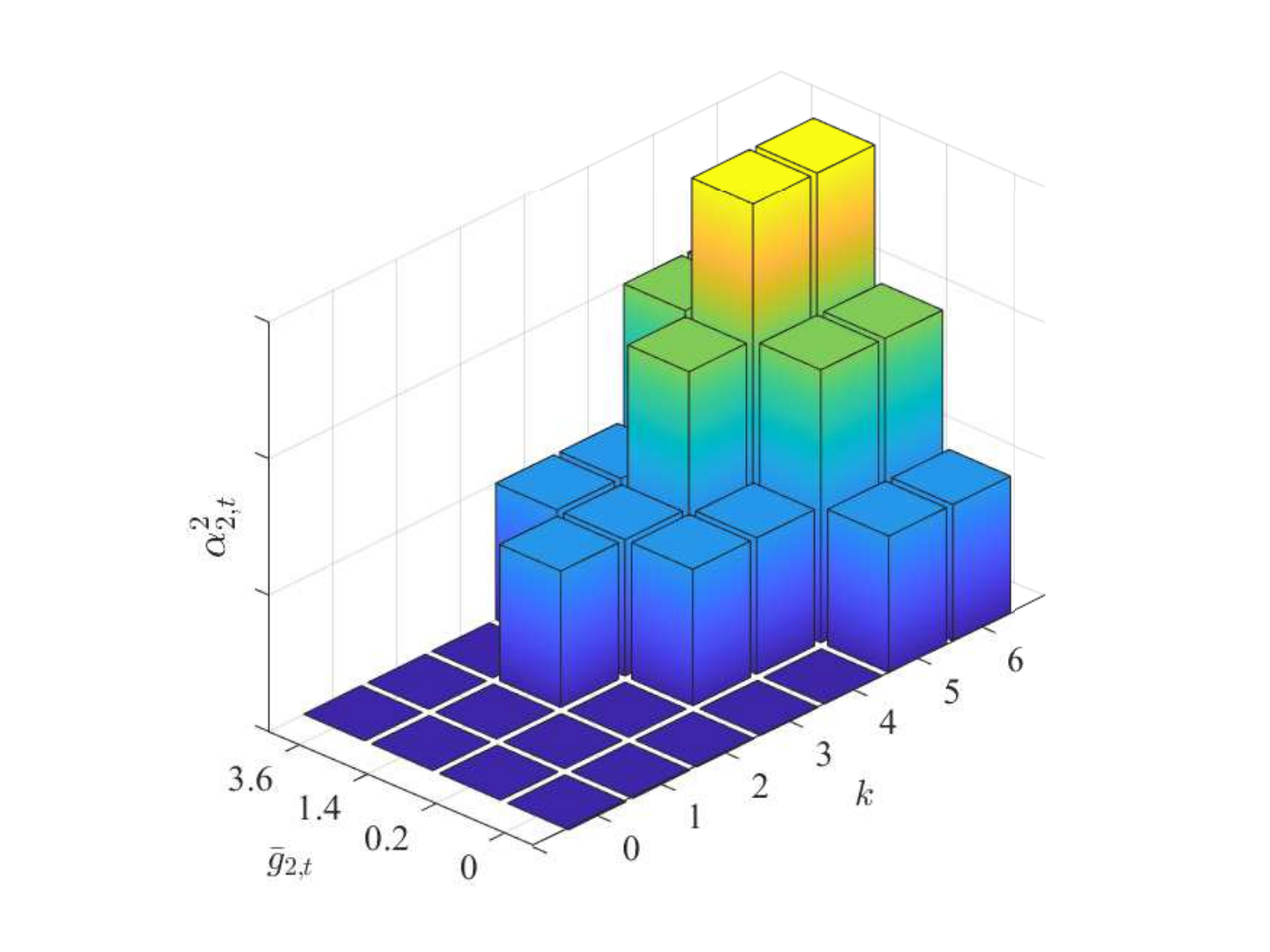}
 \caption{Optimal policy, sensor 2}
\label{f1}
\end{subfigure}
\begin{subfigure}[t]{0.24\textwidth}
  \centering
  \includegraphics[width=47mm]{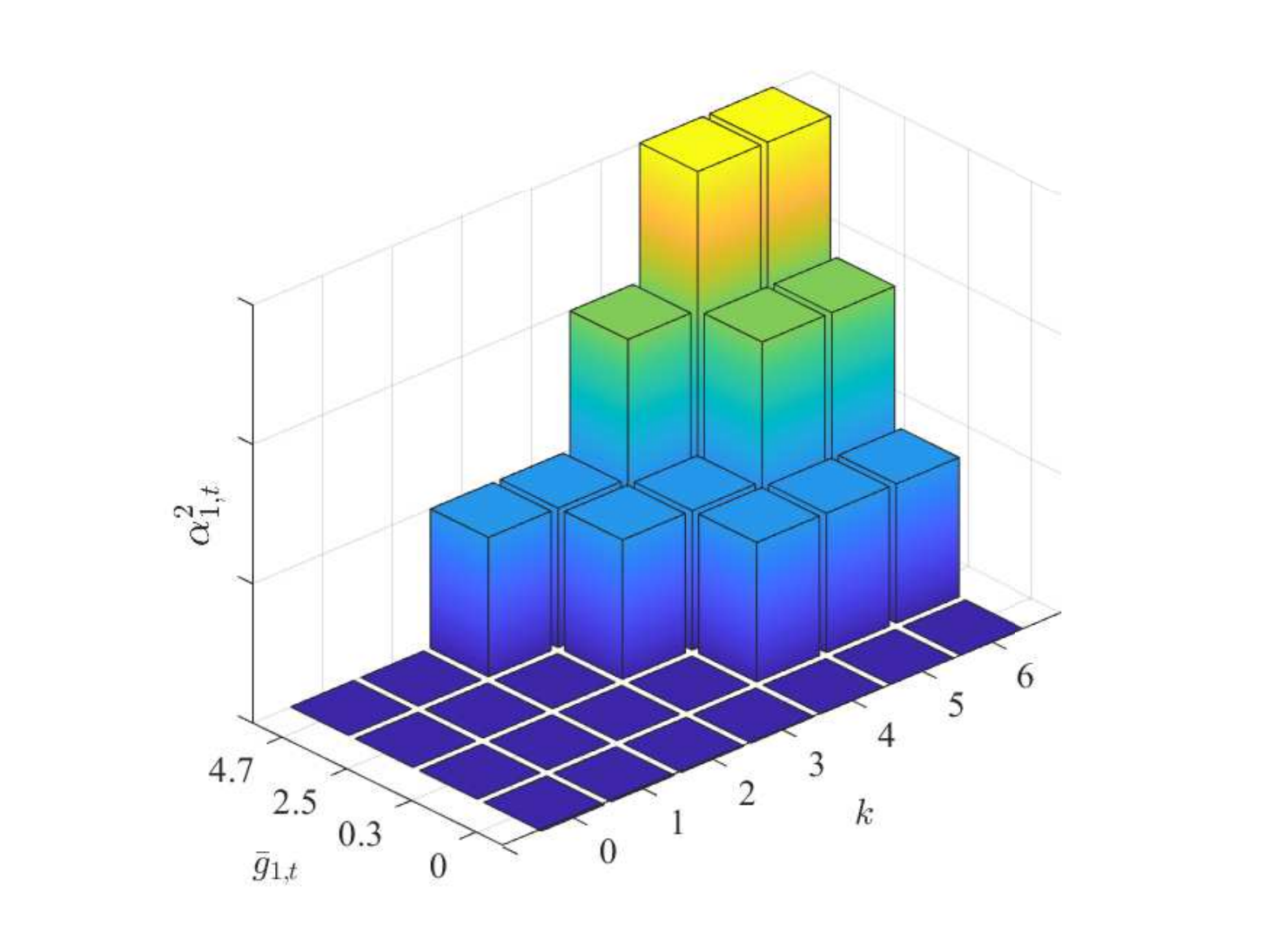}
 \caption{Sub-optimal policy, sensor1}
\label{f2}
\end{subfigure}
\begin{subfigure}[t]{0.24\textwidth}
  \centering
  \includegraphics[width=47mm]{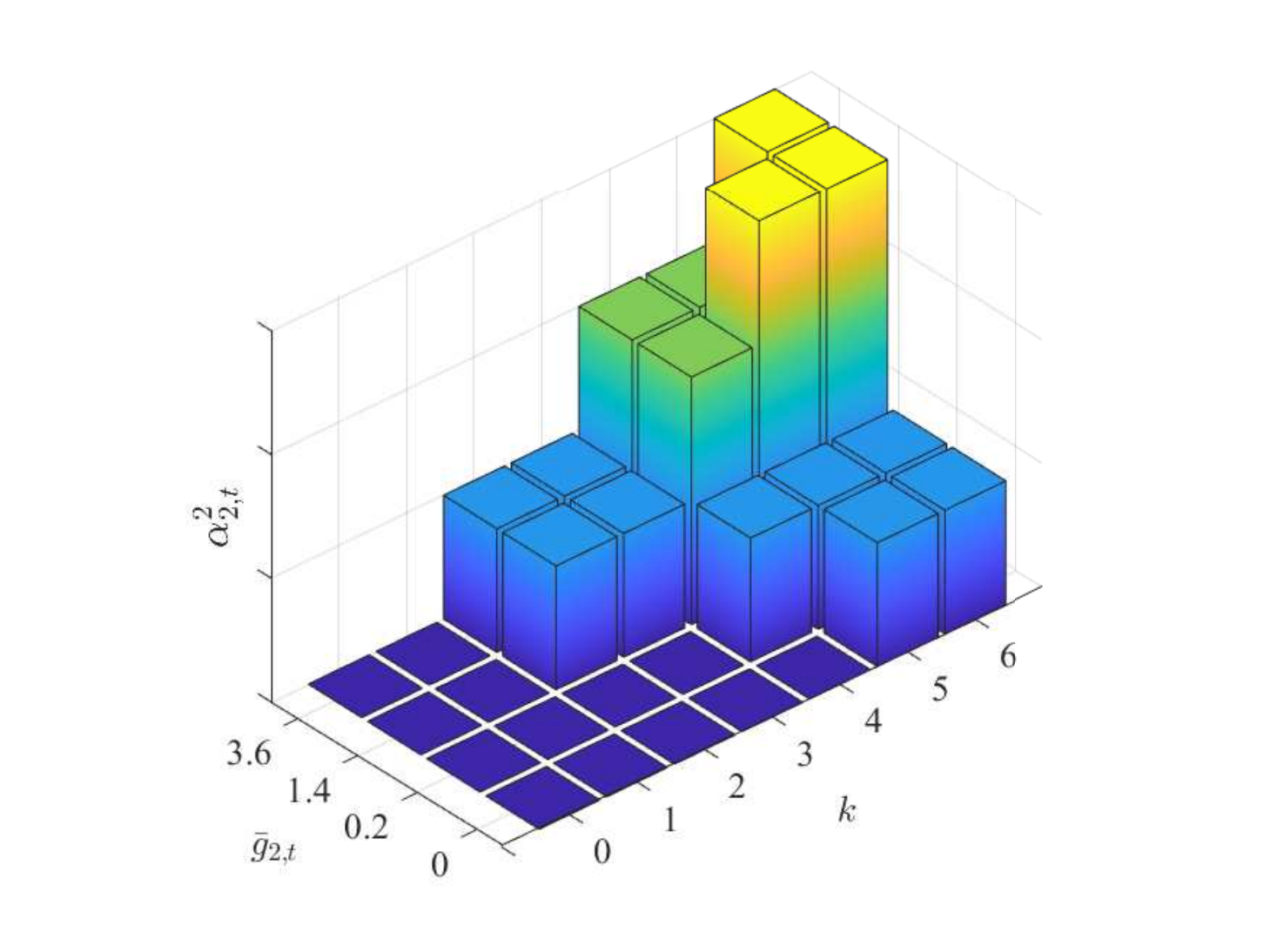}
 \caption{Sub-optimal policy, sensor 2}
\label{f3}
\end{subfigure}
\caption{  
Transmit powers $({\alpha}^2_{1,t},{\alpha}^2_{2,t})$ in optimal and sub-optimal policies for $N\!=\!2,~  \mbox{SNR}_s\!=\!3 \mbox{dB}, \mathcal{P}_{tot} \!=\! 5\mbox{mW}, K\!=\!6,b_u\!=\!0.5\mbox{mJ}, (\gamma_{g_1},\gamma_{g_2})\!=\! (1, 1.5), f_D T_s \!=\!0.04,  L\!=\!4, \bar{\mathcal{G}_1}\!=\!\{0,0.3,2.5,4.7\},\bar{\mathcal{G}_2}\!=\!\{0,0.2,1.4,3.6\},~  (\rho_1, \rho_2)\!=\!(0.4, 0.5), (e_{1,t},e_{2,t}) \!=\!(2b_u,2b_u)$.}
\label{f4}
\end{figure*}
%------------------------------------------------------
\section{simulation Results}\label{simulation}
We corroborate our analysis with MATLAB simulations and investigate: (i) the effect of policy (optimal versus sub-optimal) on transmit powers of sensors, (ii)  the achievable $J$-divergence  when we adopt the optimal, and sub-optimal, and random policies to set transmit powers of sensors,  (iii) error probability $P_e$ when we adopt the optimal and sub-optimal policies to set transmit powers of sensors,  and the trade-off between $P_e$ and consumed transmit power, (iv) the behavior of  $P_e$ as different system parameters vary, 
(v) the effect of random deployment of sensors on $P_e$. 

\begin{figure}[!t]
\centering
\includegraphics[width=70mm]{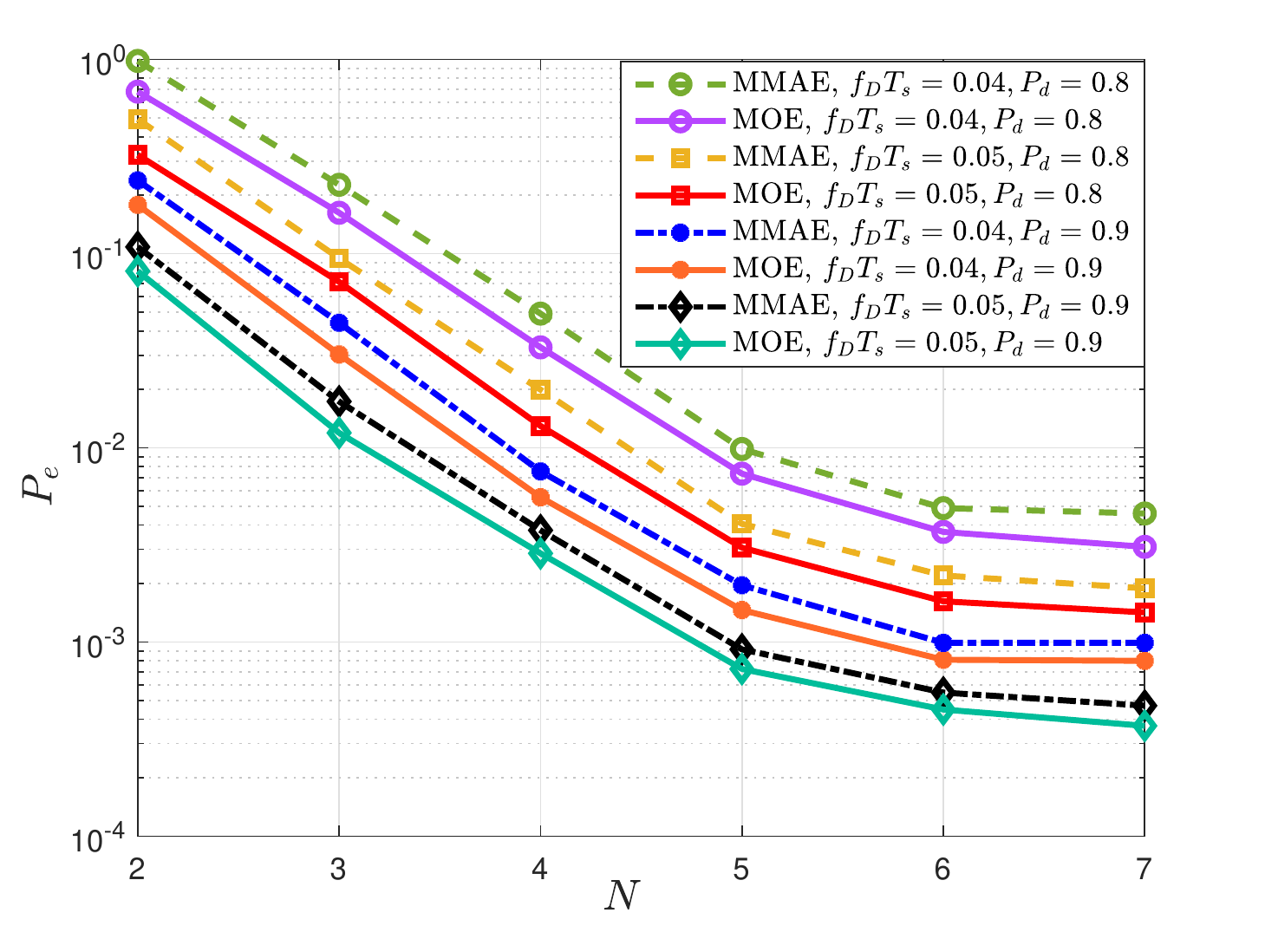}
 \caption{$P_e$ vs. $N$ for $K\! =\! 10,\mathcal{P}_{tot}\!=\!15$mW, $b_u\!=\!2$mJ $\gamma_{g}\!=\!1.5,L\!=\!4,$ SNR$_s\! =\!3$dB.}
\label{f9}
\end{figure}

\begin{figure}[!t]
\centering
\includegraphics[width=70mm]{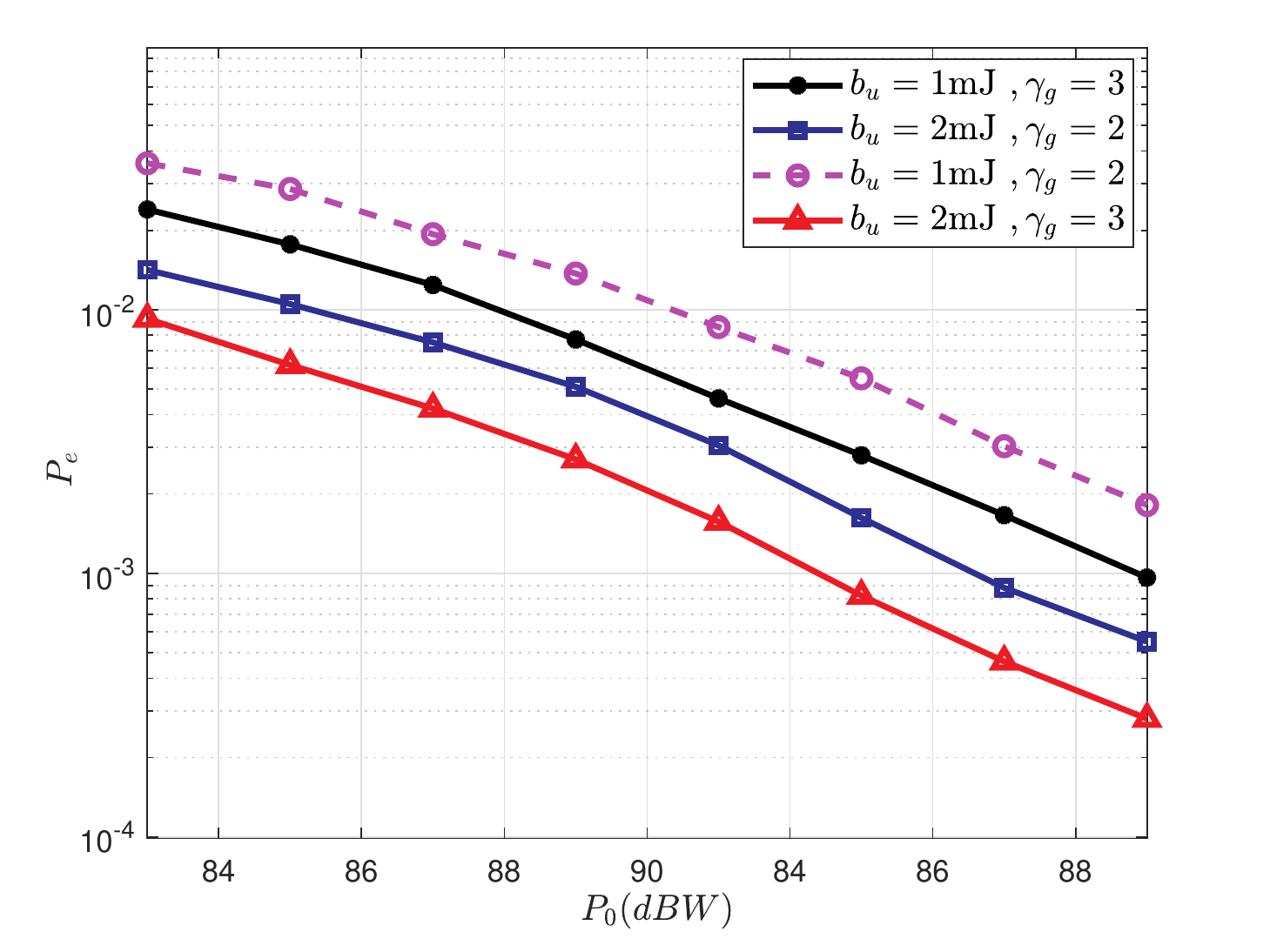}
 \caption{$P_e$ vs. $P_0$ for $K \!=\! 10,N\!=\!5, \mathcal{P}_{tot}\!=\!15$mW$, L\!=\!4,(r_0,r_1)\!=\!(1$m,$100$m). }
\label{f11}
\end{figure}

\begin{figure}[!t]
\centering
\includegraphics[width=70mm]{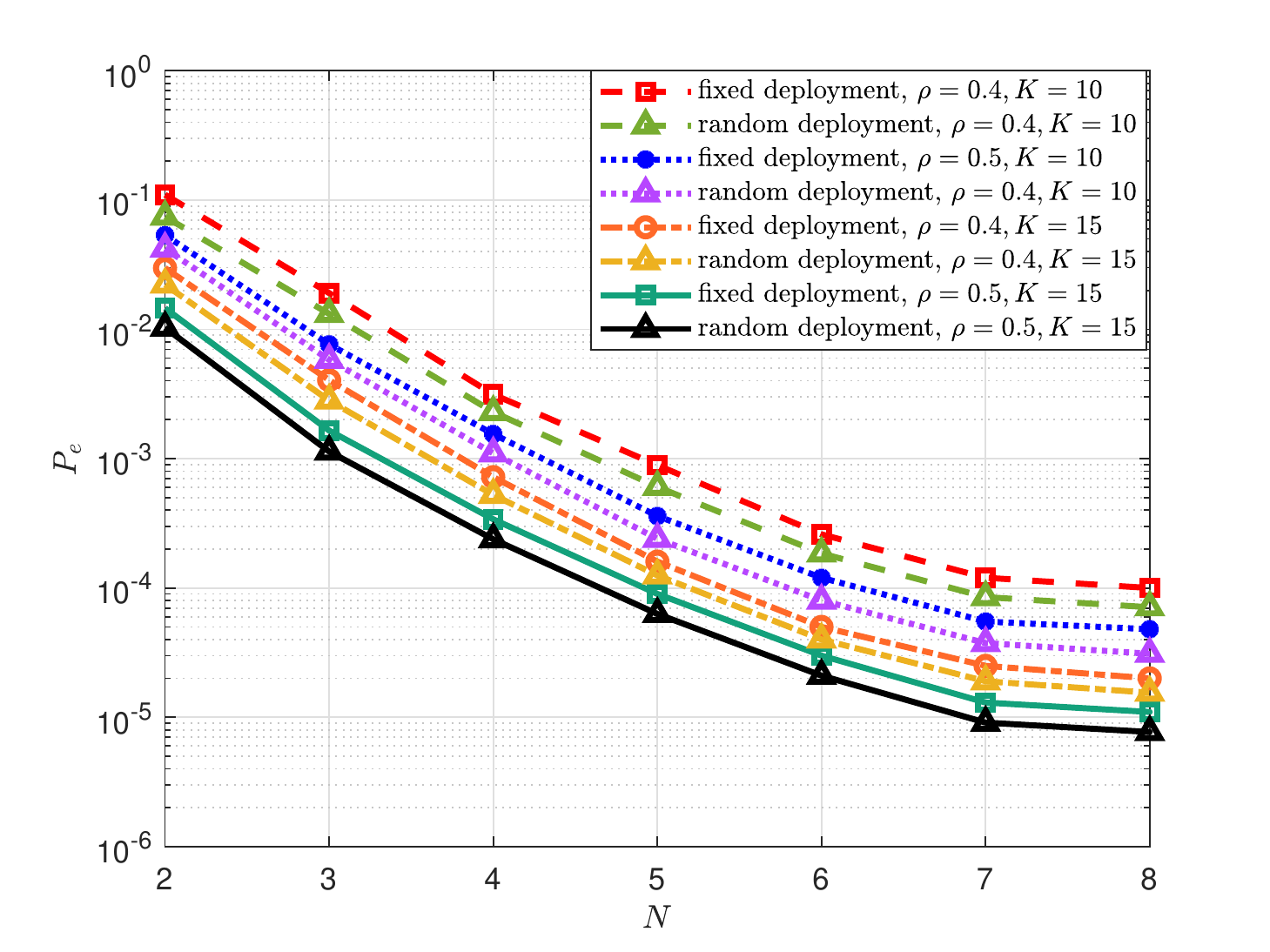}
 \caption{$P_e$ vs. $N$ for $\mathcal{P}_{tot}\!=\!15\mbox{mW},  b_u\!=\!4\mbox{mJ},\gamma_{g}\!=\!1.5, L=4$ (i) for fixed deployment SNR$_s\!=\!5$dB, $P_{\text{d}}\!=\!0.9$, (ii) for random deployment $(r_0,r_1)\!=\!(1$m,$100$m), $P_0 \!=\!84$dBW. }
\label{f12}
\end{figure}

 In all our simulations, we let $\sigma_{w_n}^2 \!=\!\sigma_w^2 \!=\!1,\forall n$ and $P_{d_n}\!=\! P_d,\forall n$. Also,  $\gamma_{g_n}^2\!=\!\gamma_g^2, \forall n$ except for Fig. 
\ref{f4}. We let $P_{d}\!=\!0.9$  except for Fig. \ref{f9}, the discount factor $\eta\!=\!0.9$ except for Fig. \ref{f11}, and $f_D T_s=0.05$ except for Fig. \ref{f9}.  %{\red what is $f_D T_s$ in the figures 6,7,8,9,11,12,13,14? it is missing from captions. Is $f_D T_s$  the same except for Fig. 10? is $\sigma_w^2=1$ in all figures?}
We assume the Gaussian observation noise variance $\sigma_{v_{n}}^2\!=\!\sigma_v^2,\forall n$  and we define the SNR corresponding to observation channel as SNR$_s\!=\!20\log(\mathcal{A}/\sigma_v)$.  
We adopt a solar-power energy harvesting model  similar to \cite{Ku}, in which the harvesting condition is classified  to $M\!=\!4$  states as ``Poor", ``Fair", ``Good", and ``Excellent". We assume $ ~\mathcal{E}\!=\!\{{ 0 },2b_u,4b_u,6b_u\}$  and the transition probability matrix $\Phi_{\mathcal{E}}$ is characterized in terms of an energy harvesting parameter $\rho$ as the following
%{\red why set has 3 elements and matrix is $4 \times 4$? can you find a model for energy harvesting that links the pdf statistics of harvesting to its transition probability, like channel gains?}
%--------------------------------
\begin{equation*}
\small
\Phi_{\mathcal{E}}=
\begin{pmatrix}
\rho & 1 - \rho & 0 & 0\\    
\frac{1 - \rho}{2}& \rho&\frac{1 - \rho}{2} & 0 \\           
0 & \frac{1 - \rho}{2}& \rho&\frac{1 - \rho}{2}\\ 
0 & 0 & 1 - \rho& \rho\\
\end{pmatrix}.
\end{equation*}
%------------------------------------
{ We let $\rho\!=\!0.5$ except for Figs. \ref{f4} and \ref{f12}.}
Our battery-related parameters are $(K,b_u)$. 
Our system setup is based on a given set of 
{ $L$ channel gain quantization thresholds $\bar{\mathcal{G}} \!=\! \{\mu_{1},\mu_{2},...,\mu_{L}\}$. To explore the effect of quantization thresholds we consider two different objective functions to obtain the quantization thresholds $\{\mu_{l}\}_{l=1}^L$.

$\bullet$ \textit{Finding $\{\mu_{l}\}_{l=1}^L$ via Minimizing Mean Absolute Error (MMAE)}:
 We consider mean of absolute quantization error (MAE), denoted as $\mathbb{E} \{|g_{n,t}-\bar{g}_{n,t}|\}$, as the objective function 
\begin{equation}
 \mathbb{E} \{|g_{n,t}-\bar{g}_{n,t}|\} = \sum_{l=0}^{L-1} \int^{\mu_{l+1} }_{\mu_{l}} (x-\mu_{l}) f_{g_{n,t}}(x)dx.
\end{equation}
%----------------------------------------------
To find $\{\mu_{l}\}_{l=1}^L$ that minimize MAE, 
we take the derivative of MAE with respect to $\mu_{l}$ and set the derivative equal to zero. 

$\bullet$ \textit{Finding $\{\mu_{l}\}_{l=1}^L$ via Maximizing output Entropy (MOE)}: We consider  the mutual information between $g_{n,t}$ and $\bar{g}_{n,t}$, denoted as $I(g_{n,t}; \bar{g}_{n,t})$, as the objective function, where
 $I(g_n;\bar{g}_{n,t}) \!=\!  H(\bar{g}_{n,t}) \!-\!  H (\bar{g}_{n,t}|g_{n,t})$,
%----------------------------------------------------------
and $H(x)$ denotes the entropy of discrete random variable
$x$. To find $\{\mu_{l}\}_{l=1}^L$ that maximize $I(g_{n,t}; \bar{g}_{n,t})$, we note that $H (\bar{g}_{n,t}|g_{n,t})$ is zero, since given $g_{n,t}$, $\bar{g}_{n,t}$ is  known.  Furthermore, $H(\bar{g}_{n,t})$ is maximized when $\bar{g}_{n,t}$ follows a uniform distribution, i.e., we set $\phi_{n,l}\!=\!\Pr(\bar{g}_{n,t} \! =\! \mu_{l})  \!=\!\Pr(\mu_{l} \! \leq \! g_{n,t} \!<\! \mu_{l+1}) \!=\! \frac{1}{L+1}$, and the threshold $\mu_{l}$ can be obtained as  $\mu_{l} \!=\!  \left(-\gamma_{g_{n}}\ln{\left(1-\frac{l-1}{L+1}\right)}\right)^{\frac{1}{2}}$.
}

Overall, our channel-related parameters are $(\gamma_{g_n},f_D T_s, L, \bar{\mathcal{G}})$, where $\bar{\mathcal{G}}$ depends on $(\gamma_{g_n},L)$ and the choice of objective function to obtain the quantization thresholds. 
The state probabilities and the entries of the transition probability matrix $\Psi_{\bar{\mathcal{G}}}^{(n)}$
can be obtained via \eqref{gaz}-\eqref{trans_gg} given  $(\gamma_{g_n},f_D T_s, L, \bar{\mathcal{G}})$.

$\bullet$ {\bf Effect of policy on transmit powers of sensors}: To show how transmit power $\alpha_{n,t}^2$ of sensor $n$ changes based on the adopted policy, we consider $N\!=\!2$ sensors. Recall for the sub-optimal policy $\alpha_{n,t}$ depends on the local state $s_{n,t}\!=\!(b_{n,t}, \bar{g}_{n,t-1}, e_{n,t-1})$, whereas for the optimal policy the local action $\alpha_{n,t}$ depends on the global state ${\boldsymbol{s}}_t\!=\!(s_{1,t},s_{2,t})$. 
%
%-----------------------------------------
\begin{comment}
For instance, for $\gamma_{g_n}\!=\!2$ and  $\gamma_{g_n}\!=\!2.5$, respectively, $\Psi_{\bar{\mathcal{G}}}^{(n)}$ are 
%
\begin{align*}
\small
&\Psi_{\bar{\mathcal{G}}}^{(n)}\!=\!\begin{pmatrix}
0.87 &   0.13   &      0     &    0\\    
 0.13  &  0.56  &  0.31    &     0 \\           
0  &  0.27  &  0.52 &  0.21\\   
 0     &    0  &  0.28 &   0.72\\
\end{pmatrix},\\
\small
&\Psi_{\bar{\mathcal{G}}}^{(n)}\!=\!\begin{pmatrix}
0.85 &   0.15   &      0     &    0\\    
 0.16  &  0.52  &  0.32    &     0 \\           
0  &  0.38  &  0.40 &  0.22\\   
 0     &    0  &  0.31 &   0.69\\
\end{pmatrix}
\end{align*}
\end{comment}
%-----------------------------------------------
%
We use Algorithm \ref{Alg1} and Algorithm \ref{Alg2}, to find and set transmit power $\alpha^2_{n,t}$ corresponding to the optimal and the sub-optimal policies, respectively. 
Fig. \ref{f4} illustrates $(\alpha_{1,t}^2, \alpha_{2,t}^2)$ when optimal and sub-optimal policies are adopted, given a set of energy harvesting, battery-related, and channel-related parameters, and assuming the quantization thresholds are obtained via MMAE. To enable the illustration, we assume the state of energy harvesting for both sensors is ``Fair'', i.e.,  $(e_{1,t-1},e_{2,t-1})\!=\!(2b_u, 2 b_u)$, and the states of battery $(b_{1,t},b_{2,t})$ and the states of quantized channel gains    $(\bar{g}_{1,t-1},\bar{g}_{2,t-1})$ are variable.
%
%We assume $K\!=\!6,~N\! =\! 2,~\gamma_{g_1}\!=\!1,~\gamma_{g_2}\!=\!1.5, ~\rho_1 = 0.4,~\rho_2 = 0.5,~ \mathcal{P}_{tot}\! =\! 5$ mW, $e_{1,t}= e_{2,t} = 2b_u \!=\! 1$ mJ.
%
%The transmit power maps show how much power the sensor should spend for its data transmission, given its battery state $b_{1,t}$, the feedback information (i.e., the quantization interval to which the channel gain $g_{1,t}$ belongs) and  energy harvesting state $e_{1,t}$.
%
%
{ %For instance, for the parameters in Fig. (\ref{f0}), when $\bar{g}_{1,t} \in \mathcal{I}_{3}$ and $b_{1,t}=6$, then ${\alpha^2}_{1,t}\!=\!1.5$ mW,  ${\alpha^2}_{2,t}\!=\!2$ mW for optimal policy and ${\alpha^2}_{1,t}\!=\!1 $ mW, ${\alpha^2}_{1,t}=3 $ mW for sub-optimal policy - seems incomplete to me, with typo, confusing Instead I sugegst the following:
For example, this figure shows that when the local states are $s_{1,t}\!=\!(7,3,2), s_{2,t}\!=\!(7,3,2)$, then transmit powers of sensors corresponding to the optimal policy is $(\alpha_{1,t},\alpha_{2,t})\!=\!(1.5\mbox{mW},2\mbox{mW})$, whereas transmit powers of sensors corresponding to the sub-optimal policy is   $(\alpha_{1,t},\alpha_{2,t})\!=\!(1\mbox{mW},3\mbox{mW})$. The Achievable $J_{tot}$ corresponding to optimal and sub-optimal policies are 11.58 and 10.43, respectively. These figures also show that, $\alpha^2_{n,t}(b_{n,t}, \bar{g}_{n,t-1}, e_{n,t-1})$ is monotonically increasing in $b_{n,t}$,  given $\bar{g}_{n,t-1}$ and
$e_{n,t-1}$. 
%
%Here $J_{tot}$ for optimal policy is 11.58 and for sub-optimal policy is 10.43.

%Question: 
%what are  the corresponding $J_{tot}$ values? 
}

%{\red I think it is better to have $N=2$ and plot power of both sensors. Also, we can show the effect of changing quantization thresholds according to MAE and MOE.  }
%--------------------------------------------
\begin{figure}[!t]
\centering
\includegraphics[width=70mm]{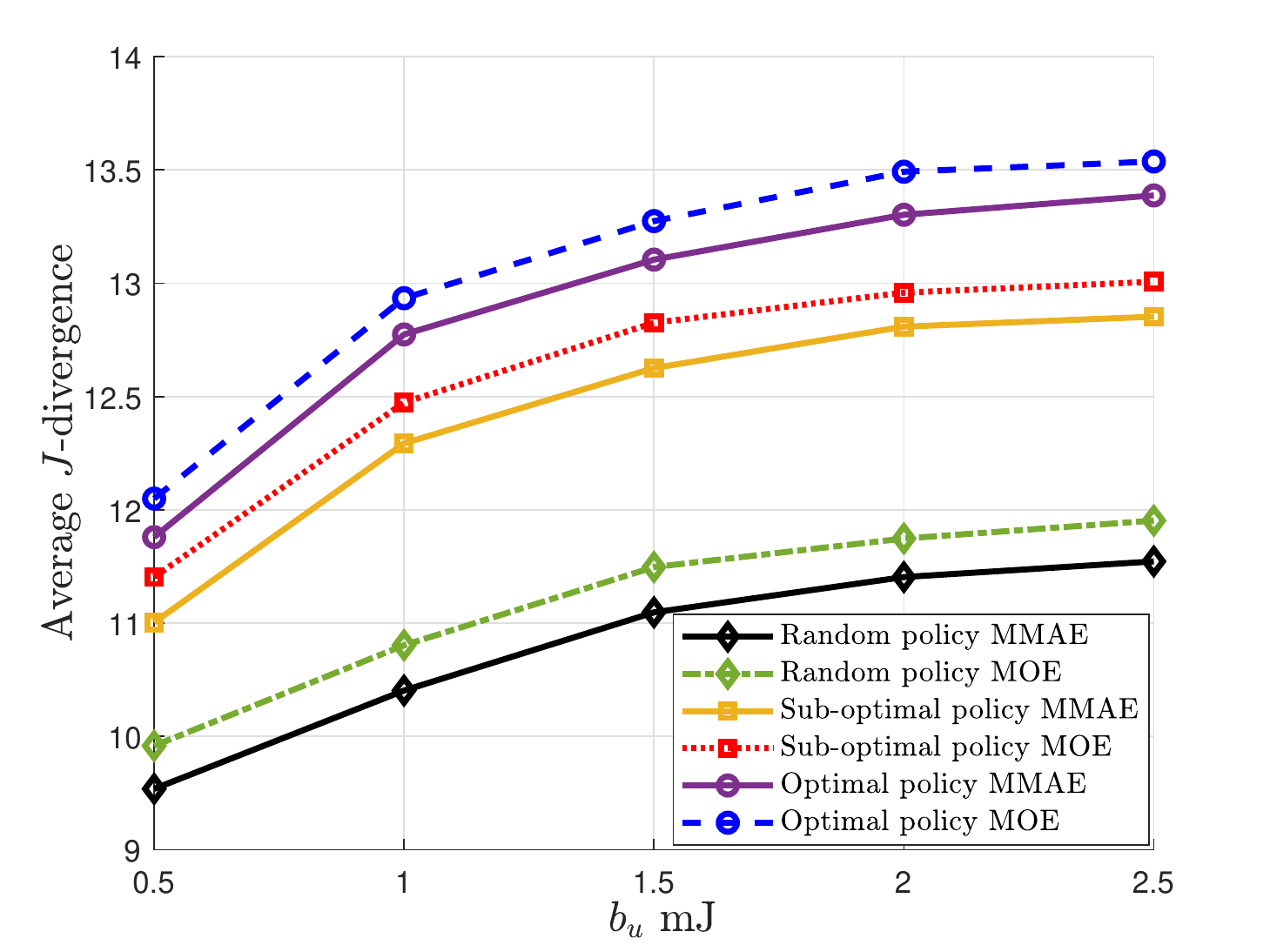}
 \caption{The average $J$-divergence versus $b_u$  for $K\!=\!5, N\!=\!3, \mathcal{P}_{tot} \!=\! 5 $mW$,\gamma_{g}\!=\!2, L\!=\!3$, SNR$_s\! =\! 3$dB.}
\label{f5}
\end{figure}
%----------------------------------------------------
%--------------------------------------------
\begin{figure}[!t]
\centering
\includegraphics[width=70mm]{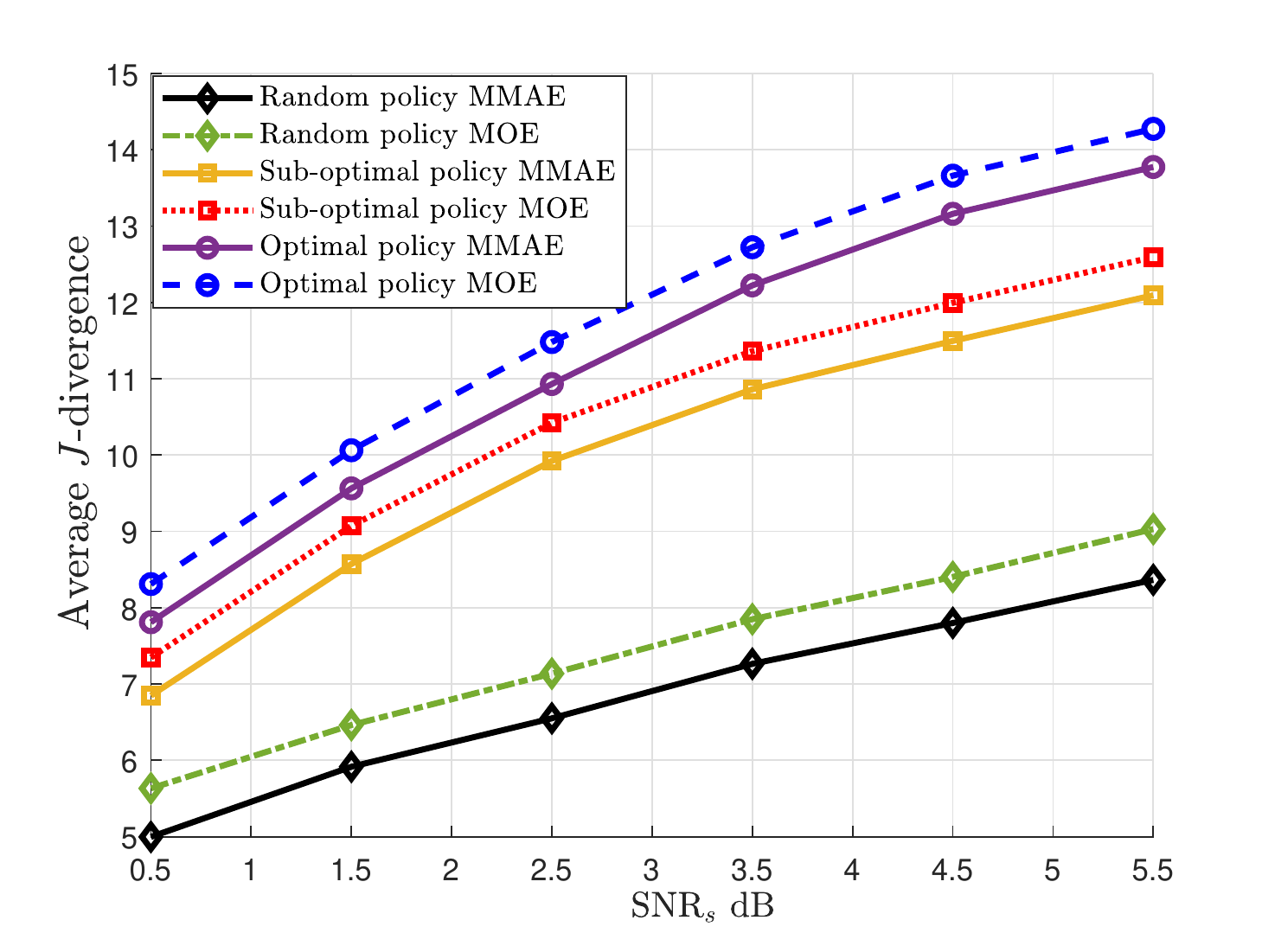}
 \caption{The average $J$-divergence versus SNR$_s$ dB for $K\!=\!5,~ N\!=\!3, \mathcal{P}_{tot} \!=\! 5 $mW$, b_u \!=\! 1\mbox{mJ}, \gamma_{g}\!=\!2, L\!=\!3.$}
\label{f6}
\end{figure}
%----------------------------------------------------
%--------------------------------------------
\begin{figure}[!t]
\centering
\includegraphics[width=70mm]{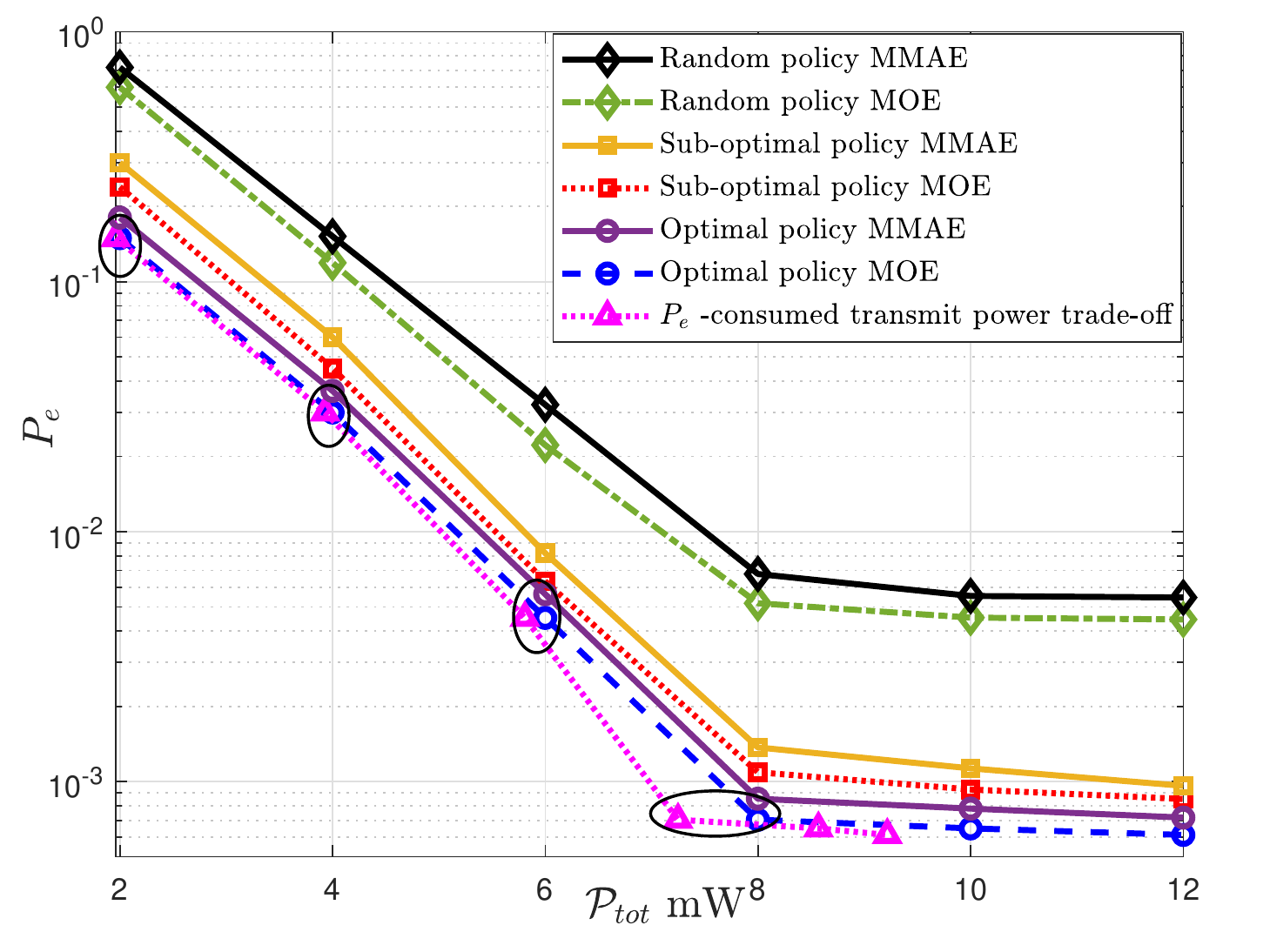}
 \caption{$P_e$ versus $\mathcal{P}_{tot}$  for $K\!=\!5,~N\!=\!3,~b_u\!=\!1\mbox{mJ},~\gamma_{g}\!=\!2,~ L\!=\!3$, ~SNR$_s\! =\! 3$dB.}
\label{f7}
\end{figure}
%----------------------------------------------------

$\bullet$ {\bf Achievable $J$-divergence corresponding to optimal, sub-optimal, and random policies}:
Fig.~\ref{f5} and Fig.~\ref{f6}  show  average $J$-divergence  versus $b_u$ and SNR$_s$ respectfully. 
 To plot the curves
 %labeled as ``optimal policy MMAE'' and ``sub-optimal policy MMAE'' we obtain the quantization thresholds via MMAE, 
%
%``Optimal policy MOE'' and  , ``Sub-optimal MOE'' first we obtain the $\{\mu_l\}$'s with MMAE and MOE method respectfully,} 
%
we set transmit power $\alpha^2_{n,t}$ corresponding to the optimal, the sub-optimal, and random policies, and then
 we  average $J_{tot,t}\!=\!\sum_{n=1}^N J_{n,t}$ over $10^4$ independent Monte Carlo runs. 
%we generate 10000 realizations of random noises and fading channels - 
%with respect to algorithm 1 or 2, which part is repeated 10000 times and taken average over?}. 
%
%In order to examine the performance of optimal and sub-optimal policy, we added a ``random policy" to our simulations in this section. 
%{\blue For the random policy curve we applied MOE method to obtain $\{\mu_l\}$'s}.
For random policy, we randomly choose $\alpha^2_{n,t}$ such that the two constraints in (\ref{valu_iter_time}) are satisfied, i.e., 
(i)
$\alpha^2_{n,t}T_s/b_u \leq b_{n,t}, ~\forall t, n$, (ii) $\sum^N_{n=1} \alpha_{n,t}^2 \leq {\cal P}_{tot}, ~\forall t$.
Fig. \ref{f5} illustrates that, given a $K$ value, the average $J$-divergence increases in $b_u$, however, it remains almost the same after $b_u$ reaches and exceeds a certain value. {This is due to the fact that, for larger $b_u$ values transmit power is not limited by energy harvesting and stored energy. Instead, it is limited by the communication channel noise variance $\sigma_w^2$.}
{ Fig. \ref{f6} 
%illustrates the average $J$-divergence versus SNR$_s$ for optimal, sub-optimal and random policy. For all the policies, we investigate MMAE and MOE methods. 
shows that  the average $J$-divergence increases in SNR$_s$.} This is due to the fact that as SNR$_s$ increases,  $P_{f_n}=P_f, \forall n$ in (\ref{pd_pf1}) decreases  (given a $P_d$ value). Decreasing $P_f$ leads into increasing the average $J$-divergence, where  $P_f$  and $J_{n,t}$ are related through 
(\ref{sim_j}) and (\ref{ABCD}).
%
%{\blue what you said about $P_f$ makes more sense. I'll run simulation to verify it. Update: increasing SNR$_s$, makes $\alpha^2$  increase, but it's not significant. Decreasing $P_f$ while SNR$_s$ is increasing, is the main reason of increasing J-divergence.} 
In both figures, average $J$-divergence achieved by the sub-optimal policy is smaller than average $J$-divergence achieved by the optimal policy, and larger than average $J$-divergence achieved by the random policy.

$\bullet$ {\bf $P_e$ corresponding to optimal and sub-optimal policies, and  $P_e$-consumed transmit power trade-off}:
Fig.~\ref{f7} shows $P_e$ versus ${\cal P}_{tot}$. 
%{\blue To plot the curve  labeled as ``Optimal policy MMAE'', ``Optimal policy MOE'' and ``Sub-optimal MMAE'' , ``Sub-optimal MOE'' first we obtain the $\{\mu_l\}$ with MMAE and MOE method respectfully,} then we obtain {\it optimal} and {\it sub-optimal} policy, set transmit power $\alpha^2_{n,t}$ accordingly, and run Monte-Carlo simulation to find $P_e$. 
%
%
To plot the curves
we set transmit power $\alpha^2_{n,t}$ corresponding to the optimal and the sub-optimal policies, and then
we  
{consider $10^4$ independent Monte Carlo runs to find $P_e$, i.e., we generate $10^4$ realizations of random noises and fading channels and count the errors, $P_e$ is the number of errors occurred divided by $10^4$.}
Fig.~\ref{f7} reveals two important points: (i) ``optimal policy MOE'' and ``optimal policy MMAE'' achieve  the lowest $P_e$, followed by ``sub-optimal policy MOE'' and ``sub-optimal policy MMAE'', followed by  ``random policy MOE'' and ``random policy MMAE'', (ii)   ``sub-optimal policy''  performs very close to ``optimal policy''. Note that for all curves,  $P_e$ decreases as ${\cal P}_{tot}$ increases, however, it reaches an error floor after ${\cal P}_{tot}$ exceeds a certain value.  This is due to the fact that for larger ${\cal P}_{tot}$ values, $P_e$ is not limited by ${\cal P}_{tot}$. Instead, it is limited by  $\sigma_w^2$. 
{
Fig.~\ref{f7} also allows us to examine the existing trade-off between the consumed transmit power and $P_e$.   Consider the curve labeled ``$P_e$-consumed transmit power trade-off'' in Fig.~\ref{f7}, which shows how much transmit power is required to provide a certain $P_e$ value. This curve is obtained from examining the points on  ``optimal policy MOE''   and checking whether the constraint
 $\sum^N_{n=1} \alpha_{n,t}^2 \leq {\cal P}_{tot}, ~\forall t$.
 is active or inactive. At a given point, when this constraint is active (inactive), the consumed transmit power is equal to (less than) ${\cal P}_{tot}$. Note that as ${\cal P}_{tot}$ increases and $P_e$ reaches an error floor, the consumed transmit power is less than ${\cal P}_{tot}$.}

Since finding the sub-optimal policy has a much lower computational complexity than that of the optimal policy, and its performance is very close to the  optimal policy, from this point forward, we focus on the sub-optimal policy. %{\red can you add both quantization methods? can you also like Fig. 5 of J1 add Pe-power tradeoff?}

\begin{figure}[!t]
\centering
\includegraphics[width=70mm]{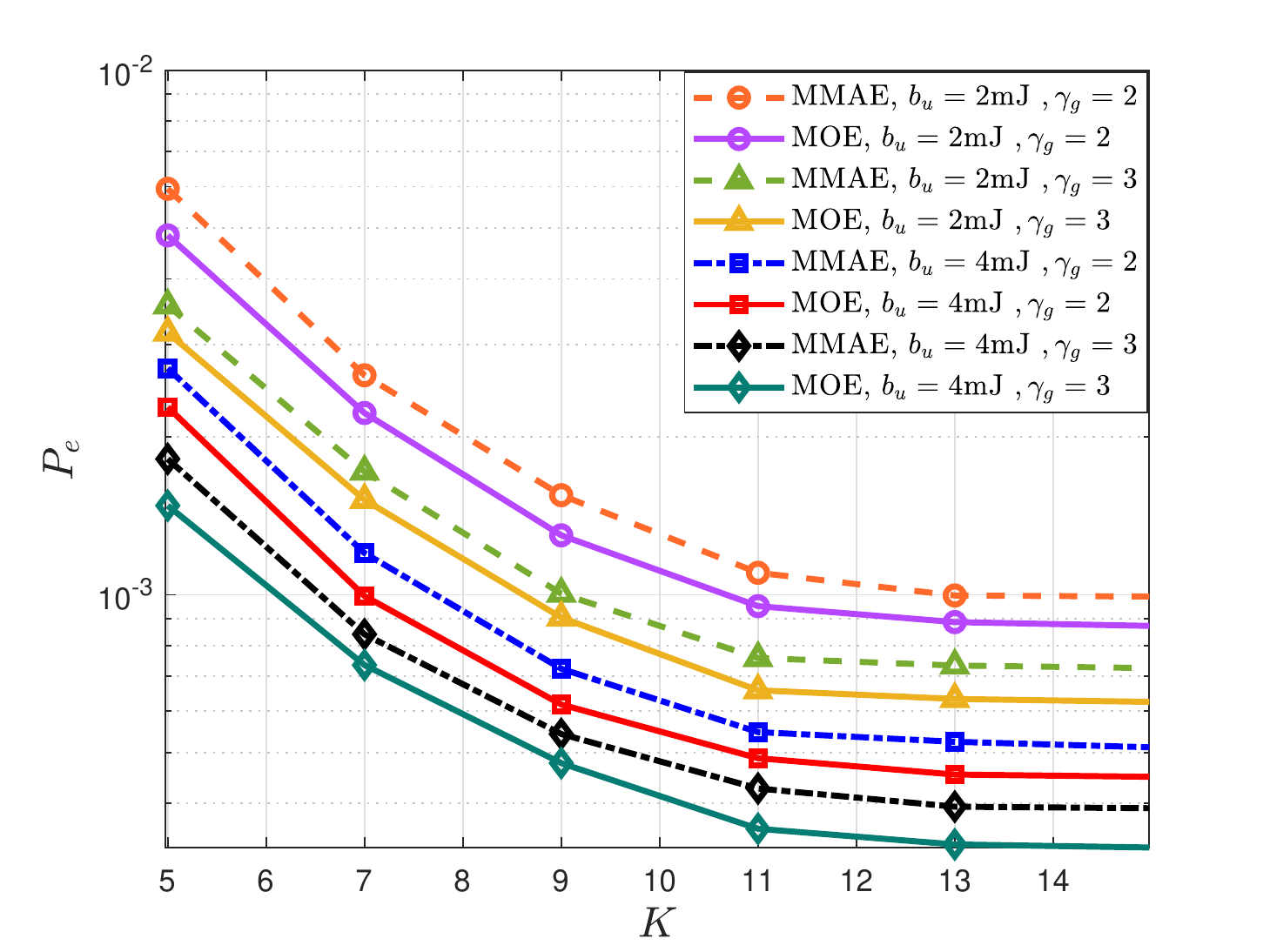}
 \caption{$P_e$ vs. $K$ for $N\!=\!10, \mathcal{P}_{tot}\!=\!15$mW, $L\!=\!4,$ SNR$_s\! =\!5$dB.} 
\label{f8}
\end{figure}

\begin{figure}[!t]
\centering
\includegraphics[width=70mm]{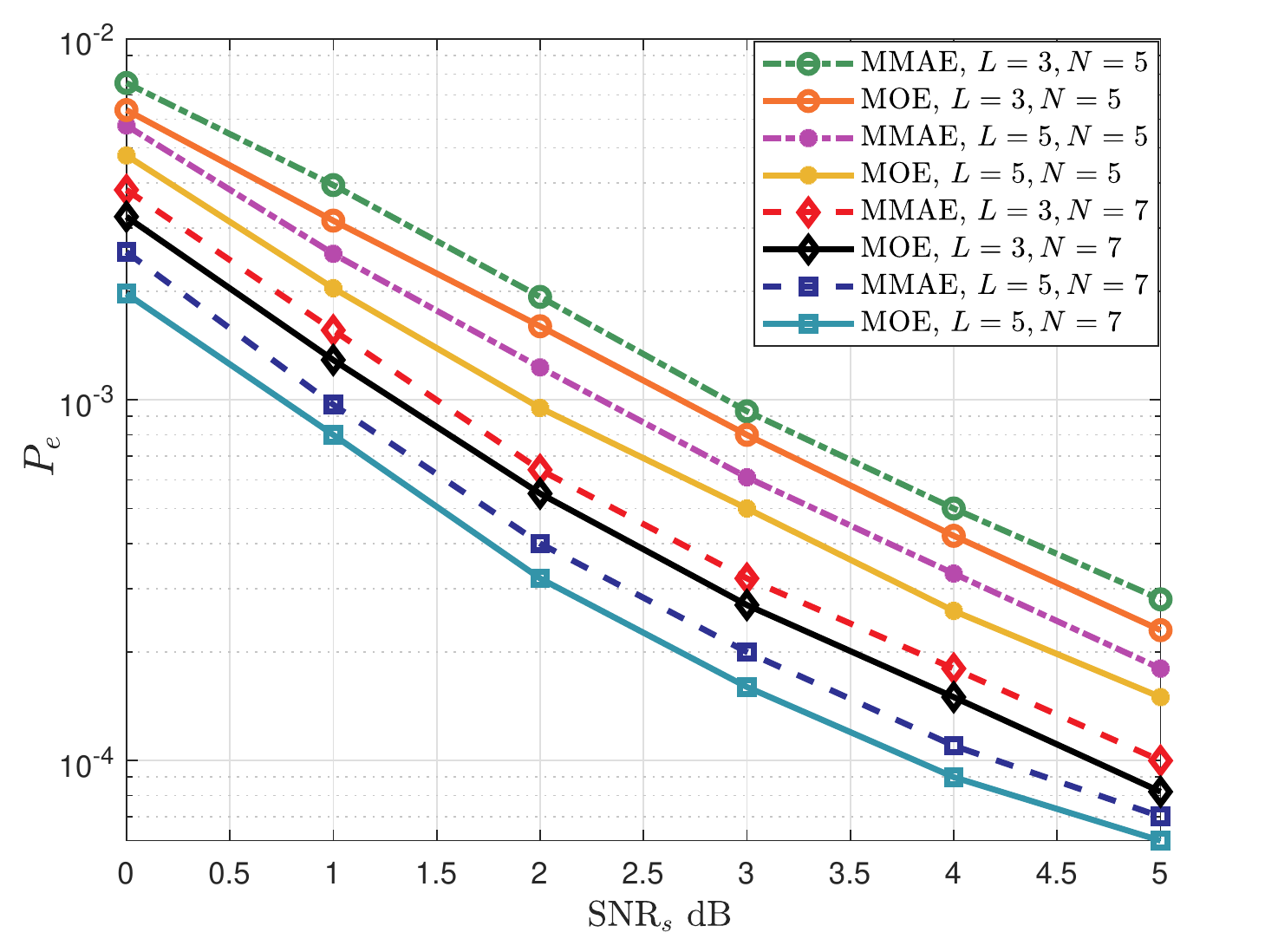}
 \caption{$P_e$ vs. SNR$_s$ for $K \!=\! 10,\mathcal{P}_{tot}\!=\!15$mW, $b_u=2$mJ, $\gamma_{g}\!=\!2$.}
\label{f10}
\end{figure}

\begin{figure}[!t]
\centering
\includegraphics[width=70mm]{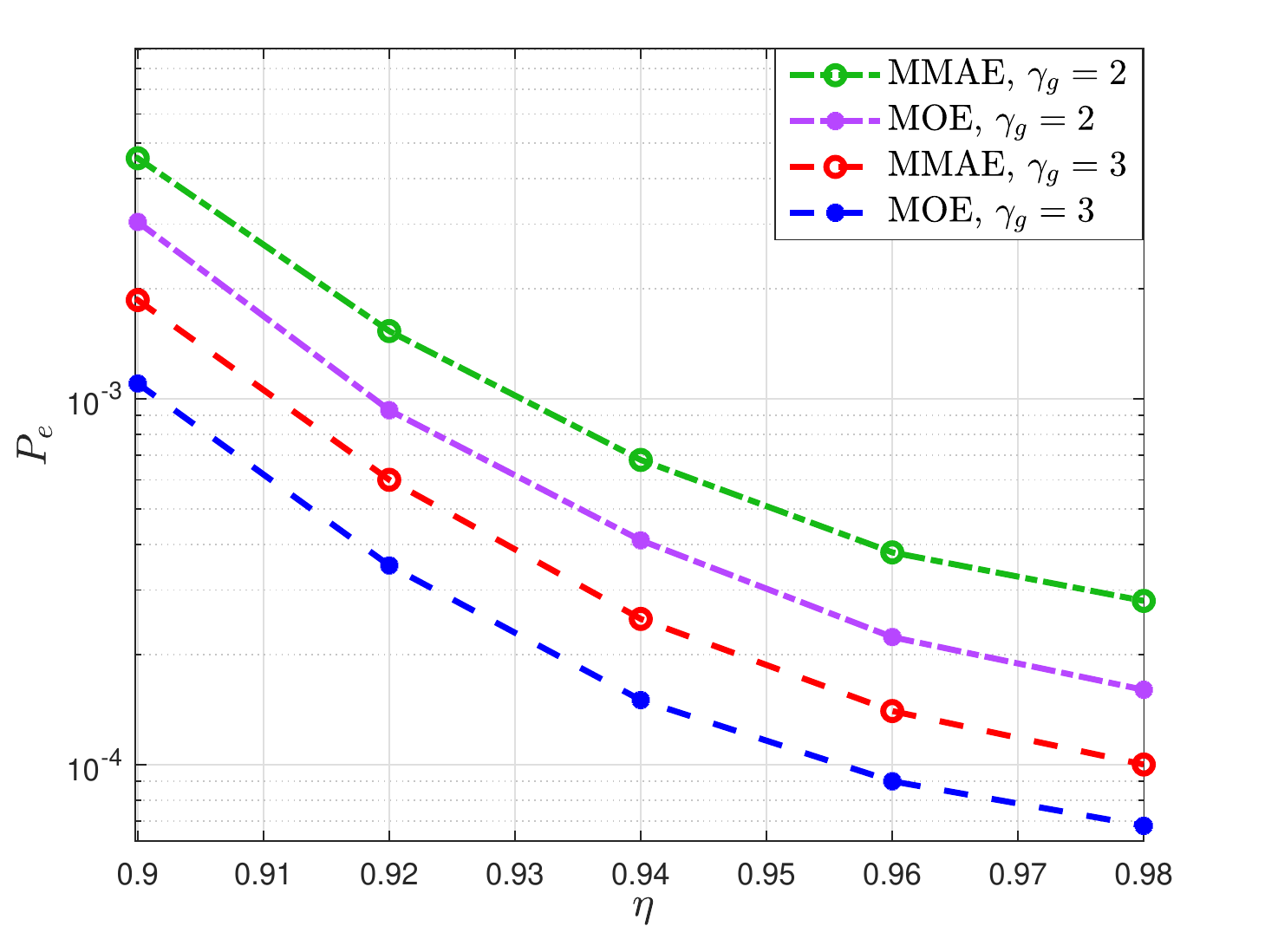}
 \caption{$P_e$ vs. $\eta$ for $K \!=\!10,N=5,\mathcal{P}_{tot}\!=\!15\mbox{mW}, b_u \!=\! 1, L\!=\!4, $SNR$_s=3$dB.}
\label{f14}
\end{figure}

$\bullet$ {\bf Dependency of $P_e$ on different system parameters}:
Fig.~\ref{f8}-\ref{f14} plot $P_e$ corresponding to sub-optimal policy in terms of different system parameters. Fig. \ref{f8} depicts $P_e$ versus $K$ as $\gamma_{g}$ and $b_u$ change.  Given  the pair ($\gamma_{g}, b_u$), $P_e$ decreases as $K$ increases, until it reaches an error floor. The error floor becomes smaller as (i) $\gamma_g$  increases, given $b_u$, (ii) $b_u$  increases, given $\gamma_g$.
The presence of error floor is due to the fact that, for larger $K$ values $P_e$ is 
no longer restricted by $K$, and instead it is restricted by   $\sigma^2_{w}$, leading to an error floor. 
%{\red given $b_u$ what is the effect of increasing $\gamma_g$?} 
%{\blue Given $b_u$ the error floor becomes smaller when $\gamma_g$ increases}
%becomes dominant and leads to an error floor.
% 

{ Fig.~\ref{f9} depicts $P_e$ versus $N$ as  $f_D T_s$ and $P_{\text{d}}$ vary. We observe that, given the pair ($f_D T_s,P_{\text{d}}$), $P_e$ reduces when $N$ increases, however, it reaches an error floor after cretin value of $N$. This is due to the fact that for larger $N$ values, $P_e$ becomes limited by ${\cal P}_{tot}$ and $\sigma_w^2$. Furthermore, we notice that $P_e$ decreases when
%
%(i) given the pair ($f_D T_s,P_{\text{d}_n}$), 
%$N$ increases, but it reaches error floor after cretin value of $N$. 
%
(i) given the pair ($N$, $P_{\text{d}}$), $f_D T_s$ increases; 
%{\red why? if it was $f_D$ only I expected increasing Doppler worsens the error, but this is $f_D T_s$. What if $T_s$ is fixed and $f_D$ increases? }
(ii) given  the pair ($N$, $f_D T_s$),  $P_{\text{d}}$ increases.

Fig. \ref{f10} shows $P_e$ versus SNR$_s$ as $L,N$ change. Examination of this figure shows that $P_e$ reduces when (i) given the pair ($L,N$), SNR$_s$ increases. This is because as SNR$_s$ increases, $P_{f_n}=P_f, \forall n$ in (\ref{pd_pf1}) decreases, (ii) given the pair (SNR$_s$, $L$), $N$ increases, (iii) given  the pair (SNR$_s$, $N$), $L$ increases. This is because as $L$ increases the feedback information from the FC to the sensors on channel gain increases. }

%-------------------------------------------
%--------------------------------------------
%----------------------------------------------------
%--------------------------------------------
%----------------------------------------------------
%--------------------------------------------
%----------------------------------------------------
%--------------------------------------------
%----------------------------------------------------
{ Fig. \ref{f14} shows $P_e$ versus $\eta$ as $\gamma_{g}$ varies. Given  $\gamma_{g}$, $P_e$ decreases in $\eta$. 
This is due to the fact that, as $\eta$ increases, the sub-optimal $\alpha^2_{n,t}$ values increases, leading to a decrease in $P_e$. 
%
%{\red The sensors' transmitted power increases as the discount factor  $\eta$ increases, therefore, $P_e$ decreases. }
%Have you verified this for yourself? or you are just saying a sentence without support?} {\blue this sentence is from Mao paper. I tested it myself too for a small setting.}
We note that there is a performance-computational complexity trade-off as $\eta$ increases. Recall the mean of the network lifetime $T$ is
$\mathbb{E}\{T\}\!=\!1/(1-\eta)$. As $\eta$ increases, 
 the number of
iterations required for the value iteration algorithm  (i.e., step 3 of Algorithm \ref{Alg2}) to converge 
increases.
%{\red which step in Algorithm 2?} {\blue step 3}
% 
%depends on the value of $\eta$. With a larger $\eta$, a larger number of iterations is required.
}
%%%%%%%%%%%%%%%%%%%%%%%%%%%%%%%%%%%%%%%%%%%%%%%%%%%
\par 
$\bullet$ {\bf Effect of random deployment of sensors on $P_e$}:
 We consider a circle field  where  the signal source with power $P_0$ is located at its center.
 Sensors are randomly deployed in the field such that the distance of sensor $n$ from the source, $r_{n}$, is uniformly distributed in the interval of $(r_0,r_1)\!=\!(1\mbox{m},100\mbox{m})$.
 %uniformly , and it is at least $r_0=1$m away from any sensor within the field. 
  We assume the quantization
thresholds are obtained via MMAE.
 Fig.~\ref{f11} illustrates $P_e$ versus $P_0$ as $\gamma_g$ and $b_u$ change. 
 We observe that $P_e$ decreases when: (i) given the pair ($\gamma_g,b_u$), $P_0$ increases,
(ii) given the pair ($P_0$, $b_u$),  $\gamma_g$ increases, (iii) given  the pair $(P_0$, $\gamma_g$),  $b_u$ increases. These observations are all expected. 
%
% We note that as $P_0$ increases, $P_e$ decreases. Also, we notice that, given channel parameter $\gamma_{g_n}$, as $b_u$ increases, $P_e$ reduces as well. 
 %
%\par Fig.~\ref{f12} demonstrates  the behavior of $P_e$  of the optimized system as different parameters change. This figure shows $P_e$ versus $K$ for $P_0=12,~14$ dBmW, $b_u = 2,~4 $ mJ. 
%We notice that,  given $P_0$ and $b_u$, as $K$ increases, $P_e$ reduces, until it reaches an error floor. Increasing $\rho$ any further, does not lower $P_e$. This behavior is similar to Fig.~\ref{f8}, data transmission power $\alpha^2_{n,t}$ is restricted by the amount of harvested energy. Hence, increasing $b_u$ reduces $P_e$. On the other hand, for large $K$ and $b_u$, data transmission power $\alpha^2_{n,t}$ is not limited by energy harvesting any more and it is limited by the communication channel noise.

{ Fig.~\ref{f12} illustrates $P_e$ versus $N$ for fixed and random deployment, as $\rho$ and $K$ vary. 
%To limit the number of curves in this figure, we only consider MMAE-based channel gain quantization thresholds. 
For the given set of parameters, the performance of fixed and random deployments is close to each other. Also, given the pair ($\rho, K$), $P_e$ reduces when $N$ increases, however, it reaches an error floor after $N$ exceeds a certain value. This is due to the fact that for larger $N$ values, $P_e$ becomes limited by ${\cal P}_{tot}$ and $\sigma_w^2$.
Furthermore, we notice that $P_e$ decreases when
%(i) given the pair ($\rho,K$), 
%$N$ increases, but it reaches error floor after cretin value of $N$. 
%
(i) given the pair ($N$, $K$),  $\rho$ increases, (ii)  given  the pair ($N$, $\rho$), $K$ increases.
}
%--------------------------------------------
%----------------------------------------------------
%--------------------------------------------
%\begin{figure}[!t]
%\centering
%\includegraphics[width=70mm]{rand_pe_vs_k.eps}
% \caption{$P_e$ vs. $P_0$ for $N= 5,~\sigma^2_{w_n}\!=\!1,~\forall n,~$.}
%\label{f12}
%\end{figure}
%----------------------------------------------------
%----------------------------------------------------
\section{Conclusions}\label{conclu}
Considering an EH-enabled WSN with $N$ sensors and  a feedback channel from the FC to the sensors, tasked with  binary distributed detection, we developed  adaptive channel-dependent transmit power
control policies  such that the detection performance is optimized, subject to total transmit power constraint.  
%We developed a power control policy for an EH-enabled WSN, that is tasked with solving a binary distributed detection problem. 
Modeling the quantized fading channel, the energy arrival, and the dynamics of the battery as homogeneous FSMCs, and the network lifetime as a random variable with geometric distribution, we formulated our power control optimization problem as a discounted infinite-horizon constrained MDP
problem, where sensors' transmit powers are functions of the battery state, quantized CSI, and the arrived energy.
We developed the optimal policy, using dynamic programming and
utilizing the Lagrangian approach to transform the constrained
MDP problem into an equivalent unconstrained MDP problem. Determining the optimal policy, however, requires the knowledge of the global state at the FC, which imposes a significant signaling overhead to the sensors.
%
%We proposed a centralized optimal solution. The optimal policy, a.k.a. centralized solution in the dynamic control literature, requires the knowledge of the global state $\boldsymbol{s}_t$ to determine the network action $\boldsymbol{\alpha}_t$.
%
To eliminate this overhead, we developed the sub-optimal policy, using a uniform Lagrangian multiplier to transform the constrained MDP problem into $N$ unconstrained MDP problems. Different from the optimal policy, in the sub-optimal policy each sensor sets its transmit power based on its local state information. We showed that the computational complexity of finding the sub-optimal policy scales linearly in $N$ and this policy has a close-to-optimal performance. 
%
%we consider finding a sub-optimal policy, a.k.a. decentralized solution, where sensor n in the network finds its action $\alpha_{n,t}$ at time slot $t$, only based on its own local state $s_{n,t}$.
%
%While the computational complexity of finding the optimal policy scales exponentially in $N$, we showed that the computational complexity of finding the sub-optimal policy scales linearly in $N$. 
%
We studies the error probability $P_e$ in terms of different system parameters, including $K, N, {\cal P}_{tot}$, SNR$_s$. Although $P_e$ decreases as each of these parameters increases, there is an error floor that ultimately depends on the communication channel noise variance and ${\cal P}_{tot}$.
%We also demonstrated that increasing $K$ or $b_u$ do not necessarily lower the detection error, and it depends on the communication channel noise variance . 
%
We expanded our work to random deployment of sensors and examined how it affects the error probability.
%extension to random deployment, for different K, $\rho$ fixed and random are close to each other, this is important from practical perspective. 
%
%sub-optimal has a close-to-optimal performance, with a much lower computational complexity that scales linearly in $N$. what parameters affect the performance more?
%
The insights obtained in this work are useful for adaptive transmit power control of EH-enabled WSNs tasked with distributed detection. 
%%%%%%%%%%%%%%%%%%%%%%%%%%%%%%%%%%%%%%%%%%%%%%%%%%%%%%%%%%%%%%%%%%%%%%%%%%%%%%
%%%%%%%%%%%%%%%%%%%%%%%%%%%%%%
\bibliographystyle{IEEEtran}
\bibliography{RefEH2}
\end{document}